\documentclass[]{aa}
\usepackage{graphicx}
\usepackage{dirtytalk}
\usepackage{currvita}
\usepackage[T1]{fontenc}
\usepackage{txfonts}
\usepackage{hyperref}
\usepackage{xcolor}
\hypersetup{
colorlinks,
linkcolor={red!80!black},
citecolor={blue!80!black},
urlcolor={blue!80!black}
}

\newenvironment{myfont}{\fontfamily{ppl}\selectfont}{\par}
\DeclareTextFontCommand{\textmyfont}{\myfont}

\newcommand{\code}[1]{\texttt{#1}}

\def\nifs{\iso{56}Ni}

\def\cofs{\iso{56}Co}

\def\cm3{cm$^{-3}$}
\def\kms{km\,s$^{-1}$}

\def\msun{$M_{\odot}$}

\def\one{\ts {\,\sc i}}
\def\two{\ts {\,\sc ii}}
\def\three{\ts {\,\sc iii}}
\def\four{\ts {\,\sc iv}}
\def\five{\ts {\sc v}}

\def\beq{\begin{equation}}
\def\eeq{\end{equation}}

\def\lesssim{\mathrel{\hbox{\rlap{\hbox{\lower4pt\hbox{$\sim$}}}\hbox{$<$}}}}
\def\gtrsim{\mathrel{\hbox{\rlap{\hbox{\lower4pt\hbox{$\sim$}}}\hbox{$>$}}}}

\def\one{{\,\sc i}}
\def\two{{\,\sc ii}}
\def\three{{\,\sc iii}}
\def\four{{\,\sc iv}}
\def\five{{\sc v}}

\def\cmfgen{{\code{CMFGEN}}}

\def\sn{SN\,2023ixf}

\def\ekin{$E_{\rm kin}$}
\def\mej{$M_{\rm ej}$}

\def\ergs{erg\,s$^{-1}$}

\def\ha{H$\alpha$}
\def\hb{H$\beta$}
\def\hg{H$\gamma$}
\def\hd{H$\delta$}

\def\pag{Pa\,$\gamma$}

\def\heinir{He\,{\sc i}\,1.083\,$\mu$m}

\def\mgiinir{Mg\two\,$\lambda\lambda$\,$1.091,\,1.095$}
\def\mgiiopt{Mg\two\,$\lambda\lambda$\,$9218,\,9244$}
\def\mgiiuv{Mg\two\,$\lambda\lambda$\,$2795,\,2802$}
\def\lya{Ly\,$\alpha$}
\def\niidoub{[N\two]\,$\lambda\lambda$\,$6548,\,6583$}
\def\niiauroral{[N\two]\,$\lambda$\,$5755$}

\def\kidoub{K\one\,$\lambda\lambda$\,$7665,\,7699$}

\def\caiidoub{[Ca\two]\,$\lambda\lambda$\,$7291,\,7323$}
\def\caiitrip{Ca\two\,$\lambda\lambda\,8498-8662$}

\def\oidoub{[O\one]\,$\lambda\lambda$\,$6300,\,6364$}
\def\oiauroral{[O\one]\,$\lambda$\,$5577$}
\def\oitrip{O\one\,$\lambda\lambda$\,$7771-7775$}

\def\oiidoub{[O\two]\,$\lambda\lambda$\,$7320,\,7330$}

\def\oiiidoub{[O\three]\,$\lambda\lambda$\,$4959,\,5007$}

\def\mgifs{Mg\one]\,0.457\,$\mu$m}

\def\naid{Na\,{\sc i}\,D}

\def\neiifs{[Ne\two]\,12.810\,$\mu$m}

\def\ariimir{[Ar\two]\,6.983\,$\mu$m}

\def\mic{\,$\mu$m}

\def\nkiinir{[Ni\two]\,1.939\,$\mu$m}
\def\nkiimir{[Ni\two]\,6.634\,$\mu$m}
\def\nkiiopt{[Ni\two]\,$\lambda$\,7378}

\def\civuv{C\four\,1550\,\AA}

\newcommand{\iso}[2]{\ensuremath{^{#1}\rm{#2}}}

\begin{document}

\title{SN\,2023ixf: ultraviolet-to-infrared radiative-transfer modeling of the nebular-phase evolution until 1000 days}

\titlerunning{Nebular-phase modeling of SN\,2023ixf}

\author{
Luc Dessart\inst{\ref{inst1},\ref{inst1a}}
\and
Wynn V. Jacobson-Gal\'an\inst{\ref{inst2},\ref{inst3}}
\and
K. Azalee Bostroem\inst{\ref{inst4},\ref{inst5}}
\and
Alexei V. Filippenko\inst{\ref{inst6}}
\and
WeiKang Zheng\inst{\ref{inst6}}
\and
Thomas G. Brink\inst{\ref{inst6}}
\and
Llu\'is Galbany\inst{\ref{inst7},\ref{inst8}}
\and
Claudia Guti\'errez\inst{\ref{inst7},\ref{inst8}}
\and
Stefano Valenti\inst{\ref{inst9}}
 }

\institute{
    Institut d'Astrophysique de Paris, CNRS-Sorbonne Universit\'e, 98 bis boulevard Arago, F-75014 Paris, France\label{inst1}
\and
    French-Chilean Laboratory for Astronomy, IRL 3386, CNRS and Instituto de Astrofísica,
       Pontificia Universidad Católica de Chile, Casilla 306, Santiago, Chile.\label{inst1a}
\and
  Cahill Center for Astrophysics, California Institute of Technology, Pasadena, CA 91125, USA\label{inst2}
  \and
  NASA Hubble Fellow\label{inst3}
\and
Steward Observatory, University of Arizona, 933 North Cherry Avenue, Tucson, AZ 85721-0065, USA\label{inst4}
\and
LSST-DA Catalyst Fellow\label{inst5}
\and
Department of Astronomy, University of California, Berkeley, CA 94720-3411, USA\label{inst6}
\and
Institute of Space Sciences (ICE, CSIC), Campus UAB, Carrer de Can Magrans s/n, 08193, Barcelona, Spain\label{inst7}
\and
Institut d'Estudis Espacials de Catalunya (IEEC), 08860, Castelldefels, (Barcelona), Spain\label{inst8}
\and
Department of Physics and Astronomy, University of California, Davis, 1 Shields Avenue, Davis, CA 95616-5270, USA\label{inst9}
}

   \date{}

   \abstract{We present non-local thermodynamic equilibrium radiative-transfer modeling of \sn\ during the nebular phase out to 1000\,d, using the same ejecta that matched its photospheric evolution, namely a partially stripped red-supergiant star of initially 15\,\msun\ whose terminal explosion yielded ejecta with 7--8\,\msun, kinetic energy of $1.2\,\times\,10^{51}$\,erg, and \nifs\ mass of 0.05\,\msun, augmented with a cold dense shell (CDS) of 0.2\,\msun\ at 8000\,\kms. Interaction with circumstellar material persists at all epochs, powering the ultraviolet (UV) flux at all times, but dominating the optical only after $\sim$\,600\,d. Matching the $V$-band light curve requires invoking both enhanced $\gamma$-ray escape and dust formation after $\sim$\,200\,d, first in the CDS and eventually in the inner ejecta as well. Depending on where they form relative to the dust, emission lines are uniformly attenuated or skewed with a blue-red asymmetry. Our models suggest a rising dust mass (chosen as an C-rich and Si-rich mixture) in the CDS and inner ejecta, possibly reaching $10^{-4}$\,\msun\ at 700\,d, while an external cold dust component is required to match the mid-infrared emission. The UV radiation, largely unaffected by dust, is influenced by the emission and absorption from Fe lines, together with strong, blueshifted emission from \lya\ and \mgiiuv, both present at $\gtrsim$\,200\,d and with a strengthening fractional flux thereafter. Optical-depth effects play a critical role for the UV flux, and most notably on \lya\ whose strength depends strongly on the CDS structure (mass and extent) and the treatment of power injection. The CDS is continuously slowing down from 8000\,\kms\ at 112\,d to $\sim$\,6500\,\kms\ at 998\,d, suggesting a growth in mass of several 0.1\,\msun. \sn\ shares many similarities with SN\,1993J at 1--3\,yr, but it is eventually fainter due to dust extinction and cooler (i.e., weak [N\two] and no [O\three] lines) likely as a result of greater CDS and ejecta masses.
}


   \maketitle


\section{Introduction}
\label{sect_intro}

\sn\ is a nearby, H-rich (Type II) supernova (SN) that showed early signs of interaction with circumstellar material (CSM). Its proximity at 6.85\,Mpc, and its discovery \citep{itagaki_23} as well as classification \citep{perley_23ixf_23} within hours of shock breakout, have made possible an intense study across the electromagnetic spectrum, from X-rays \citep{grefenstette_23ixf_23,chandra_23ixf_24,nayana_23ixf_25} to ultraviolet \citep[UV;][]{teja_23ixf_23,bostroem_23ixf_24,zimmerman_23ixf_24,bostroem_uv_25}, optical (see, e.g., \citealt{bostroem_23ixf_23}; \citealt{jacobson_galan_23ixf_23}; \citealt{zheng_23ixf_25}), near-infrared (NIR; \citealt{park_23ixf_25,derkacy_23ixf_26}), mid-infrared (MIR; \citealt{derkacy_23ixf_26}), and radio ranges \citep{iwata_23ixf_25,nayana_23ixf_25}, including high-resolution spectroscopy \citep{smith_23ixf_23,dickinson_23ixf_25} and spectropolarimetry \citep{vasylyev_23ixf_23,singh_23ixf_24,shrestha_23ixf_25,vasylyev_23ixf_26}. These data cover from the earliest times following first light through the photospheric and nebular phases, and continue today as \sn\ evolves into a young SN remnant. \sn\ is an extraordinary transient in which a full coverage of the electromagnetic spectrum can be, and has been, obtained \citep{wynn_sed_25}.

Pre-explosion imaging has set some constraints on the progenitor star. There is a broad consensus that the progenitor is a luminous, dusty red supergiant (RSG) with an abnormally large wind mass-loss rate and some modest level of photometric variability \citep{dong_23ixf_23,jencson_23ixf_23,kilpatrick_23ixf_23,niu_23ixf_23,soraisam_23ixf_23,neustadt_23ixf_24,qin_23ixf_24,ransome_23ixf_24,vandyk_23ixf_24b,vandyk_23ixf_24}. There is, however, much disparity in the inferred stellar luminosity and thus progenitor mass, with values covering most of the available range for RSG stars, from 11\,\msun\ \citep{kilpatrick_23ixf_23} to 20\,\msun\ \citep{soraisam_23ixf_23}.

Estimates of the progenitor mass of \sn\ have also resulted from the inference of the ejecta mass obtained with radiation-hydrodynamics simulations and comparison with the bolometric or multiband light curves, combined in some (though not all) cases with important constraints provided by the evolving photospheric velocity. Here, too, there is much disparity in the results. To reproduce the shorter-than-standard photospheric-phase duration of \sn, some studies require a partially stripped progenitor \citep{fang_23ixf_25,forde_23ixf_25,hsu_23ixf_25,kozyreva_23ixf_25,zheng_23ixf_25}, but others invoke a standard RSG progenitor \citep{bersten_23ixf_24,moriya_23ixf_24,singh_23ixf_24,vinko_23ixf_25} though in some cases with a higher-than-standard explosion energy. This scatter may reflect differences in the methods (gray versus multigroup, flux-limited diffusion or solution of the moment equations, local thermodynamic equilibrium (LTE) versus nonLTE, opacity floors, etc.), but it is also deeply rooted in the fundamental degeneracies of SN light curves \citep{d19_sn2p,goldberg_sn2p_19}. Such degeneracies are further aggravated when the constraint from the photospheric velocity is ignored (i.e., when photometry is the only constraint used). Even if there were a consensus, inferences from light-curve modeling are compromised by the assumptions made on the exploding star, such as assuming a progenitor in hydrostatic equilibrium (see, e.g., \citealt{bronner_rsg_25,laplace_23ixf_26}), a spherically symmetric progenitor (see, e.g., \citealt{goldberg_3d_rsg_22,goldberg_sbo_22,ma_rsg_25}), or ignoring ejecta clumping \citep{d18_fcl,dessart_audit_rhd_3d_19}.

Archival nebular-phase models have been compared with observations of \sn\ at 200--500\,d and suggest a 12--15\,\msun\ progenitor \citep{ferrari_23ixf_24,folatelli_23ixf_25,kumar_23ixf_25,wynn_iii_25,wynn_sed_25}. However, these studies relied on standard RSG progenitors and typically ignored the potential influence of interaction power or dust. Here, we cure these deficiencies by using as initial conditions for our nebular-phase modeling the tailored model x6p0 presented by \citet{dessart_23ixf_phot_26} and found suitable to describe the photospheric evolution of \sn\ from 20 to 115\,d.

This paper is organized as follows. In the next section, we summarize the observational dataset used here, covering the UV, optical, NIR, and MIR from 112 to 998\,d. In Section~\ref{sect_setup}, we present the numerical setup for the radiative-transfer calculations with \cmfgen\ \citep{HD12,dessart_csm_22}. This is done in a succinct manner since the method is identical to that used by \citet{dessart_23ixf_phot_26} --- the principles were presented by \citet{dessart_shuffle_20} and \citet{dessart_csm_22} --- apart from the additional treatment of dust with the methodology described by \citet{dessart_dust_25}. Section~\ref{sect_pwr} discusses the various approaches of treating the interaction power and the impact on SN properties. In Section~\ref{sect_rt}, we present the main results from our radiative-transfer modeling of \sn\ from 112 to 998\,d after explosion, including constraints from across the electromagnetic spectrum. We first discuss epochs prior to 200\,d when dust seems to have no impact on the SN optical radiation, and then subsequent epochs, one at a time, during which dust in the inner ejecta or in the cold dense shell (CDS) is required. We summarize in Section~\ref{sect_lc} the results of the previous section by comparing the corresponding model $V$-band light curve to that observed for \sn. Numerous sections complement these core results to provide insights on the processes and dependencies. Section~\ref{sect_form} discusses how the relative importance of decay power and interaction power at different SN ages impact the SN spectrum. We then describe in more detail in Section~\ref{sect_ir} the IR properties of \sn, not so much in relation to emission by dust but more in connection to metal-line emission from the inner ejecta of \sn. Comparisons are made to models with different ejecta or initial mass. Section~\ref{sect_uv} explores the UV properties of \sn\ and the successes and failures of the model in matching them. Comparison to alternative models including that of \citet{dessart_late_23} is used to gauge the sensitivity of results to model assumptions. Section~\ref{sect_dust} summarizes the results for the dust properties inferred from the models presented in Section~\ref{sect_rt}. The following two Sections~\ref{sect_ha} and \ref{sect_hei} zoom-in on the spectral evolution in the \ha\ and \heinir\ regions. Section~\ref{sect_dep} explores additional sensitivities of model results to assumptions, focusing now on the CDS composition or mass.
Section~\ref{sect_93J} presents a comparison of SN\,1993J and \sn, at nebular times, highlighting the similarities of their spectra despite their distinct SN types. We present our conclusions in Section~\ref{sect_conc}. Further information is provided in the Appendix.

\section{Observations}
\label{sect_obs}

Essentially all the observations used in this study have been published in past works. Specifically, the optical spectra and photometry are from \citet{zheng_23ixf_25} and \citet{wynn_sed_25}, as well as augmented with a spectrum taken at 998\,d with the Gran Telescopio Canarias (see details in Appendix~\ref{sect_gtc_data}). The UV data are a little scattered in time but can be grouped around epochs of about 200, 310, 619, and 723\,d, and are from \citet{bostroem_uv_25}. The NIR data are from \citet{park_23ixf_25} at 199\,d and from \citet{wynn_sed_25} at 260, 310, 370, 658, and 693\,d. Finally, the IR data covering the NIR and MIR ranges with the James Webb Space Telescope (JWST) are from \citet{medler_23ixf_25} at epochs 253, 374, 600, 720\,d. Full details on these observations are to be found in those references. A summary of the epochs used for the modeling and the sources of these data is given in Table~\ref{tab_obs}. For simplicity, we named each epoch that we modeled by the date at which the optical data were taken since the optical range contains a large fraction of the total SN flux (though progressively less true as time advances).

The optical spectra obtained by \citet{zheng_23ixf_25} were flux calibrated using the optical photometry. When multiple datasets are used for the same epoch, some flux mismatches may occur. Part of the offset may arise from the different post-explosion times, which can be up to a few weeks in some cases. The spectral evolution is, however, slow at those nebular epochs (as can be seen, for example, with the small change between the observations at 620 and 708\,d; \citealt{wynn_sed_25}), so these time mismatches are not a concern. Whenever flux mismatches are encountered, we applied a global scaling to the flux level to match that obtained by the Hubble Space Telescope (HST) or JWST, or to the (flux-calibrated) optical spectra of \citet{zheng_23ixf_25}.

Throughout the work, we adopted for \sn\ a distance of 6.85\,Mpc \citep{riess_h0_22}, a redshift of $z = 0.000804$ \citep{perley_23ixf_23}, a total visual extinction $A_V=$\,0.127\,mag ($E(B-V)=$\,0.041\,mag; \citealt{jacobson_galan_23ixf_23}), and a time of first light at MJD $=$\,60082.788 \citep{li_23ixf_24}. To avoid offsets in resolution between different datasets and between observations and models (the latter have a huge resolution of about 5\,\kms), all spectra were rebinned to a fixed resolution of 300\,\kms. This also helps in spectral regions where the data are very noisy (e.g., away from strong lines in the UV). When comparing with the observed IR line profiles of \sn, we rebinned the model to degrade it to the low resolution of about 100 (i.e., $\sim$\,3000\,\kms) of the MIRI observations.

\begin{table}
\caption{Spectral ranges, observational epochs, and sources for the \sn\ data that were modeled here (see Sec.~\ref{sect_obs} for details).}
\label{tab_obs}
\begin{center}
\begin{tabular}{l@{\hspace{2mm}}|c@{\hspace{2mm}}|c@{\hspace{2mm}}}
\hline
Range   &     Days since first light        & Source   \\
\hline
UV      &  183-214, 308-314, 619, 723-724   & [1]	 \\
optical &  112, 147, 175, 208, 265, 300     & [2] \\
optical &  329, 379, 442, 620               & [2] \\
optical &  708                              & [3]  \\
optical &  998                              & This work  \\
NIR     &  199                              & [4]  \\
NIR     &  260, 310, 370, 658, 693          & [3]     \\
 IR     &  253, 374, 600, 720               & [5] \\
\hline
\end{tabular}
\end{center}
{\bf Sources:} [1]: \citet{bostroem_uv_25}; [2]: \citet{zheng_23ixf_25}; [3]: \citet{wynn_sed_25}; [4]: \citet{park_23ixf_25};  [5]: \citet{medler_23ixf_25}.
\end{table}

\begin{figure}
\centering
\includegraphics[width=0.9\hsize]{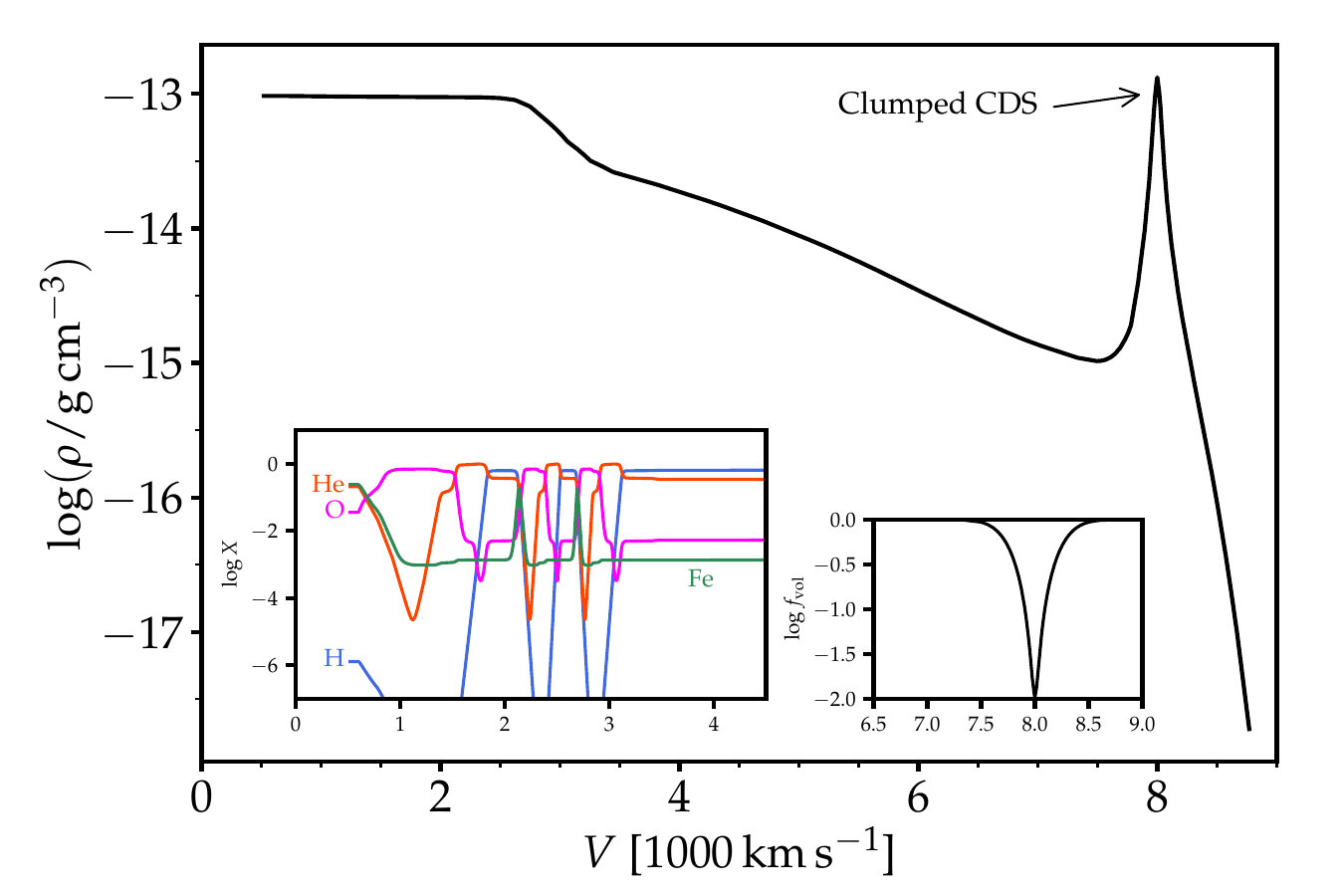}
\caption{Profiles of the density (main panel), composition (left inset), and volume filling factor (right inset) versus velocity at 112\,d for the ejecta model x6p0 used in this work.
\label{fig_init}
}
\end{figure}

\section{Numerical setup}
\label{sect_setup}

The approach we use is for the most part identical to that of \citet{dessart_23ixf_phot_26} on the photospheric-phase modeling of \sn, but here the focus is on the later evolution out to 1000\,d. Because at nebular times the influence of interaction power is significant (e.g., at 120\,d) and eventually dominates (at $\gtrsim$\,600\,d) over decay power, only model x6p0 with a CDS and interaction power is discussed here. We used the same progenitor and explosion model as presented by \citet{dessart_23ixf_phot_26}, and modeled the influence of interaction with CSM as discussed by \citet{dessart_23ixf_phot_26} and in more detail by \citet{dessart_csm_22} and \citet{dessart_late_23} --- further discussion on the treatment of shock power is provided in the next section. As a reminder, our model x6p0 corresponds to a 15\,\msun\ star that evolved with an enhanced wind mass-loss rate during the RSG phase \citep{HD19}. It  reached core collapse with a total mass of 9.09\,\msun\ and an H-rich envelope mass of 5.04\,\msun. Its explosion produced a $\sim$\,7.5\,\msun\ ejecta with a kinetic energy of $1.2 \times 10^{51}$\,erg and  $\sim$\,0.05\,\msun\ of \nifs\ \citep{dessart_23ixf_phot_26}. This ejecta model was then augmented with a dense shell at 8000\,\kms, as suggested by observations \citep{jacobson_galan_23ixf_23}. Simulations of the SN~IIn phase suggest a dense shell of order 0.1\,\msun\ --- we chose 0.2\,\msun\ \citep{d17_13fs,dessart_wynn_23}, with a maximum clumping factor of 100 as in \citet{dessart_csm_22}. The composition of this ejecta model was imported from the 15.2\,\msun\ model of \citet{sukhbold_ccsn_16} in order to benefit from a very detailed description of all isotopes up to nickel. A shuffled-shell technique was used to introduce macroscopic mixing without microscopic mixing, following the procedure presented by \citet{dessart_shuffle_20}. While the method seems coarse, it captures the essence of the physics of nebular-phase spectra of core-collapse SNe \citep{dessart_sn2p_21,dessart_snibc_21}, and will prove its worth here again for \sn. A summary of some of these initial conditions is shown in Figure~\ref{fig_init} (the shuffled-shell composition is illustrated in the lower-left inset).

In the radiative-transfer calculations with \cmfgen, we accounted for five two-step decay chains whose parent isotope is \nifs, \iso{57}Ni, \iso{52}Fe, \iso{48}Cr, and \iso{44}Ti, together with a proper description of all isotopes associated with a given species. Molecules are ignored in the calculation so we make no prediction on the IR emission --- for example, for the CO first overtone and fundamental. Finally, a careful gridding was employed to finely resolve the CDS, generally assumed to be clumped, by allocating $\sim 70$ points. About 200 points were used for the underlying ejecta and in particular to resolve the rapid variations in composition in the inner, metal-rich regions. We used the same model atom as presented by \citet{dessart_23ixf_phot_26}, which, unlike in \citet{dessart_csm_22}, include higher-ionization stages for all species and in particular stages \three --\four\ for C, N, O, Ne, Mg, Si, and S (the model atom does not include N\five\ and thus cannot make predictions for the associated resonance transition at 1240\,\AA\ and its suspected presence in \sn; \citealt{bostroem_uv_25}). These various requirements imply a very large computational burden.

\begin{figure}[h!]
\centering
\includegraphics[width=\hsize]{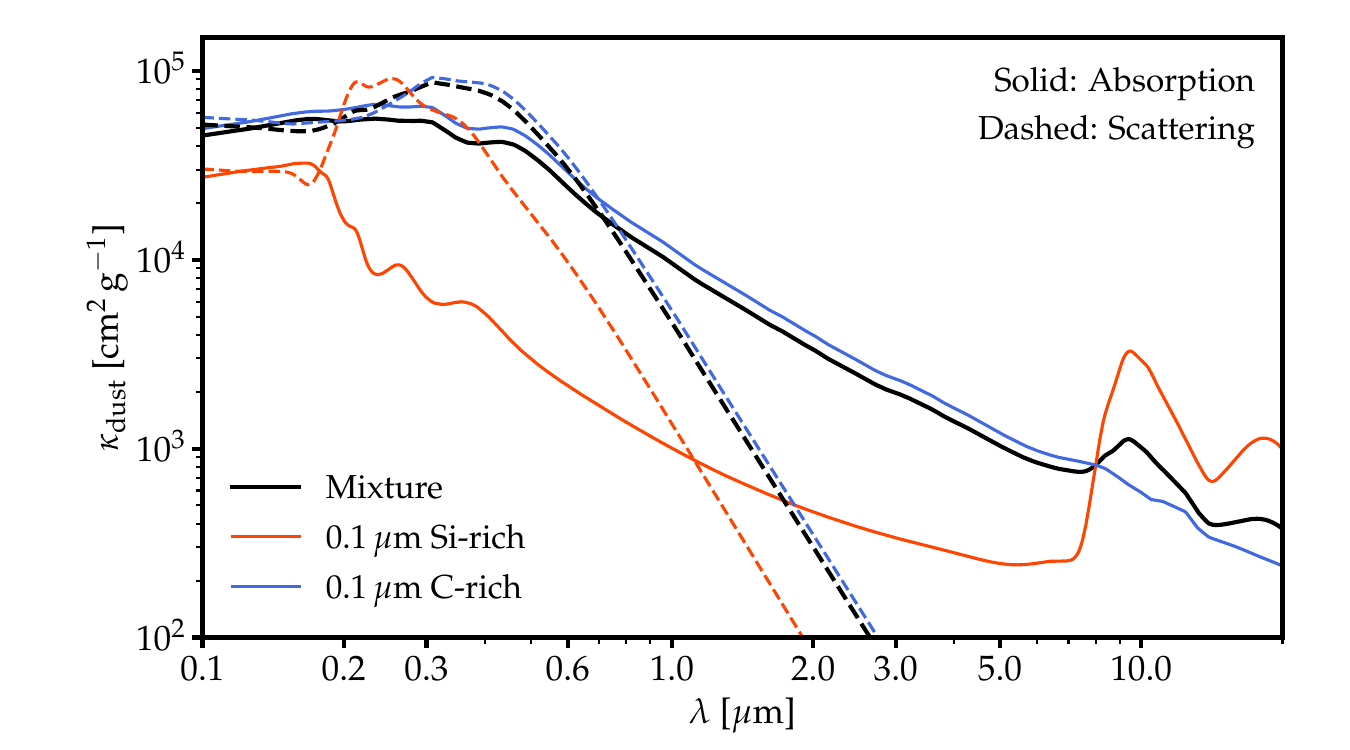}
\caption{Comparison between the dust absorptive (solid) and scattering (dashed) opacities for 0.1\mic\ Si-rich (red) and 0.1\mic\ C-rich (blue) grains, and the mixture of both with proportions 1:4.5 that we adopt for \sn\ (black).
\label{fig_dust}
}
\end{figure}

An important component of this work is the treatment of dust. Because we used the same ejecta model as \citet{dessart_23ixf_phot_26} in which a good match to spectra and light curves was obtained throughout the photospheric phase and the onset of the nebular phase, this model should continue to match observations at late times unless the injected interaction power changes (reflecting changes in CSM density) or if the formation of molecules and dust kicks in. And indeed, starting at 200\,d and thus when radiation emanating from the decay-powered part of the ejecta still dominates the optical brightness, the model starts to be overly bright relative to \sn. This occurs despite $\gamma$-ray escape, which is present at 200\,d in the model and increases with time.

We thus introduced dust in all radiative-transfer calculations of \sn\ starting at 200\,d. We used the same formalism as in \citet{dessart_dust_25}. Practically, this still requires performing the simulations with \cmfgen\ without dust first, and then introducing in a post-processing step a dust component, specifying the spatial distributions, type, and temperature. For this work, the code was augmented to allow for the presence of dust in more than one location, so that we could invoke dust formation both in the inner ejecta (e.g., below 3000\,\kms) and within the CDS, each with its own temperature. The same dust type is, however, used for both, which here is made of a mixture of Si-rich and C-rich dust in order to match the weak 10\mic\ silicate emission bump observed in \sn\ \citep{medler_23ixf_25,wynn_sed_25} ---  opacities for these dust grains are adopted from \citet{sarangi_dust_22}, and originally from \citet{draine_li_07}, \citet{weingartner_draine_01} and \citet{zubko_dust_04}. This was achieved by combining the opacities of each dust type with a proportion 1:4.5 (Fig.~\ref{fig_dust}). We adopted 0.1\mic\ dust grains to limit the dust absorption to wavelengths below 1\,\mic\ (with silicates only, a larger grain size of perhaps 1\mic\ would be needed to achieve a similar effect). In this work, we constrain the dust present in the ejecta and in the CDS primarily through its absorptive (and scattering) influence on the spectral energy distribution, thus considering the impact of dust on all line profiles between the UV and $\sim 1$\mic. Given the complexity of the modeling, we do not claim we have a robust constraint on the dust-grain size, nor on the specific chemistry, but the broad overall match of our models to the observations gives some level of confidence in the results.

Here, we support the interpretation of \citet{singh_23ixf_26} for the need of an external, echo-like, cool dust emission component, which essentially contributes emission in the IR with no impact on shorter wavelengths. Our radiative-transfer calculations account for dust in the ejecta (typically located in regions below 3000\,\kms) and in the CDS (around 8000\,\kms), each with a specific temperature. In general, this internal dust is insufficient to reproduce the brightness of \sn\ at 10\mic\ and beyond, and thus, at the plotting stage, we introduce an external cool, dust emission component.


\begin{figure*}[h!]
\centering
\includegraphics[width=\hsize]{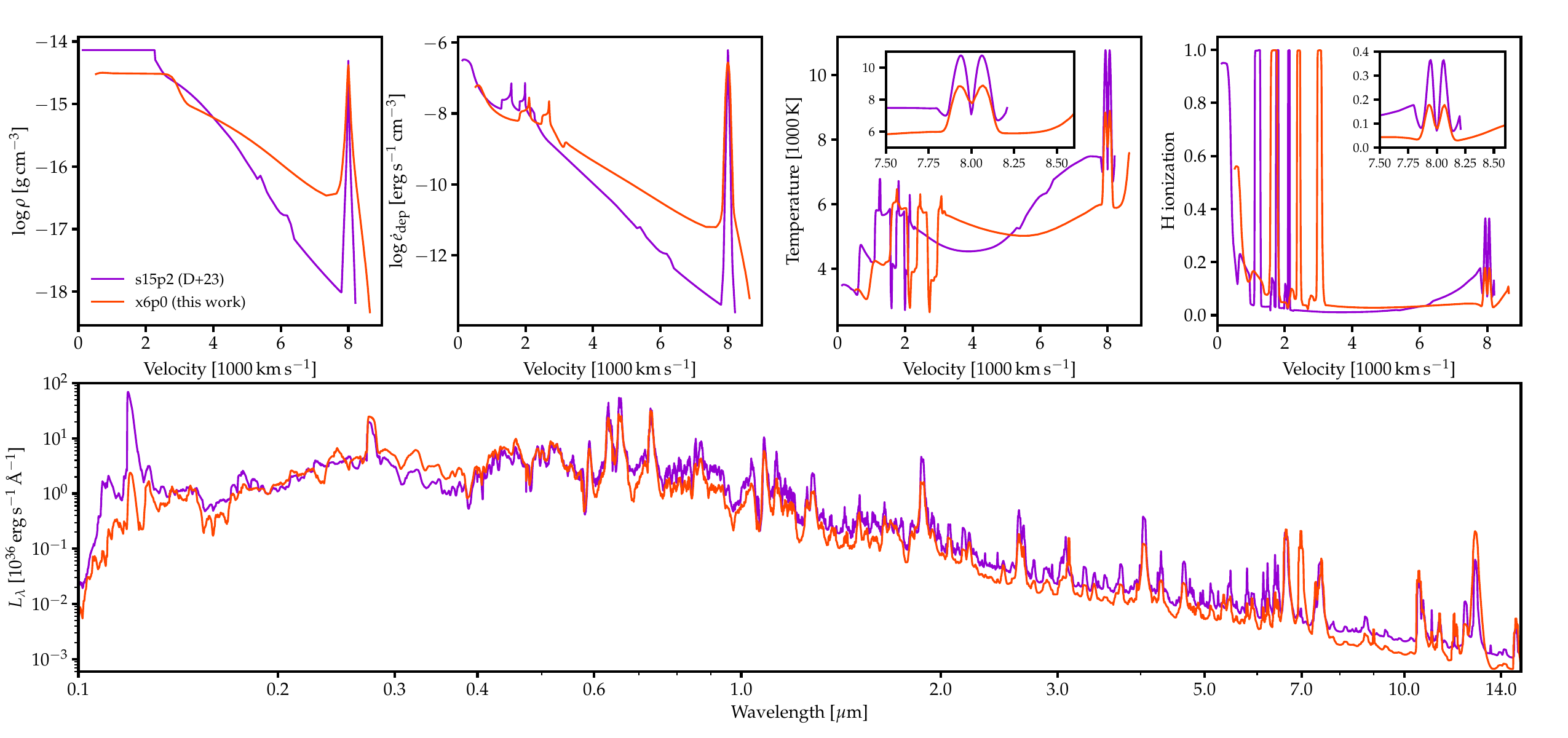}
    \caption{Comparison of gas and radiation properties for the interacting s15p2 model from \citet{dessart_late_23} at 350\,d (violet) and the interacting model x6p0 used in this work for \sn\ at 320\,d (red). Both models are subject to the same interaction power of 10$^{40}$\,\ergs, but differ in \ekin/\mej. From left to right and top to bottom, we show the mass density (uncorrected for clumping), the total energy deposition from radioactive decay and injected shock power (uncorrected for clumping), the gas temperature, the H ionization (zero corresponds to neutral H), and finally the luminosity spectrum.
\label{fig_comp_x6p0_d23}
}
\end{figure*}

\section{Discussion on the treatment of shock power}
\label{sect_pwr}

As written in the preceding section, we follow the basic ansatz of \citet{dessart_csm_22} for the treatment of CSM interaction in \cmfgen\ when modeling \sn. The key aspect is that hydrodynamics is ignored for the benefit of treating the full radiative-transfer problem in nonLTE. This is a major gain since the gas is entirely out of LTE, cools through strong lines, may be optically thick at X-ray or UV wavelengths but thin in the IR, or may be optically thick in lines but thin in the continuum. No such complexity can be treated within a hydrodynamics code, not even in one dimension (1D). We thus ignore the dynamical evolution of the CDS (i.e., change in time of its velocity, mass, clumping, etc.). The original goal of this method was to study standard SNe II from standard RSG progenitors like SN\,2017eaw \citep{weil_17eaw_20}, which we did \citep{dessart_late_23}, but the success of the method warrants extending its use to more extreme events with CSM interaction such as \sn. The interaction with CSM in this approach is thus limited to injected shock power into the \cmfgen\ grid at the interface between ejecta and CSM. There are, however, complications with this aspect, and two methods are used here.

The first approach, documented by \citet{dessart_csm_22}, is by injecting high-energy particles (i.e., electrons) directly within the CDS and with a prescribed profile. This approach focuses on the fractional shock power absorbed (i.e., thermalized) within the CDS rather than escaping. It is practically convenient since the process can be treated in \cmfgen\ in the same way as we treat the high-energy electrons that scattered with $\gamma$-rays arising from radioactive decay. The main advantage of this option is that it focuses on the dense material that absorbs the bulk of the shock power, and hence should yield a good prediction of the flux contribution from the colder gas taking place in the interaction. By design, it limits the production of highly ionized material (a daunting task in radiative transfer since that requires handling many ionization stages for all species, with associated energy levels, and treating accurately the presence of ionization jumps, with notorious convergence issues). The disadvantage is that the energy deposition profile is assumed. As we illustrate in Section~\ref{sect_dep}, changing the deposition profile alters the results, which is not surprising, and so does changing the density profile (and clumping) of the CDS. Significant improvement in this sector will require high-resolution multidimensional hydrodynamics simulations, a major endeavor deferred to a forthcoming study. By default, the \cmfgen\ simulations in this work that used this option employed a Gaussian deposition profile centered on 8000\,\kms\ and with a standard deviation of 35\,\kms\ (see Sec.~\ref{sect_dep} for alternate choices).

\begin{figure*}
\centering
\includegraphics[width=\hsize]{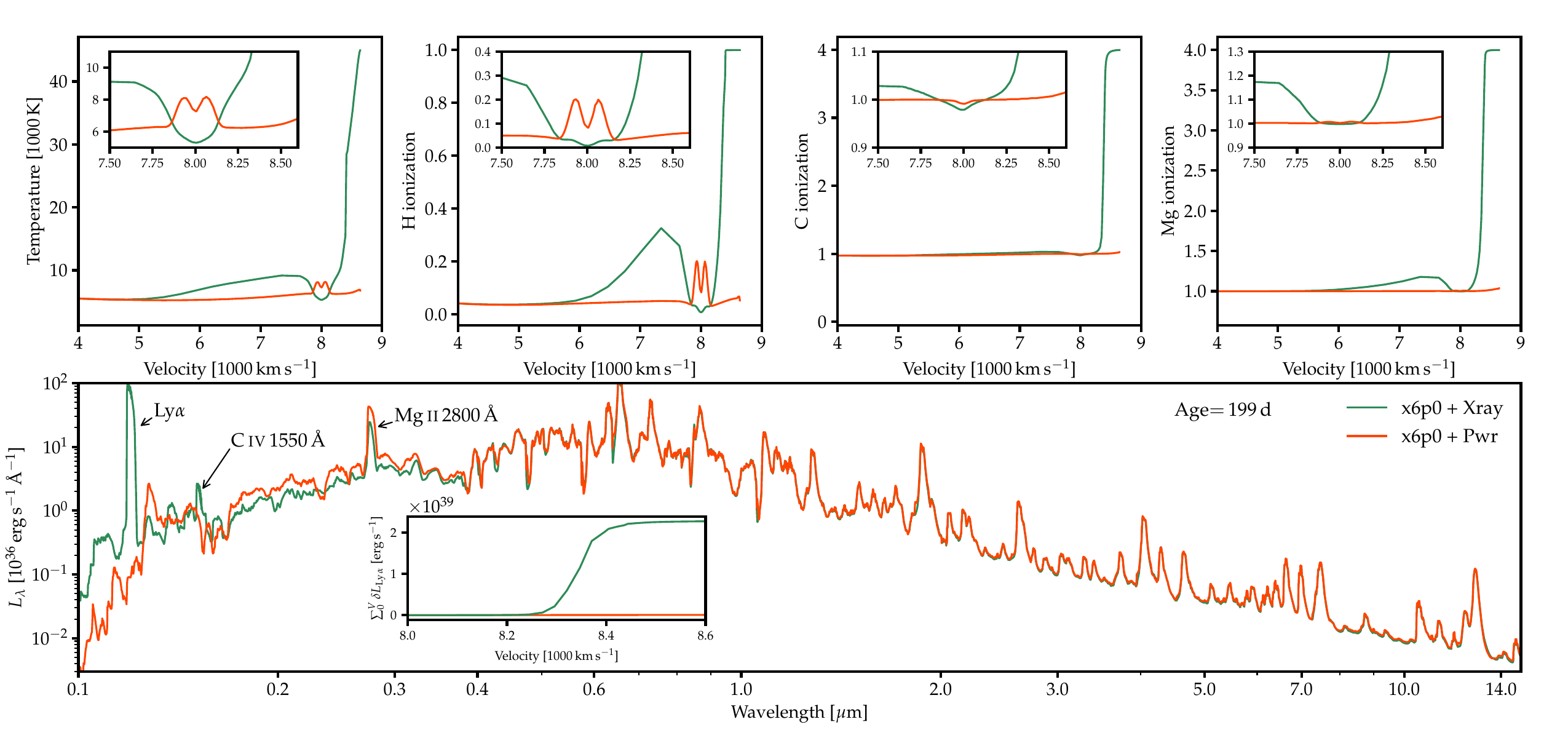}
    \caption{Comparison of gas and radiation properties for the tailored model x6p0 with an interaction power of 10$^{40}$\,\ergs\ but injected as high-energy particles within the CDS at 8000\,\kms\ (red) or in the form of X-rays throughout the outer ejecta and CDS regions (green). From left to right and top to bottom, we show the gas temperature, the H, C, and Mg ionization (zero corresponds to a neutral state), and finally the luminosity spectrum. The inset in the bottom panel gives the cumulative sum of the \lya\ luminosity as a function of velocity, from the innermost ejecta layers to the maximum velocity at 8600\,\kms.
\label{fig_xray}
}
\end{figure*}

The other option, not attempted by \citet{dessart_csm_22} nor \citet{dessart_late_23} in this context, is to inject the shock power as X-rays (the method is presented in the Appendix of \citet{dessart_pm_24}, as well as \citealt{HM98}). The main advantage then is that \cmfgen\ solves the radiative transfer for these X-rays just like for the rest of the radiation on the grid and thus X-ray photons can and do travel, as controlled by optical-depth effects, etc. The downside is the need to inject these X-rays at the reverse and forward shocks, which are by nature infinitely thin layers impossible to resolve on a radiative-transfer grid. In that case, we have approached the problem by simply injecting X-rays as an emissivity source with a broad profile that straddles the CDS and is given by $\exp(-(v_0/v)^3$, with $v_0$ set to 8400\,\kms. The volume integral of that X-ray emissivity is then scaled to match the desired injected power. Other choices were sporadically tested, for example with narrow Gaussian profiles for the X-ray emissivity specifically centered on the location of the forward and reverse shocks. Being a stretch on an already long endeavor (and paper), we kept this latter aspect largely untouched for now.

Before embarking onto a proper description of the modeling of \sn, we present two different model comparisons that highlight the complexity of modeling transients like \sn. Figure~\ref{fig_comp_x6p0_d23} shows a comparison of the present model x6p0 (ejecta mass \mej\ is 7.5\,\msun, ejecta kinetic energy \ekin\ is $1.2 \times 10^{51}$\,erg, \nifs\ mass is 0.05\,\msun, and CDS mass is 0.2\,\msun) with the model s15p2 (\mej\ is 11.49\,\msun, \ekin\ is 10$^{51}$\,erg, \nifs\ mass is 0.06\,\msun, and CDS mass is 0.1\,\msun) of \citet{dessart_late_23}, both influenced by an interaction power of 10$^{40}$\,\ergs\ injected within the CDS at 8000\,\kms\ --- the SN age is $\sim$\,350\,d. Although the two models have a comparable \nifs\ mass, the larger \ekin/\mej\ in model x6p0 leads to a larger expansion rate, smaller density, greater $\gamma$-ray escape (which explains the distinct profiles of the decay power absorbed in the inner ejecta regions), and consequently smaller flux from the decay-powered part of the ejecta (i.e., the optical and the IR) and broader lines (e.g., \oidoub\ or \neiifs). But the two models differ in more subtle ways in the UV, wherein the bulk of the interaction power is eventually channeled after being reprocessed by the ejecta and CDS. In the new x6p0 model, the UV spectrum shows stronger \mgiiuv\ emission and a faster drop toward the far-UV with essentially nonexistent \lya\ emission. In contrast, the corresponding model s15p2 shows weaker \mgiiuv\ emission but strong \lya\ emission (in fact, the strongest emission line in the entire spectrum). This difference arises from a combination of effects, related to the different density structure in both ejecta and CDS (top-left panel of Fig.~\ref{fig_comp_x6p0_d23}), the smaller model atom, and the smaller maximum radius used by \citet{dessart_late_23}. These alter the temperature, ionization, and emissivity within the CDS, as well as the blanketing external to it. We provide more details about these aspects in Section~\ref{sect_dep}.

We also explored how different the results would be if we were to inject the power in the form of X-rays. The method was succinctly described above (for a more detailed presentation, see \citealt{dessart_pm_24}). Figure~\ref{fig_xray} shows the results, this time focusing on the temperature and ionization (top row) and the spectrum from UV to IR (bottom row). Because this exploration is based on the same ejecta model x6p0 (i.e., same \nifs\ mass, etc.), the optical and IR fluxes are unchanged. But the UV part of the spectrum is dramatically different, with the production of strong \lya, the presence of \civuv, and somewhat weaker \mgiiuv\ in the model with X-rays. With the approach using X-rays, the properties within the CDS are hardly changed, but both below and above the CDS, the temperature and ionization are boosted, in particular in the low-density region at the largest velocities. This fosters both emission from high-ionization lines (e.g., \civuv), but also shuts off any blanketing above the CDS, facilitating the escape of \lya\ and far-UV photons. The inset also indicates that the \lya\ luminosity arises from the outer parts of the CDS, rising from zero at about 8300\,\kms\ --- \lya\ photons emitted deeper are absorbed within the CDS.

These two tests give some measure of the sensitivity of our results to the adopted procedure for the treatment of shock power. This should be kept in mind when assessing the quality of the fits in this study, in particular for diagnostics within the UV range. In the model comparisons to observations, we will thus show the results for different model assumptions. Irrespective of the treatment and its limitations, interaction with CSM must be present in the models in order to capture the UV properties of \sn\ throughout its evolution.


\section{Multiepoch radiative-transfer modeling of \sn\ in the nebular phase}
\label{sect_rt}

This section presents the results of our radiative-transfer modeling of \sn\ from 112 until 998\,d. Our simulations are based on a shuffled-shell ejecta structure for the model x6p0 of \citet{dessart_23ixf_phot_26} (which matches well the photospheric-phase evolution of \sn) with a varying interaction power, either introduced in the form of high-energy electrons within the CDS or X-rays in and around the CDS. We grouped together the epochs prior to 200\,d (for which optical spectra display little evolution and no evidence for dust) and the subsequent epochs, which also include UV or IR data, sometimes both. The following sections explore in more detail the characteristics of these simulations and what they tell us about the properties of \sn.

\begin{figure}[h!]
\centering
\includegraphics[width=\hsize]{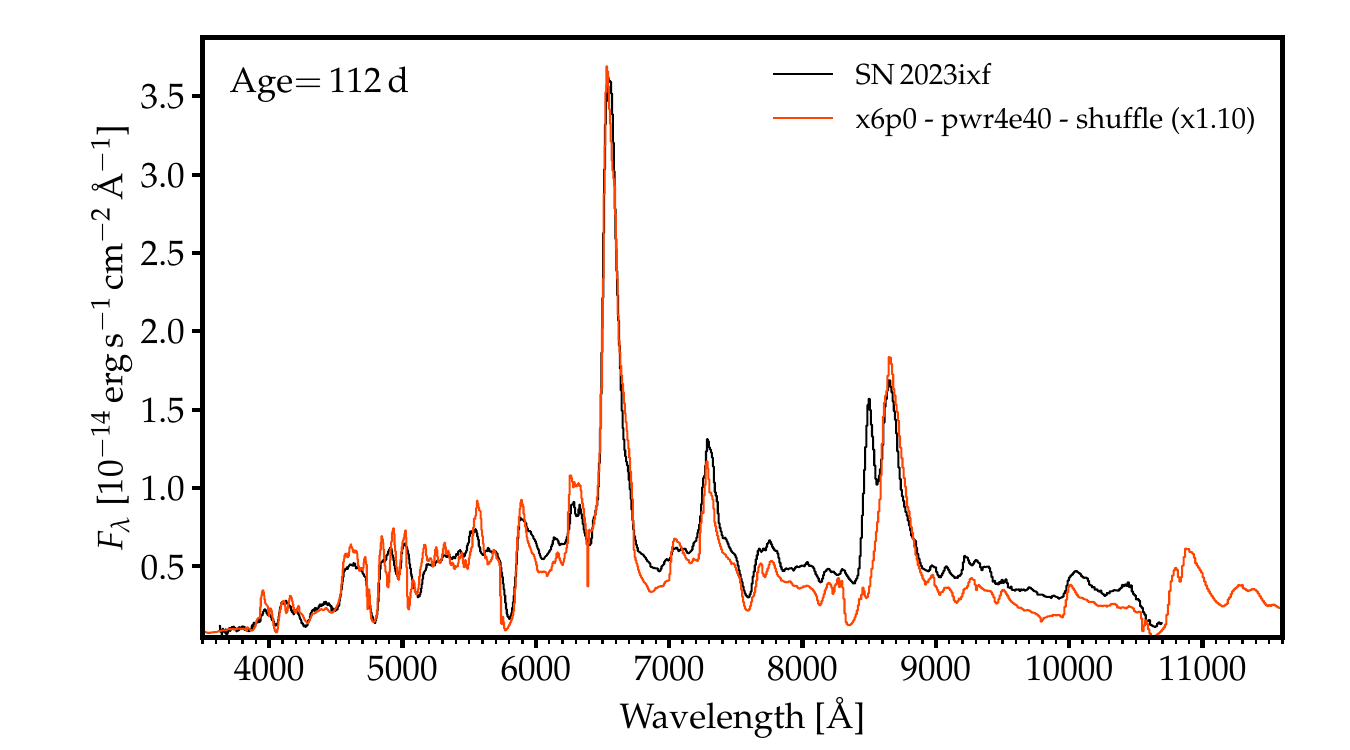}
\includegraphics[width=\hsize]{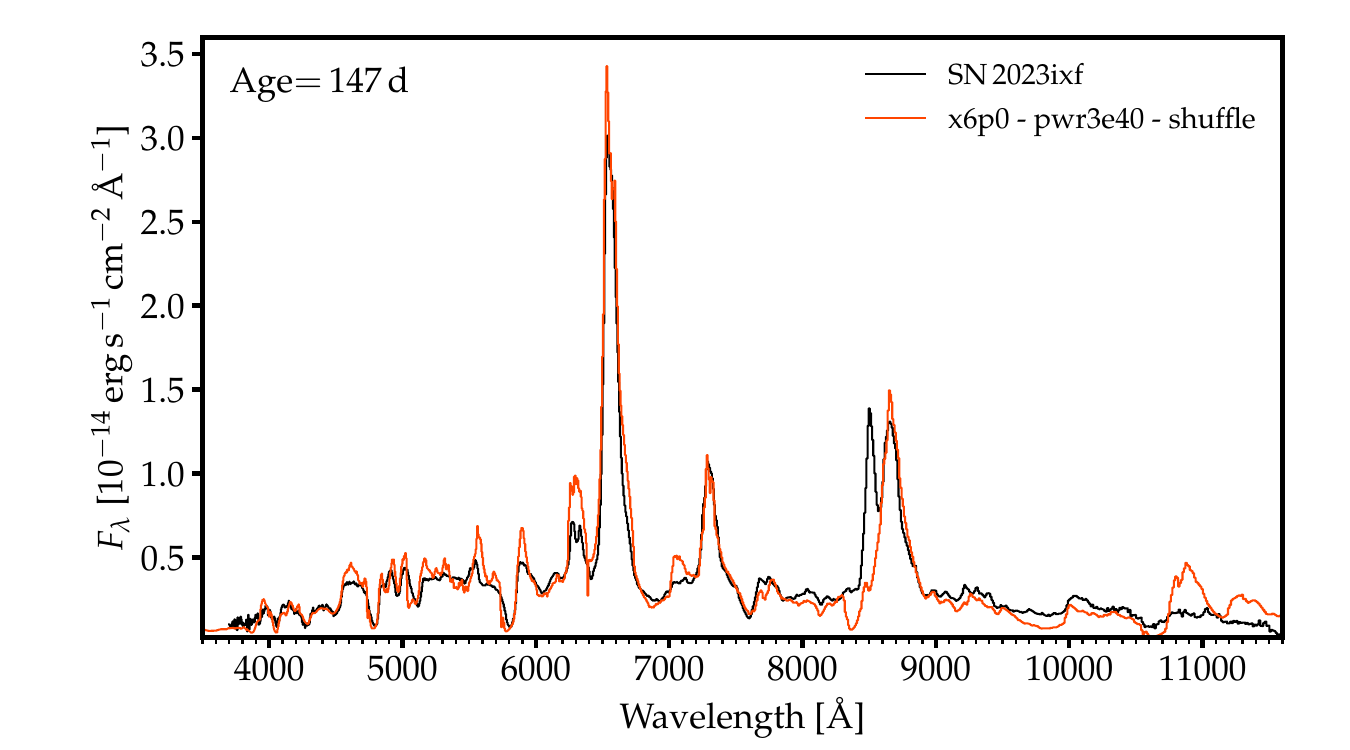}
\includegraphics[width=\hsize]{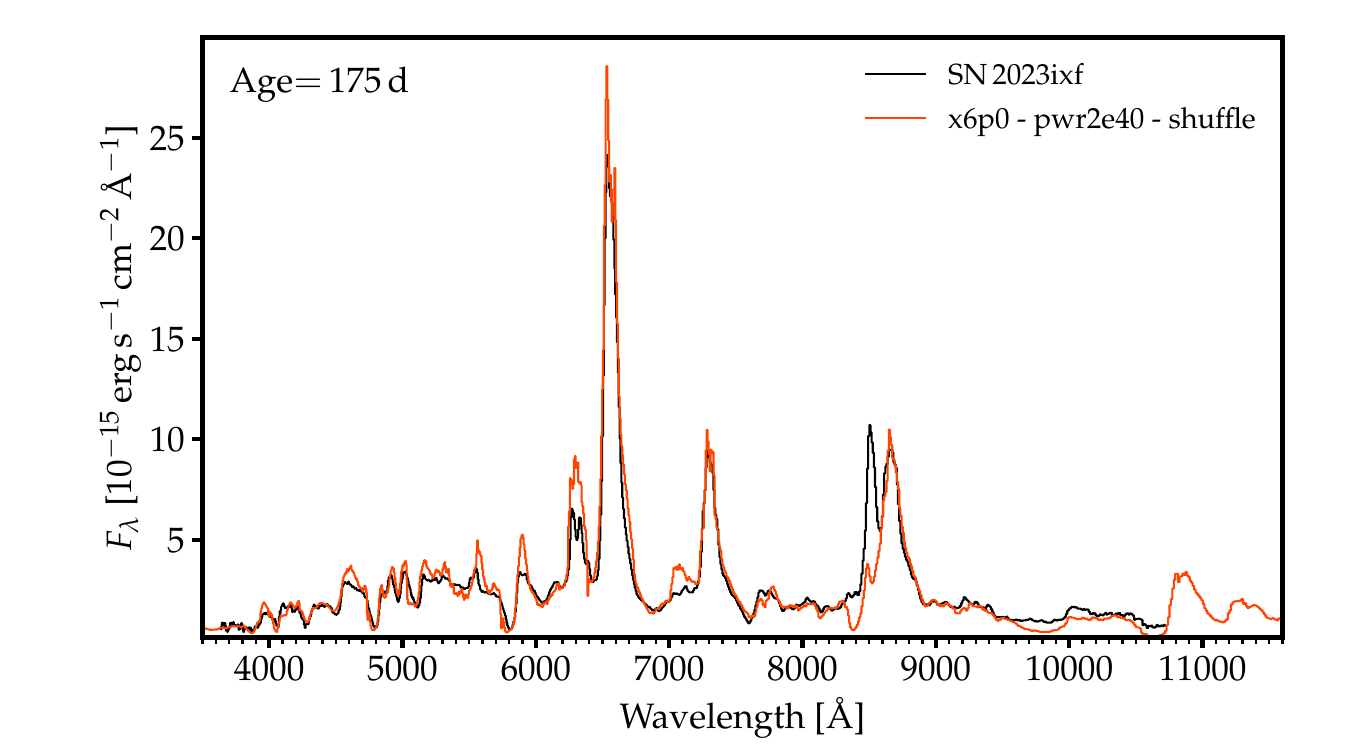}
\caption{Comparison between optical observations (black) of \sn\ at 112 (top), 147 (middle), and 175\,d (bottom) and (dust-free) model x6p0 with an interaction power of 4, 3, and $2 \times 10^{40}$\,\ergs\ (red), respectively. The model at 112\,d has been scaled up by 10\,\%. Observations were corrected for reddening and redshift, and the models were scaled to the distance of \sn. See Section~\ref{sect_pre_dust} for discussion.
\label{fig_pre_dust}
}
\end{figure}

\subsection{Evolution prior to dust formation}
\label{sect_pre_dust}


Figure~\ref{fig_pre_dust} shows a comparison between \sn\ at 112, 147, and 175\,d, and our model x6p0 with a decreasing interaction power from 4, 3, to $2 \times 10^{40}$\,\ergs. Although not obvious, interaction power alters the overall brightness of the model at those times, increasing the $V$-band brightness by about 0.3\,mag (see \citealt{dessart_23ixf_phot_26}), meaning that at the onset of the nebular phase, the SN optical brightness is not just arising from decay power and thus that interaction impacts the inferred \nifs\ mass. The only obvious sign of interaction is the broad and boxy \ha\ emission with a ledge on the red side and no P-Cygni absorption on the blue side \citep{dessart_csm_22}. All simulations with no interaction power show a broad \ha\ trough and narrower emission (see \citealt{dessart_23ixf_phot_26} and \citealt{dessart_csm_22}). Because the separation between \ha\ and the red component of \oidoub\ is $\sim$\,9150\,\kms, the \ha\ emission from the CDS at 8000\,\kms\ alters the strength of the \oidoub, and thus potentially the estimate of the progenitor mass or oxygen yield from the \oidoub\ strength (this region is also affected by overlapping emission from Fe lines). This \ha\ CDS emission also has inherently a broad, boxy profile \citep{dessart_csm_22}, and thus should not be modeled by means of a Gaussian profile, as is often done in SN emission line studies.

The pseudocontinuum is primarily due to Fe emission (a combination of Fe\two\ and Fe\one) and the H\one\ Balmer continuum nearer the Paschen edge, together with a number of strong lines. We identify the usual permitted transitions including \ha, \hb, \hg, \hd\ (all H\one\ lines below 5000\,\AA\ are affected by metal-line blanketing), He\one\,7065\,\AA\ (other lines at 6678 and 5875\,\AA\ are present in the model but overlap with \ha\ and \naid\ and are thus hard to infer in \sn), \naid, \oitrip\ (and \kidoub\ overlapping on its blue side), \caiitrip, as well as forbidden lines associated with \oiauroral\ and \oidoub, and \caiidoub. The mismatch of the \caiitrip\ is not related to a missing line in the model but rather an overestimate of the absorption of the blue components of the triplet by the red components due to the ejecta expansion (i.e., blue photons are redshifted and absorbed by the redder components at other locations in the ejecta). Asymmetry or 3D clumping would reduce this absorption.

The presence of well-developed forbidden lines, especially for \oidoub\ (which is somewhat overestimated by the model), is atypical of SNe II at a similar age (see, for example, SN\,1999em and SN\,2017eaw; \citealt{leonard_99em}; \citealt{vandyk_17eaw_19}). In standard SNe II, the relatively large ejecta mass and slow expansion rate imply relatively large densities in the metal-rich inner ejecta regions at 100--150\,d, which inhibits forbidden-line formation (the corresponding excited states collisionally rather than radiatively de-excite). In \sn, the reduced \mej\ but standard \ekin\ inferred from the shorter than standard photospheric-phase duration is supported by this spectral peculiarity: the faster expansion and lower density of the ejecta foster the formation of forbidden lines at early epochs in the nebular phase.

The subsequent evolution at 147 and 175\,d displays little change. The brightness drops, following the decrease in decay power --- the interaction power was also reduced to track the changes in the \ha\ emission profile. For unclear reasons, the flux offset between model and observations in the red part of the optical at 112\,d is essentially gone by 175\,d. Most spectral features and the pseudocontinuum are well fitted by the model, with only a slight overestimate of the \oidoub\ and He\one\,7065\,\AA. For the latter, the reason is unclear, although simulations with \cmfgen\ and the shuffled-shell technique tend to overestimate this He\one\ line. For \oidoub, a drop in strength would be expected were we to allow for CO formation \citep{liljegren_co_20,mcleod_co_24}, which is inferred from NIR observations of the CO first overtone at 199\,d by \citet{park_23ixf_25} --- no published NIR observation is available at earlier, nebular times.


\begin{figure*}[h!]
\centering
\includegraphics[width=\hsize]{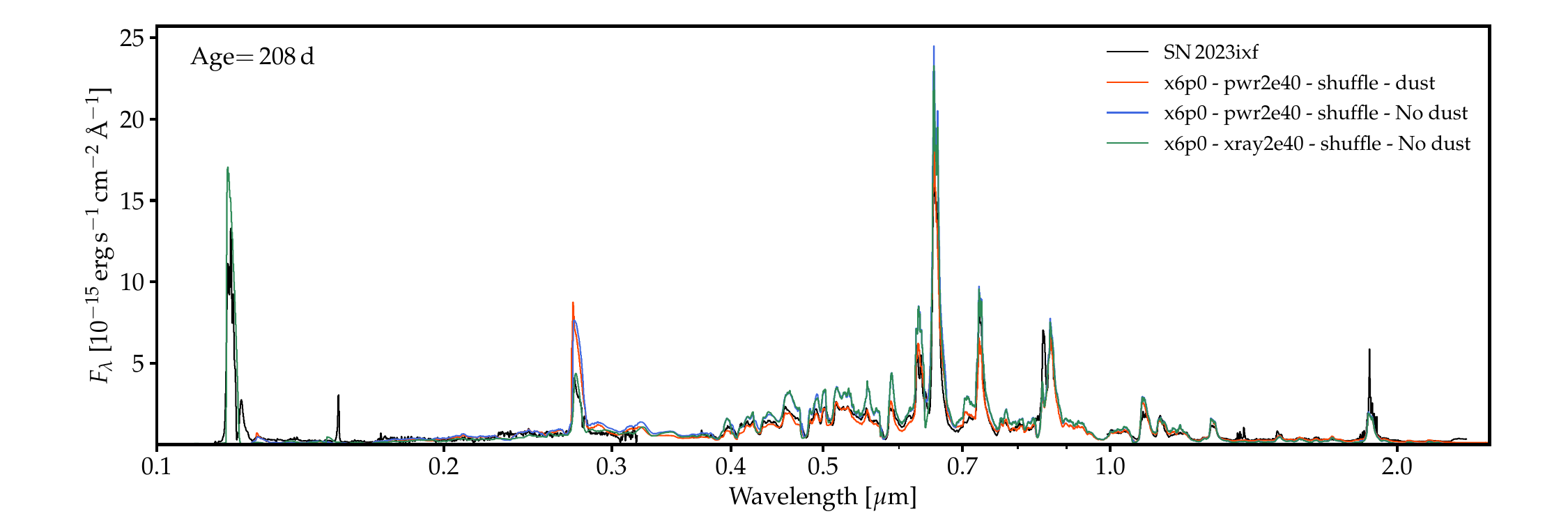}
\includegraphics[width=\hsize]{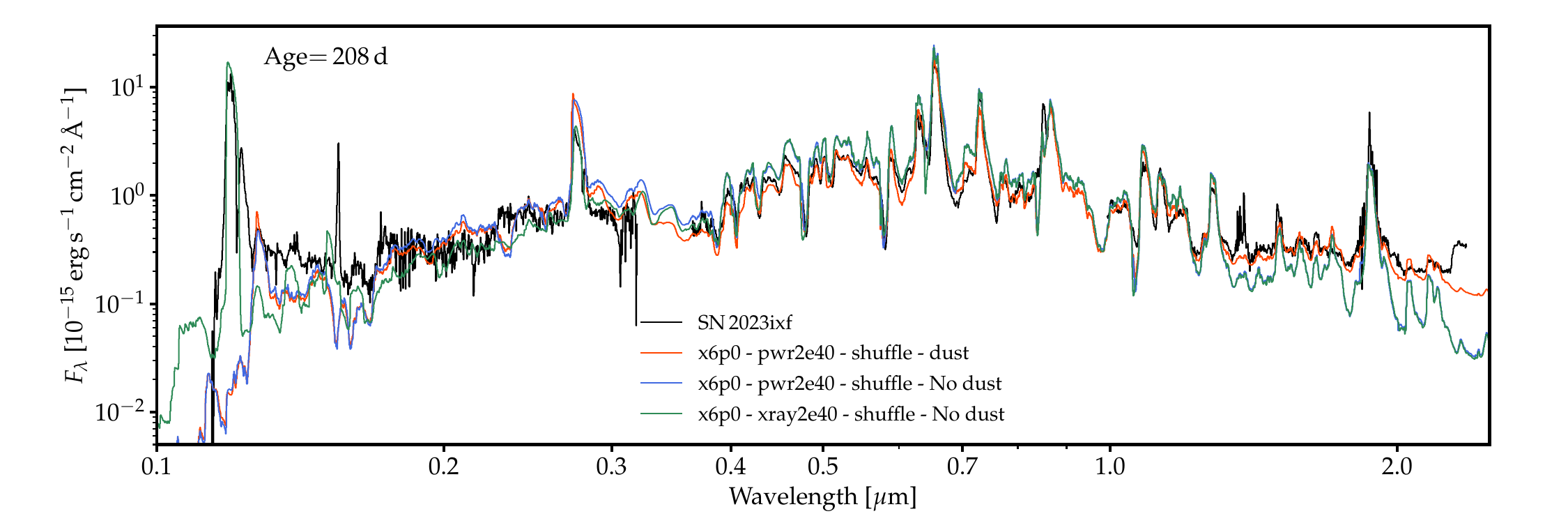}
\caption{Comparison between UV, optical, and NIR observations of \sn\ at 208\,d (black) and model x6p0 with interaction power of $2 \times 10^{40}$\,\ergs, injected within the CDS (blue and red curves) or in the form of X-rays (green). Dust is included is one of the models (red curve). The top (bottom) panel used a log-linear (log-log) scale to better emphasize local and global agreements and mismatches. See Section~\ref{sect_208d} for discussion.
\label{fig_spec_208d}
}
\end{figure*}

\subsection{\sn\ at 208\,d}
\label{sect_208d}



Figure~\ref{fig_spec_208d} compares the UV, optical, and NIR observations of \sn\ at about 208\,d (this is the epoch of the optical observations; see Table~\ref{tab_obs} for details) with variants of the x6p0 model with an interaction power of $2 \times 10^{40}$\,\ergs. Surprisingly, when evolving the model from the preceding time steps to 208\,d, the resulting model spectrum overestimates the observed flux throughout the optical (i.e., by about 0.3\,mag in the $V$ band; see blue and green curves in Fig.~\ref{fig_spec_208d}). At this time, 88\,\% of the decay power emitted is absorbed within the ejecta, and thus the escape of $\gamma$-rays is too small to explain this offset. Instead, we surmise that this extra fading is due to dust forming in the ejecta and in the CDS. This possibility is supported by the NIR excess at that time, reported by \citet{park_23ixf_25}, and deemed necessary to match our model flux in that region. So, without dust, the model overestimates the optical flux but underestimates the NIR flux. Further, the flux overestimate concerns not just emission lines forming in the inner, metal-rich ejecta but also the Fe\two\ emission arising from H-rich material forming farther out in the ejecta. Hence, dust external to this emission must be invoked, and the most likely location is the CDS.

We have thus explored various dust configurations. The model with dust presented in Figure~\ref{fig_spec_208d} includes warm dust ($5 \times 10^{-6}$\,\msun\ and 1330\,K) within the CDS and cool dust in the inner ejecta below 3000\,\kms\ ($10^{-6}$\,\msun\ at 500\,K). With this choice, the optical and NIR spectra are well matched, at a level comparable in the optical to what was achieved at epochs 112, 147, and 175\,d. Other dust configurations are shown in Figure~\ref{fig_dust_other}. Dust present in the inner ejecta impacts all emission lines forming therein, yielding a blue-red asymmetry of the corresponding line profiles. In contrast, dust in the CDS produces a blue-red asymmetry only for lines forming in the CDS, whereas lines forming deeper in the ejecta suffer mostly a wavelength-independent, uniform attenuation.

The introduction of dust has essentially no impact on the UV flux because the UV flux forms in the outer parts of the CDS and is already subject to strong attenuation by the CDS even with a dust-free CDS \citep{dessart_csm_22,dessart_dust_25}. However, the model with interaction power injected within the dense shell fails to predict the strong observed \lya\ emission line, which in \sn\ at that time is the second-strongest line in the entire spectrum. Many attempts were made to resolve this issue by modifying the deposition profile for the interaction power (see also Sec.~\ref{sect_dep}). Here, we show the alternate approach with the introduction of interaction power in the form of X-rays. While this hardly modifies the gas properties within the CDS (see Sec.~\ref{sect_pwr}), it dramatically alters the ionization ahead of the CDS, and to a lower extent below the CDS. At this relatively early time, the CDS is still optically thick to X-rays and UV photons, so that \lya\ forms in the outer parts of the CDS. But here with X-ray injection, the greater ionization ahead of the CDS limits the blanketing of CDS emission, in particular for \lya. Overall, this boosts the flux in \lya\ at the expense of the UV continuum and \mgiiuv, and even leads to the presence of \civuv. This line is observed in \sn\ as a symmetric, few 100\,\kms\ broad feature \citep{bostroem_uv_25}. Following our setup with only fast, homologously expanding flow and no unshocked CSM, the model produces instead a broad and blueshifted emission, with essentially the same morphology as \lya. This C\four\ line, which likely forms through X-ray photoionization, arises from the outer parts of the CDS in the model but instead from the unshocked CSM in \sn. Not predicted by the model (which includes N\one\ to N\four) but possibly present in \sn\ is the strong resonance line of N\five\ at 1240\,\AA\ \citep{bostroem_uv_25}.

This modeling thus gives evidence for the early formation of warm dust in \sn, with a mass that allows a simultaneous fit to the optical, where it causes $\sim$\,0.3\,mag extinction, and to the NIR, where it contributes significant emission (Fig.~\ref{fig_spec_208d}) --- this emission was also reported and discussed by \citet{park_23ixf_25} without a firm conclusion on its origin. The need of dust for matching the optical flux at this time but not earlier indicates that it is internal to the \sn, and located primarily in the CDS. With this location, it causes no blue-red asymmetry on emission-line profiles arising from deeper layers of the ejecta and powered by the decay of \cofs\ (e.g., \oidoub).


\begin{figure*}[h!]
\centering
\includegraphics[width=\hsize]{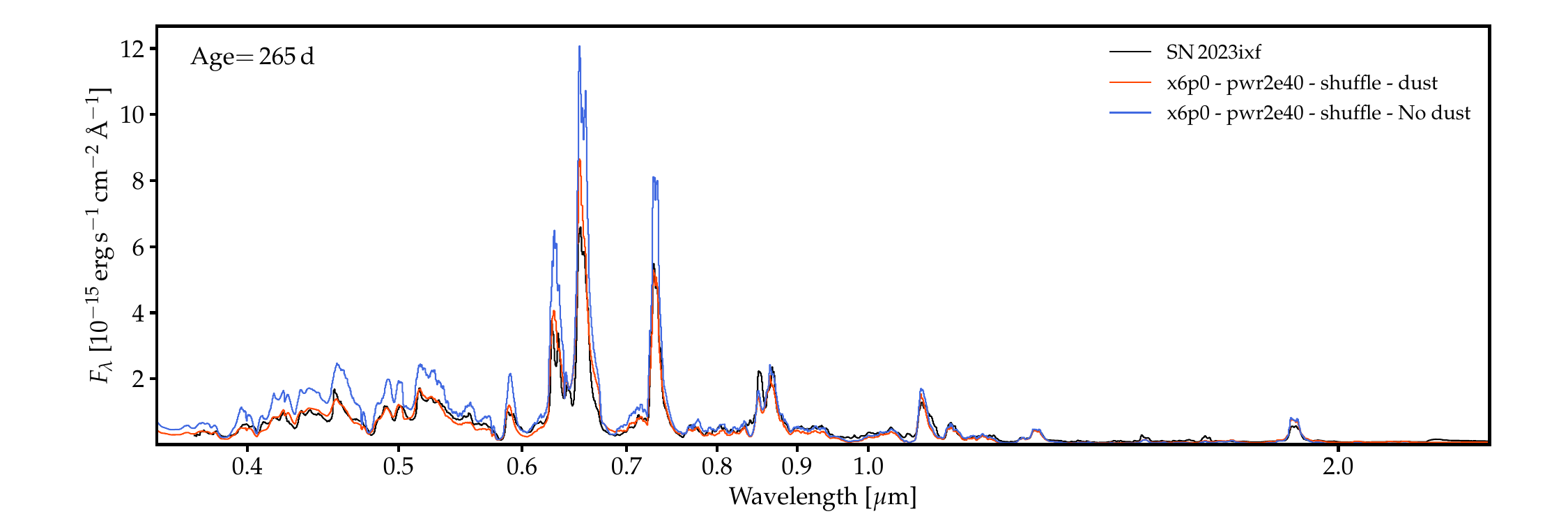}
\includegraphics[width=\hsize]{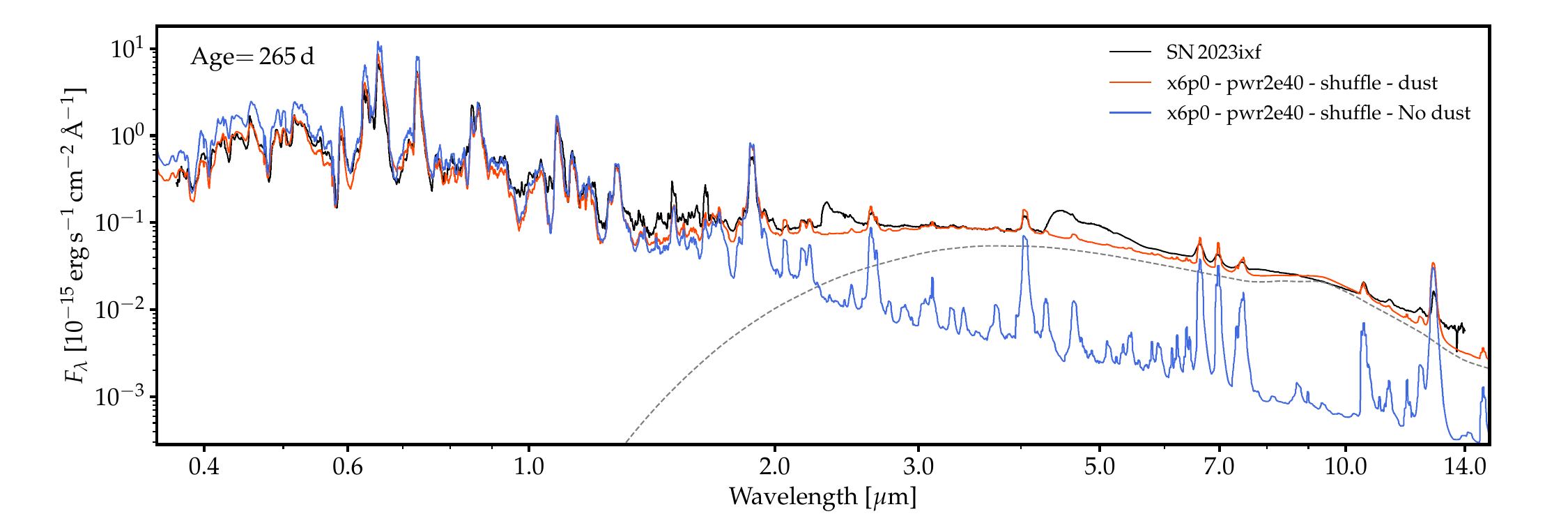}
\caption{Comparison between optical and IR observations of \sn\ at 265\,d (black) and model x6p0 with interaction power of $2 \times 10^{40}$\,\ergs, including dust (red) or not (blue). For the dusty model, the radiative-transfer calculations account for both cool dust in the inner ejecta and warm dust in the CDS, but a match to the IR emission requires an additional, external 600\,K $3 \times 10^{-4}$\,\msun\ dust emission component (dashed gray line). The top (bottom) panel used a log-linear (log-log) scale to better emphasize local and global agreements and mismatches. See Section~\ref{sect_265d} for discussion.
\label{fig_spec_265d}
}
\end{figure*}

\subsection{\sn\ at 265\,d}
\label{sect_265d}



Figure~\ref{fig_spec_265d} shows a comparison between optical and IR observations of \sn\ at 265\,d with model x6p0 and interaction power of $2\,\times\,10^{40}$\,\ergs\ injected within the CDS as high-energy electrons. In the dusty model, $2\,\times\,10^{-6}$\,\msun\ of dust at 500\,K was uniformly distributed below 3000\,\kms\ and $10^{-5}$\,\msun\ of dust at 1000\,K was placed within the CDS --- this CDS dust is important to attenuate uniformly the entire flux coming from the deeper ejecta layers. Overall, the models yield a good match to the optical (attenuation by dust) and NIR flux (mostly emission by dust), but the model is much too faint at IR wavelengths. To cure this, an additional dust emission component is required, which we introduced by invoking external, optically thin dust with a temperature of 600\,K and a mass of $3 \times 10^{-4}$\,\msun\ --- this external dust mass is too large to be accommodated within the CDS or the inner ejecta since it would cause too strong an attenuation of the emission from these regions (see also \citet{singh_23ixf_26}, who argue for similar external dust contributing in the form of a light echo). Results for alternate dust choices are discussed in Section~\ref{sect_dust}.

The model produces a satisfactory match to all emission lines from optical to IR, including the higher series of hydrogen (Paschen and Brackett), or the metal lines in the IR, such as H\one\,2.625\mic, H\one\,4.051\mic, \nkiinir, \nkiimir, \ariimir, H\one\,7.458\mic, [Co\two]\,10.52\mic, [Ni\one]\,11.304\mic, H\one\,12.369\mic, and \neiifs. These model lines are similar to those obtained in the noninteracting model s15p2 used to model SN\,2024ggi at a similar age  \citep{dessart_24ggi_25}.


\begin{figure*}[h!]
\centering
\includegraphics[width=\hsize]{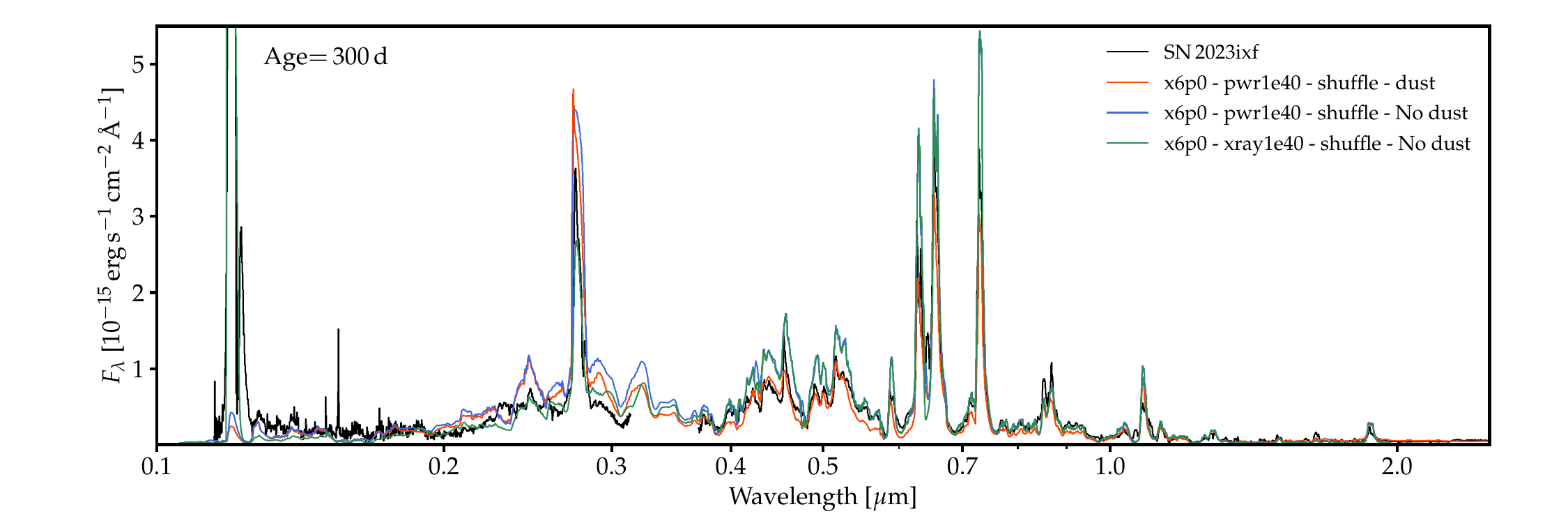}
\includegraphics[width=\hsize]{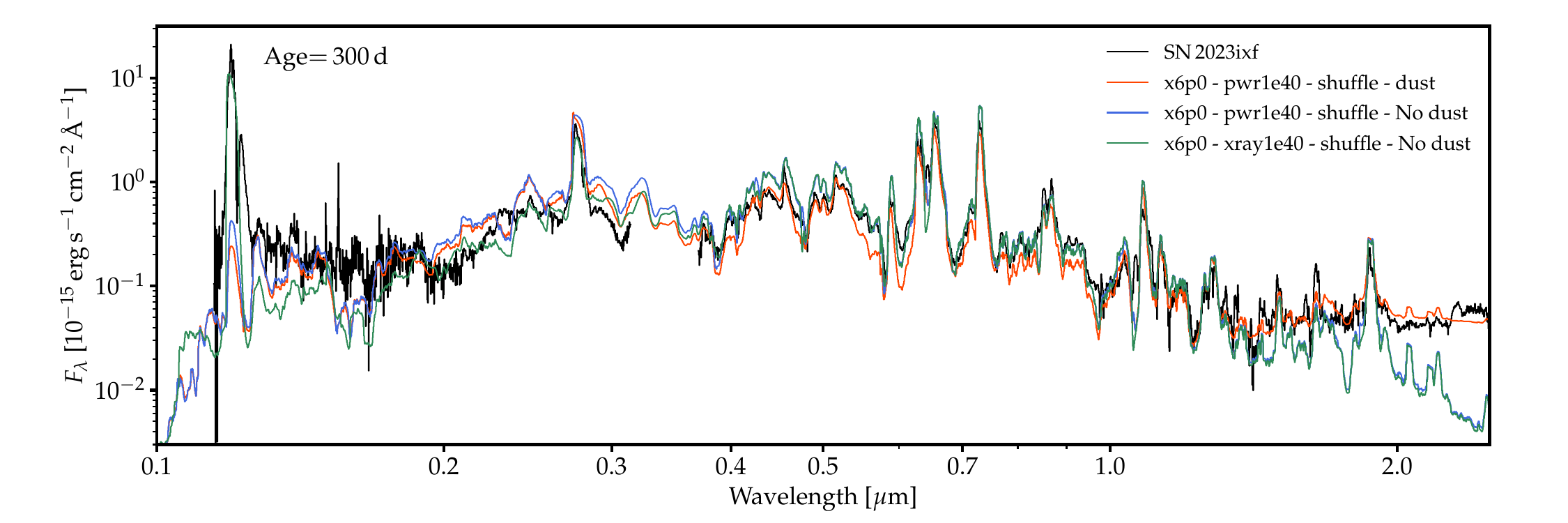}
\caption{Comparison between UV, optical, and NIR observations of \sn\ at 300\,d (black) and model x6p0 with interaction power of $10^{40}$\,\ergs, injected within the CDS (blue and red curves) or in the form of X-rays (green). For the dusty model (red), the radiative-transfer calculations account for both cool dust in the inner ejecta and warm dust in the CDS. The top panel cuts the very strong \lya\ for better visibility of the rest of the spectrum. See Section~\ref{sect_300d} for discussion. \label{fig_spec_300d}
}
\end{figure*}

\subsection{\sn\ at $\sim$\,300\,d}
\label{sect_300d}


Figure~\ref{fig_spec_300d} shows a comparison between UV, optical, and NIR observations of \sn\ at 300\,d and model x6p0 with interaction power of $10^{40}$\,\ergs, either injected as high-energy electrons within the CDS or as X-rays in the outer ejecta and CDS region. When dust is included in the model, $5 \times 10^{-6}$\,\msun\ of dust at 500\,K was uniformly distributed below 3000\,\kms\ and $10^{-5}$\,\msun\ of dust at 1000\,K was placed within the CDS. With this choice of dust parameters, the model matches satisfactorily the observations from UV to IR, although it is hard to make a strict claim on the dust mass because of the difficulty of obtaining a perfect fit to all lines. Results for alternate dust choices are presented in Figure~\ref{fig_dust_other}.

As for the modeling at 200\,d (Fig.~\ref{fig_spec_208d} and Sec.~\ref{sect_208d}), the models with power injected within the CDS underestimate the observed strength of \lya\ in \sn, although the models no longer predict negligible flux in that line. By treating the shock power in the form of X-rays, the model flux remains essentially unchanged in the optical and IR but the UV is altered, most notably with a weakening of \mgiiuv\ (i.e., by nearly a factor of two) and the UV background flux (i.e., mostly due to Fe emission), but with a strengthening of \lya\ by more than a factor of ten. In the model with X-rays, the \lya\ emission matches closely the observed strength in \sn, with the missing emission of the red side possibly due to N\five\ at 1240\,\AA\ or to an enhanced escape of \lya\ photons from the limbs and receding part of the CDS.


\begin{figure}[h!]
\centering
\includegraphics[width=\hsize]{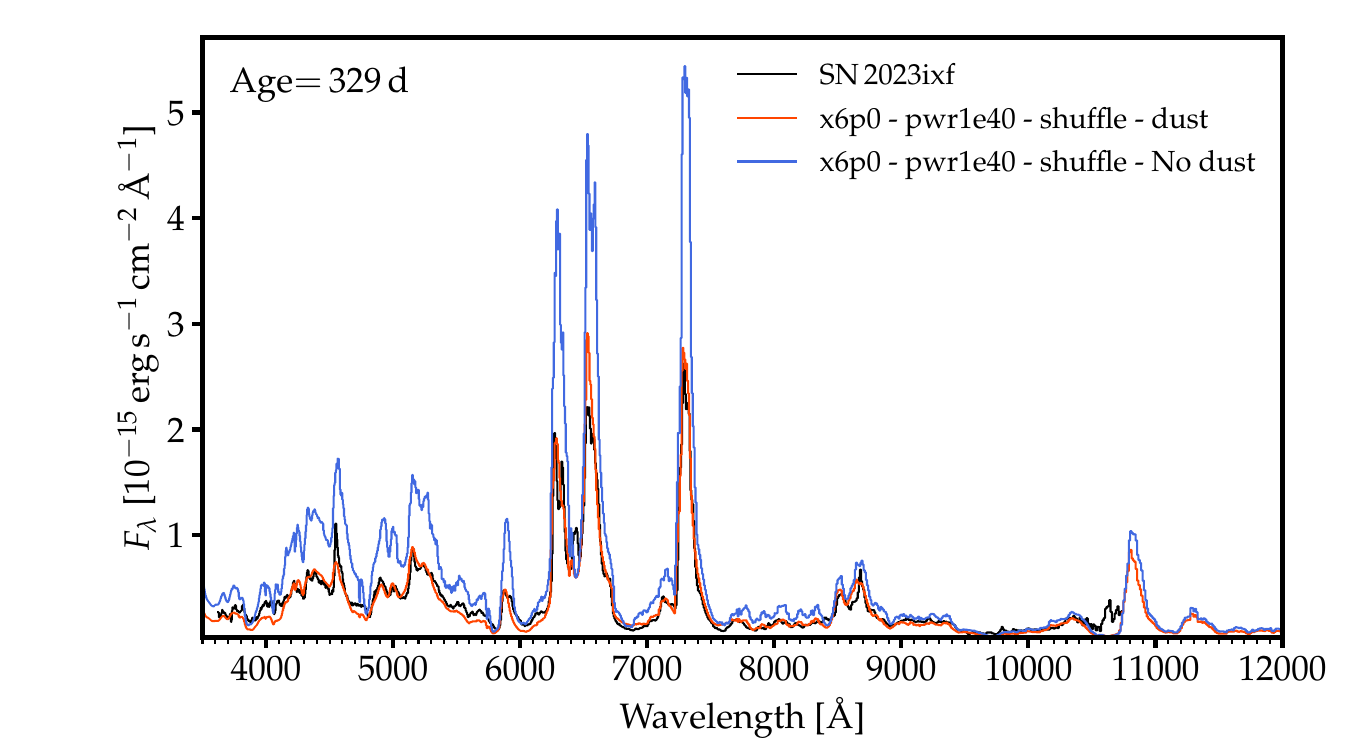}
\caption{Comparison between optical observations of \sn\ at 329\,d (black) and model x6p0 with interaction power of $10^{40}$\,\ergs, including dust (red) or not (blue). We used the same dust parameters in the dusty model as at 300\,d apart from an increase in the CDS dust mass to $2 \times 10^{-5}$\,\msun. See Section~\ref{sect_329d} for discussion.
\label{fig_spec_329d}
}
\end{figure}

\subsection{\sn\ at 329\,d}
\label{sect_329d}


Figure~\ref{fig_spec_329d} shows a comparison between optical observations of \sn\ at 329\,d (black) and model x6p0 with interaction power of $10^{40}$\,\ergs. Without dust, the model strongly overestimates the flux throughout the optical. To produce the required attenuation, dust must be introduced. The dusty model presented in Figure~\ref{fig_spec_329d} (red curve) employed $5 \times 10^{-6}$\,\msun\ of dust at 500\,K and uniformly distributed below 3000\,\kms, as well as $2 \times 10^{-5}$\,\msun\ of dust at 1000\,K and placed within the CDS. Dust in the CDS does not impact the morphology of lines formed in the inner ejecta but causes a global attenuation in the optical range. This photometric fading of \sn\ could also be interpreted as enhanced $\gamma$-ray escape, but this process is too small in our model to account for the drop in brightness (the fraction of the decay power absorbed by the ejecta is 61\% in this model at 329\,d).


\begin{figure*}[h!]
\centering
\includegraphics[width=\hsize]{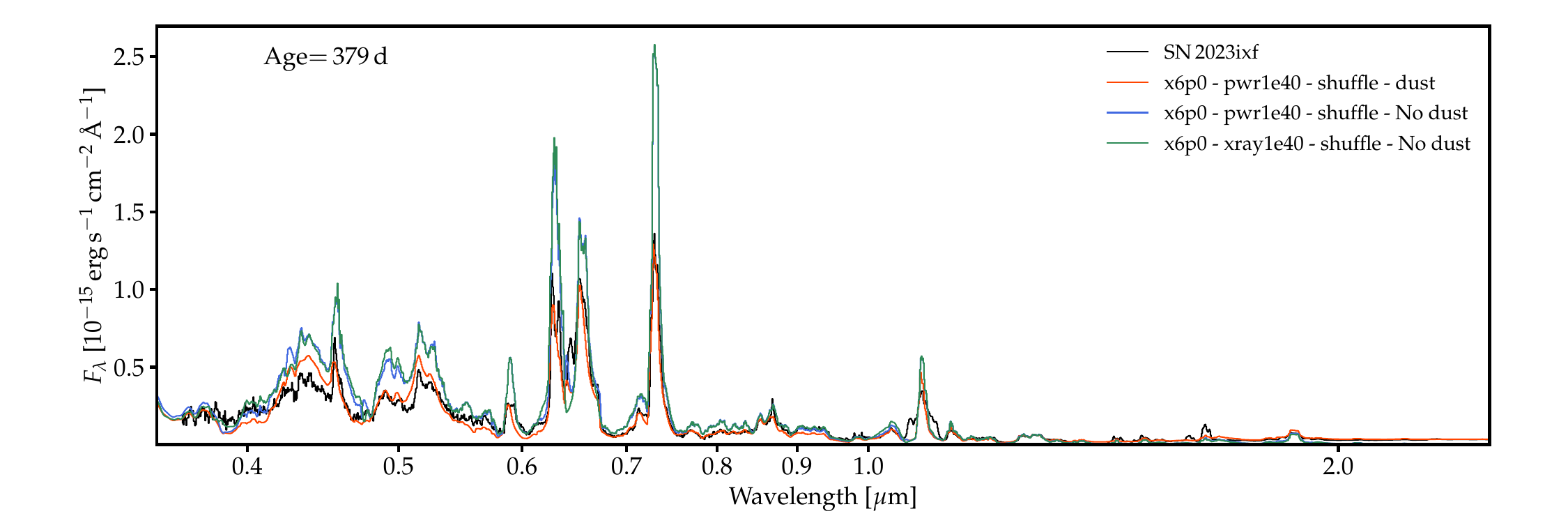}
\includegraphics[width=\hsize]{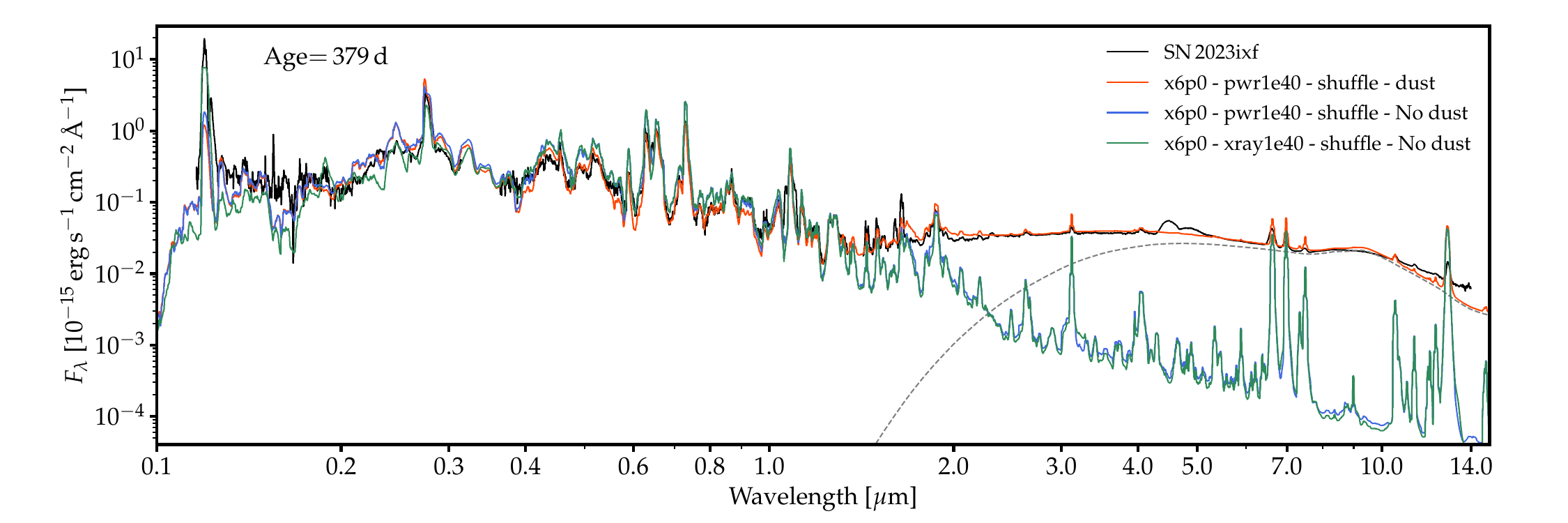}
\caption{Comparison between UV, optical, and IR observations of \sn\ at 379\,d (black) and model x6p0 with interaction power of $10^{40}$\,\ergs, injected within the CDS (blue and red curves) or in the form of X-rays (green). The dusty model has a $10^{-5}$\,\msun\ 500\,K dust uniformly distributed in the inner 3000\,\kms\ and $7 \times 10^{-6}$\,\msun\ 1000\,K dust confined within the CDS. An external, optically thin dust component ($6 \times 10^{-4}$\,\msun\ and 480\,K) is added {\it a posteriori} to match the IR flux beyond a few microns (dashed gray line). See Section~\ref{sect_379d} for discussion.
\label{fig_spec_379d}
}
\end{figure*}


\subsection{\sn\ at 379\,d}
\label{sect_379d}

Figure~\ref{fig_spec_379d} compares the UV, optical, and IR observations of \sn\ at 379\,d to model x6p0 with an interaction power of $10^{40}$\,\ergs\ injected within the CDS (dusty model shown in red and dust-free model shown in blue), or in and around the CDS in the form of X-rays (green). The dusty model has $10^{-5}$\,\msun\ 500\,K dust uniformly distributed in the inner 3000\,\kms\ and $7 \times 10^{-6}$\,\msun\ 1000\,K dust confined within the CDS. An external, optically thin dust component ($6 \times 10^{-4}$\,\msun\ and 480\,K) is added {\it a posteriori} to match the IR flux beyond a few microns (dashed gray line). This external cool dust component is necessary because adding it to the ejecta or to the CDS would cause too much attenuation.

With this choice of parameters, we obtain a good match to the observations. Dust is essential to obtain an overall agreement with the flux throughout the optical. In the UV, where the dust plays little role since it resides interior to the UV-formation region, the model with X-rays underestimates the flux in \lya\ by a few tens of percent but continues to achieve better than the model with power injected within the CDS --- the \mgiiuv\ line is well matched by all three models. Overall, the UV flux seems underestimated by the model, pointing for the need of a higher interaction power \citep{dessart_csm_22}. Most lines continue to be well matched in the optical, but the \heinir\ starts to exhibit a broad boxy profile in \sn, which the model fails to capture. That is, the model predicts some He\one\ emission from the CDS but much weaker than observed at that time (H\one\,1.094\mic\ is a weak contributor in the model relative to \heinir). Furthermore, the metal lines observed in the IR continue to be well matched by the model, apart from the overestimate of \neiifs. This might point to ejecta with a lighter O/Ne/Mg shell, or a lower ionization of the Ne-rich material, perhaps as a consequence of clumping (see Sec.~\ref{sect_ir} for discussion).


\begin{figure}[h!]
\centering
\includegraphics[width=\hsize]{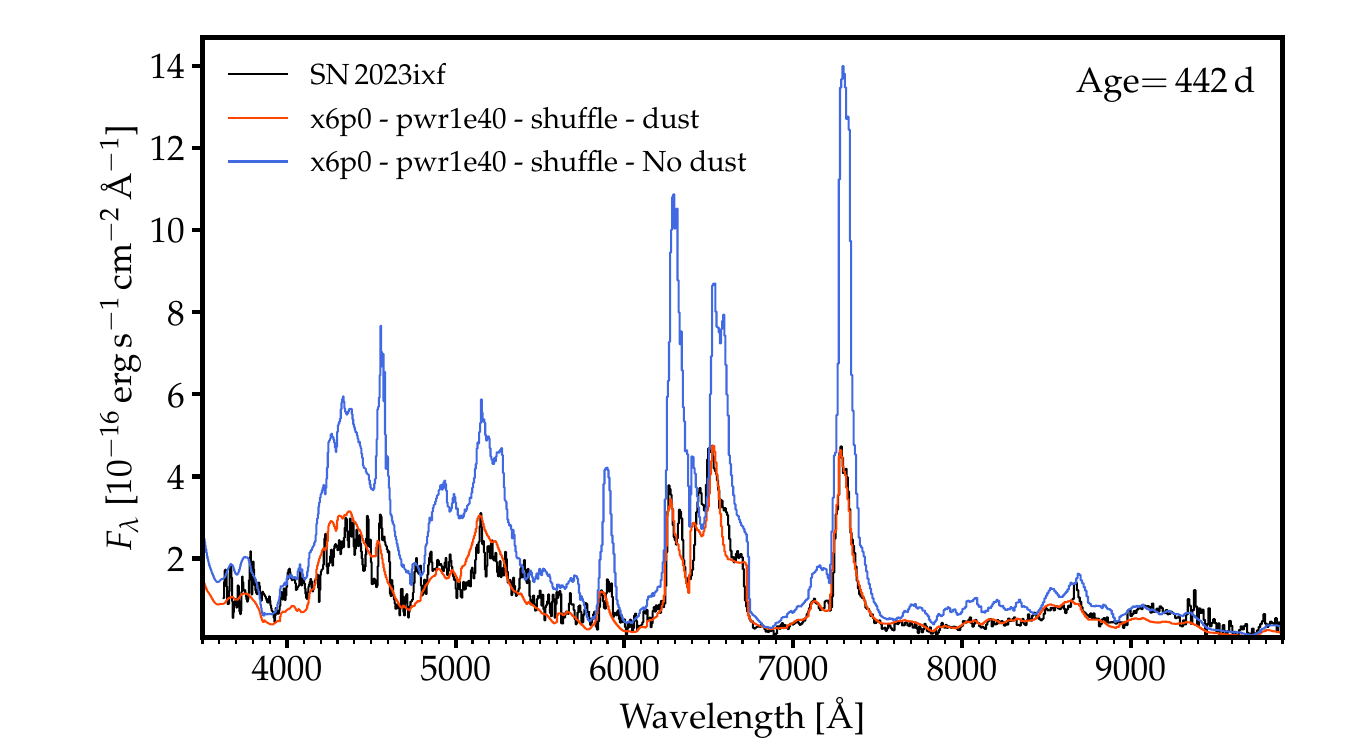}
\caption{Comparison between optical observations of \sn\ at 442\,d (black) and model x6p0 with an interaction power of $10^{40}$\,\ergs, including dust (red) or not (blue). The dusty model has a $2 \times 10^{-5}$\,\msun\ 500\,K dust uniformly distributed in the inner 3000\,\kms\ and $3 \times 10^{-5}$\,\msun\ 500\,K dust confined within the CDS. See Section~\ref{sect_442d} for discussion.
\label{fig_spec_442d}
}
\end{figure}

\subsection{\sn\ at 442\,d}
\label{sect_442d}


\begin{figure*}
\centering
\includegraphics[width=\hsize]{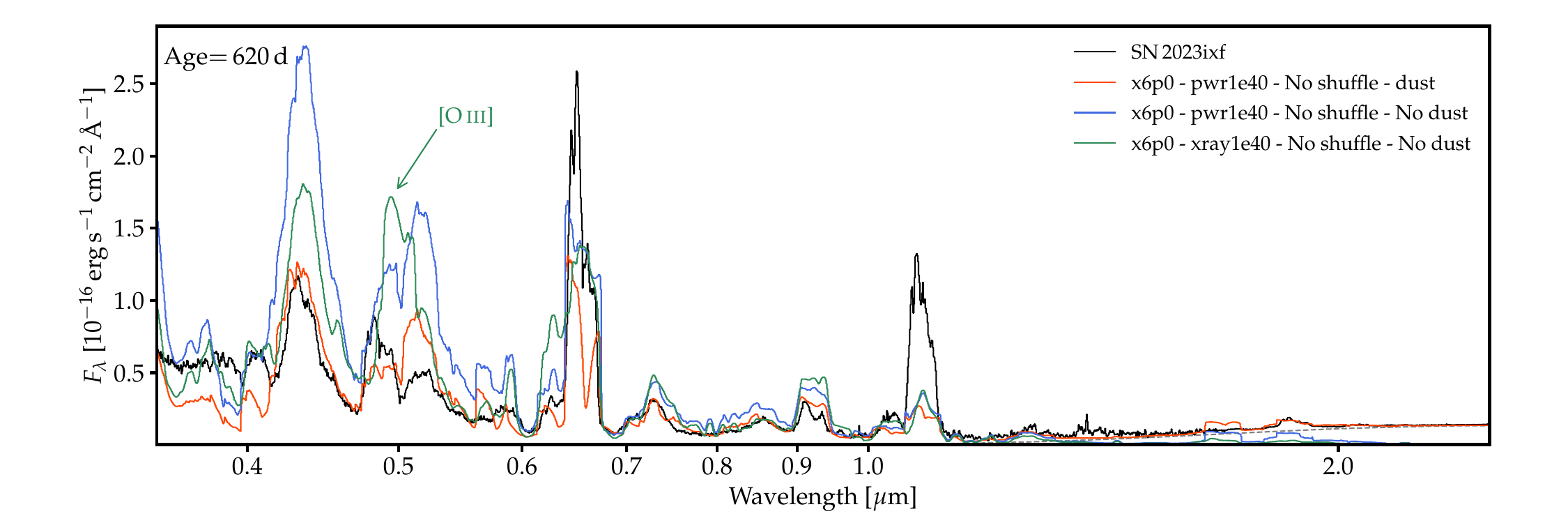}
\includegraphics[width=\hsize]{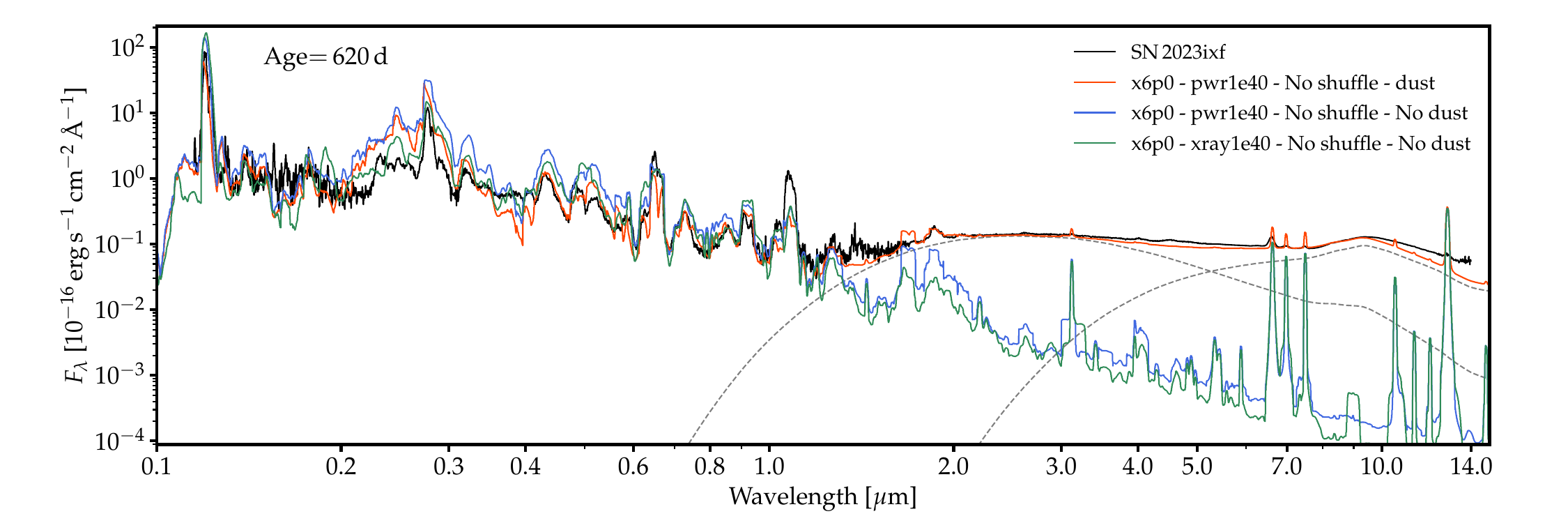}
\caption{Comparison between UV, optical, and IR observations of \sn\ at 620\,d (black) and model x6p0 with interaction power of $10^{40}$\,\ergs, injected within the CDS (blue and red curves) or in the form of X-rays (green). The dusty model (red) has a $10^{-4}$\,\msun\ 350\,K dust uniformly distributed in the inner 3000\,\kms\ and within the CDS. A full description of the bound-bound transitions contributing to the (dust-free) model spectrum ``x6p0 - pwr1e40 - No Shuffle'' is shown in Figure~\ref{fig_species_620d}. See Section~\ref{sect_620d} for discussion.
\label{fig_spec_620d}
}
\end{figure*}

Figure~\ref{fig_spec_442d} shows a comparison between optical observations of \sn\ at 442\,d (black) and model x6p0 with interaction power of $10^{40}$\,\ergs. Dust is again essential to reproduce the optical flux. By this time, the overestimate is not so obvious from the weak, pseudocontinuum flux, but more obviously from the emission lines which are now very strong. Without dust, the model exhibits strong Fe\two\ emission below about 5500\,\AA, and strong individual emission lines due to \naid, \oidoub, \ha, and \caiidoub, the last of these being the strongest line in the optical. By introducing dust in the CDS, we attenuate the flux throughout the optical, and by additionally introducing dust in the inner ejecta, we further quench the emission from \naid, \oidoub, and \caiidoub, introducing as well a blue-red profile asymmetry. Overall, the introduction of dust leads to a reduction by a factor of about three of all emission lines and is required. In some sense, this effect is more dramatic and obvious than potentially subtle blue-red asymmetries on complicated line profiles possibly sitting on a slanted continuum (e.g., \mgifs). Here, the adopted dust temperatures are, however, mostly ``place-holders" since there is no NIR or MIR data at that time to constrain the dust emission. The dust properties are thus constrained here from the attenuation needed throughout the optical to bring the model into agreement with the observations.


\subsection{\sn\ at 620\,d}
\label{sect_620d}

 At 620\,d, \sn\ enters a phase wherein shock power starts to dominate over decay power as evidenced, for example, by the flattening of the $V$-band light curve \citep{wynn_sed_25}. Physically, it corresponds to a profound change since the SN radiation arises preferentially from the CDS (i.e., located in the outermost and thus fastest ejecta regions, and at the primordial metallicity) rather than from the inner ejecta (i.e., the metal-rich regions below about 3000\,\kms), as described by \citet{dessart_late_23} --- see also \citet{bostroem_uv_25}. Combined with the production of dust, radiation from the inner ejecta is strongly attenuated and hard to disentangle from the flux contribution (except for sufficiently isolated broad lines), such that the SN properties are more difficult to assess. Practically, this is a complicated phase to model and a good match to observations has proven difficult to obtain, likely as a consequence of our simplistic treatment of the interaction. Because the inner ejecta contribute little, we adopted a standard chemical mixing for the ejecta (rather than the shuffled-shell structure adopted above), allowing us to reduce the number of grid points and accelerate the simulations.

Figure~\ref{fig_spec_620d} compares the UV, optical, and IR observations of \sn\ at 620\,d and model x6p0 with an interaction power of $10^{40}$\,\ergs, injected within the CDS (with or without dust) or in the form of X-rays. At this epoch, we resorted to using cold dust in the radiative transfer model and simply add at the plotting stage the necessary optically thin dust emission to match the overall IR flux level. In the best-match model, the dusty model used $10^{-4}$\,\msun\ 350\,K dust uniformly distributed in the inner 3000\,\kms\ and within the CDS, to which we added two IR emission sources corresponding to 0.0012\,\msun\ of dust at 330\,K and $6 \times 10^{-6}$\,\msun\ of dust at 900\,K (the same dust mixture is used in all cases). The amount of cold dust implied by the IR emission is ten times larger than adopted in the \cmfgen\ model to match the inner ejecta emission; thus, all or most of this cold dust should be external. For the warmer dust, the mass is much below what we adopted for the CDS dust and thus may be internal.

\begin{figure*}
\centering
\includegraphics[width=\hsize]{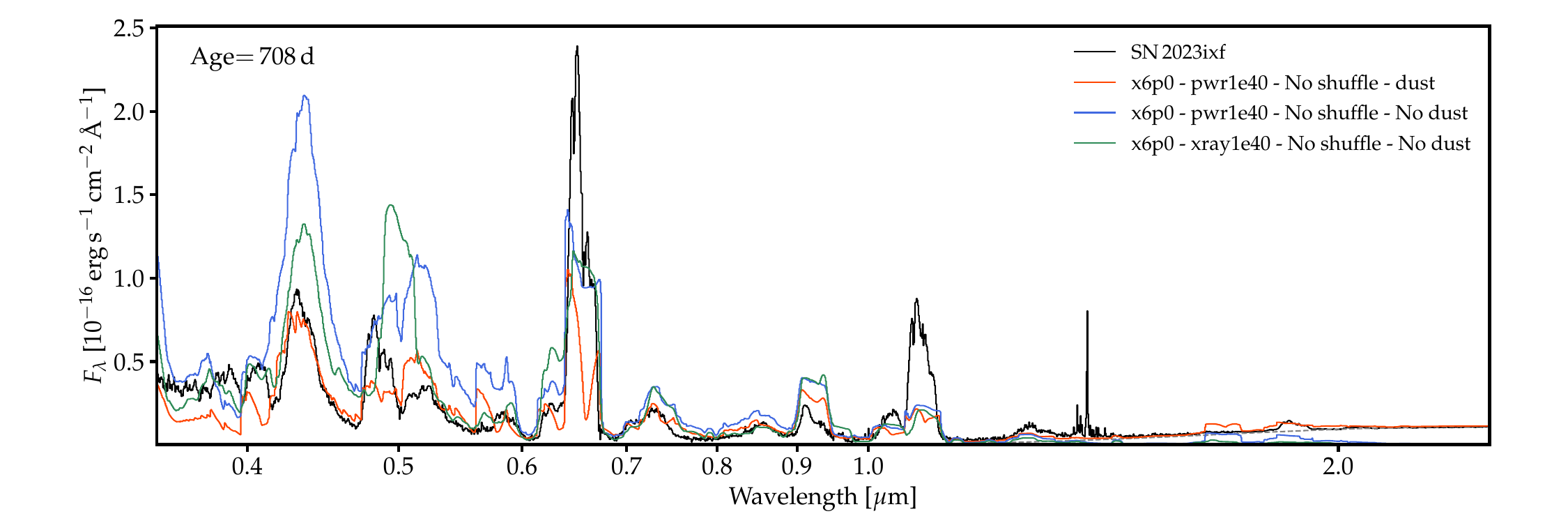}
\includegraphics[width=\hsize]{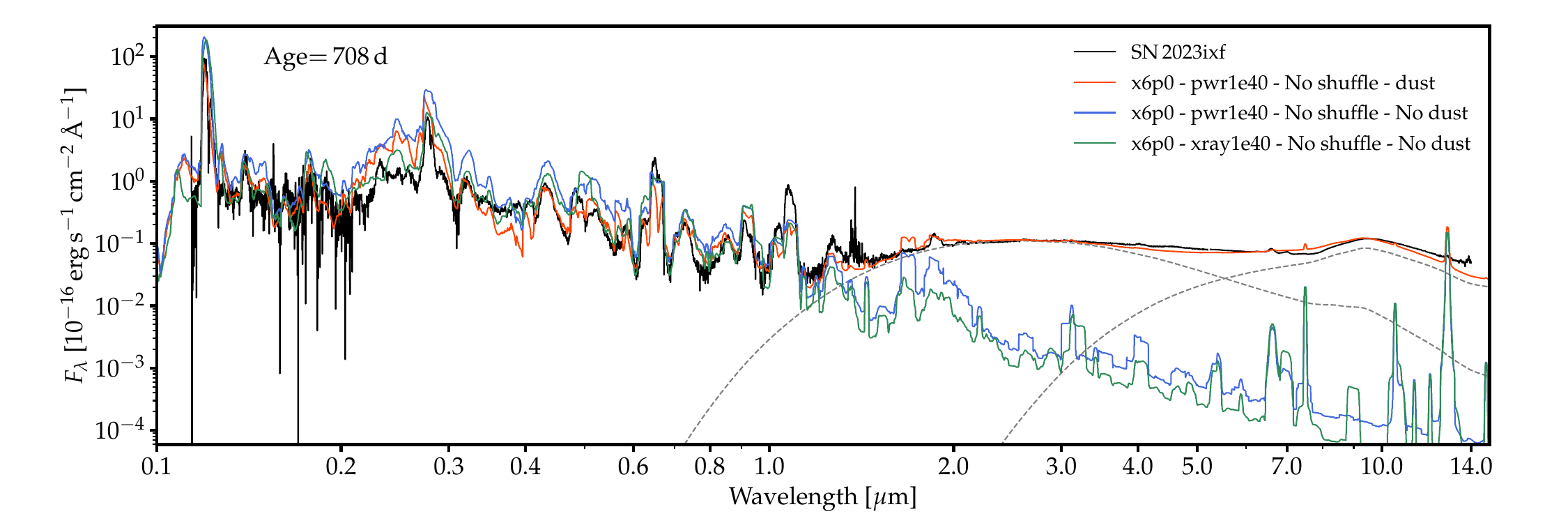}
\caption{Comparison between UV, optical, and IR observations of \sn\ at 708\,d (black) and model x6p0 with interaction power of $10^{40}$\,\ergs, injected within the CDS (blue and red curves) or in the form of X-rays (green). The dusty model has a $10^{-4}$\,\msun\ 350\,K dust uniformly distributed in the inner 3000\,\kms\ and $2 \times 10^{-4}$\,\msun\ 350\,K dust confined within the CDS. For the IR emission, an additional contribution is added at the plotting stage with two external optically thin dust emission components, namely 0.0017\,\msun\ of dust at 300\,K and $5 \times 10^{-6}$\,\msun\ of dust at 900\,K (dashed lines). See Section~\ref{sect_708d} for discussion.
\label{fig_spec_708d}
}
\end{figure*}

Unlike at previous epochs, the models with interaction power all yield a satisfactory match to \lya, but \mgiiuv\ and its vicinity are better matched by the model with X-ray injection. This modest difference likely arises from the lower density, lower optical depth, and larger temperatures in and around the CDS. Irrespective of dust, these line profiles are asymmetric and strongly attenuated on the red side. In the optical, all lines observed in \sn\ are predicted by the model, and generally better matched by the addition of dust in the ejecta and CDS. We can identify the strong lines of \ha\ and \heinir. There are no carbon lines in the optical, but we predict both \niiauroral\ and \niidoub\ (about a tenth of the \ha\ strength in the model, both broad boxy) as well as \oiauroral, \oidoub, and \oiidoub\ (all three broad boxy, thus primarily from the CDS). A notable difference is the prediction in the X-ray model of \oiiidoub, which appears as broad and blueshifted emission, whereas the observations exhibit little evidence for it in \sn\ at this time. Also present is \neiifs\ but overestimated. Narrow emission (i.e., from the inner ejecta) is predicted for \naid\ (the observations exhibit a broader feature, perhaps indicating \naid\ emission from the CDS as well). In the red part of the optical are \mgiiopt\ and \mgiinir. [S\two] is predicted and observed at 1.034\mic, as well as possibly 4078\,\AA. The emission from \caiidoub\ is both narrow (inner ejecta) and weak, and the \caiitrip\ is mostly gone. Fe emission dominates below 6000\,\AA\ in the optical (which completely swamps \hb\ and \hg), together with broad boxy emission associated with [Fe\two]\,7155\,\AA\ and around 9100\,\AA\ (in the NIR, the model also predicts relatively strong Fe\two\ line at 1.687\,\mic, which is not observed). Broad \nkiiopt\ emission from the CDS is also predicted by the model, whereas forbidden lines of Ni in the IR are dominated by emission from the inner ejecta. In the IR, the model lines are all quite weak relative to the dust emission and the \neiifs\ is notably overestimated by the model.

Two strong lines are not so well matched by those models. The \ha\ profile is too weak (even without dust), and when dust is introduced a strong attenuation results at line center. This peculiarity likely arises from the assumption of spherical symmetry, whereas in practice the CDS would likely break up into clumps, thereby reducing the effective optical depth (see also discussion of \citealt{dessart_dust_25}). When this possibility is allowed, the central dip is indeed reduced (see, for example, the discussion by \citealt{flores_shell_22}). The second deficiency of the model is the underestimate of the \heinir\ emission line. The model does qualitatively follow the observed evolution for that line in \sn, which is narrow until about 300\,d and strengthens and broadens thereafter, but it does so much faster in the observations. The peak flux is a factor of four weaker in the model than observed at 620\,d. There is some complicated, nonlinear physics at play here that the model does not capture accurately enough.


\subsection{\sn\ at 708\,d}
\label{sect_708d}

For completeness, and because there are both UV and IR observations of \sn\ taken at $\sim 708$\,d, we show in Figure~\ref{fig_spec_708d} a comparison between these observations and our model with an interaction power of $10^{40}$\,\ergs, injected within the CDS or in the form of X-rays. In the dusty model, the CDS dust mass has been raised to $2 \times 10^{-4}$\,\msun, and the two external optically thin dust emission components have 0.0017\,\msun\ of dust at 300\,K and $5 \times 10^{-6}$\,\msun\ of dust at 900\,K. The model achieves a satisfactory match from the UV to the IR, although the latter is dominated by external dust contribution not included in the \cmfgen\ model. As at the previous epoch, the model with X-ray power injection exhibits strong \oiiidoub\ emission, more symmetric than at the previous epoch, and still overestimating whatever [O\three] emission there may be in \sn\ at this epoch.


\begin{figure}[h!]
\centering
\includegraphics[width=\hsize]{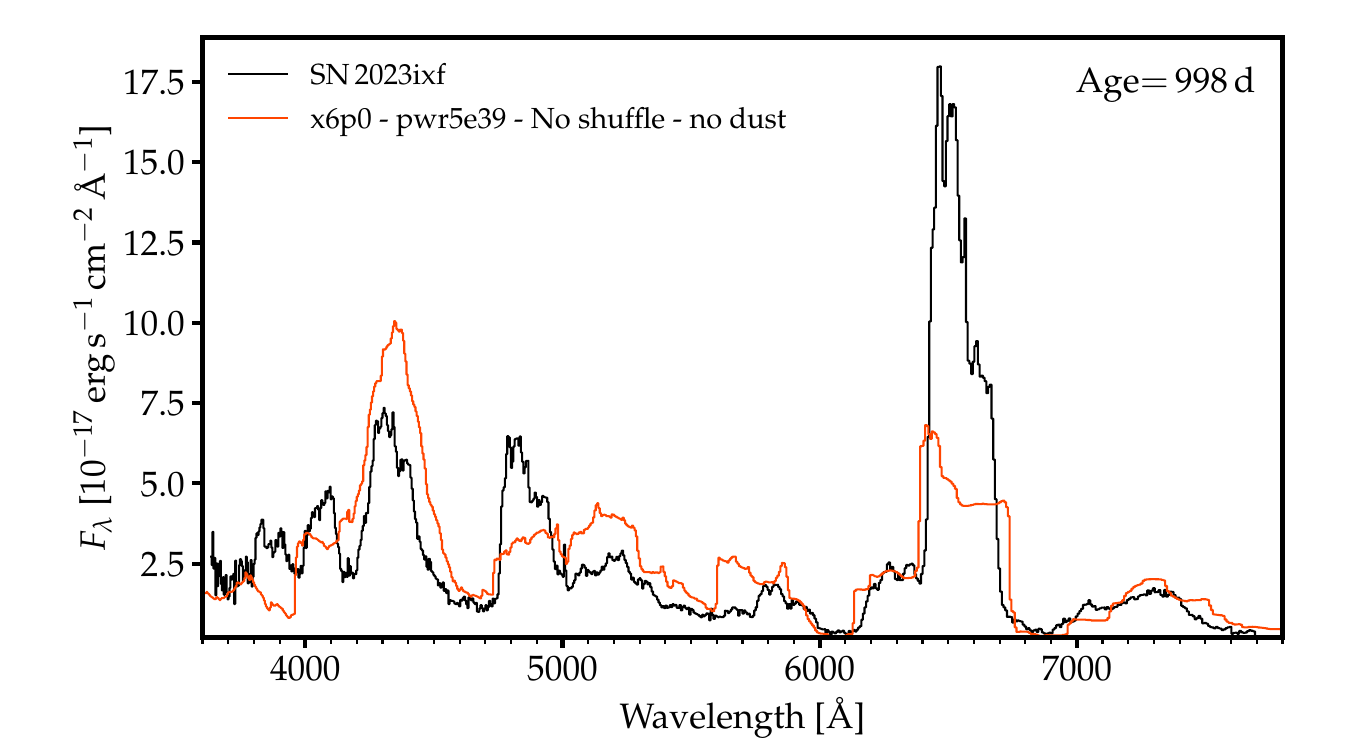}
\caption{Comparison between optical observations of \sn\ at 998\,d (black) and model x6p0 at 1000\,d with an interaction power of $5 \times 10^{39}$\,\ergs\ without dust. See Section~\ref{sect_998d} for discussion.
\label{fig_spec_988d}
}
\end{figure}

\subsection{\sn\ at 998\,d}
\label{sect_998d}


The final epoch of observations in this work is at 998\,d with an optical spectrum obtained from the Gran Telescopio Canarias (for details, see Sec.~\ref{sect_gtc_data}). In Figure~\ref{fig_spec_988d}, we compare these observations to model x6p0 with an interaction power of $5 \times 10^{39}$\,\ergs\ and at 1000\,d. The model matches quite well the observations with a few notable exceptions. Numerous emission features in \sn\ now appear narrower than in the model, with a reduction in extent both on the blue and red sides, indicating that it cannot be exclusively related to dust attenuation. This concerns the broad, boxy lines of \hb, \oidoub, and \ha, as well as the widespread Fe\two\ emission around 4500\,\AA. The model \ha\ line is also weaker than observed and with a poor match to the blue-red asymmetry. The asymmetry in the model is nearly exclusively driven by the overlap with the red component of \oidoub\ arising from the CDS --- the size of the jump across the \ha\ profile corresponds to the height of the 6364\,\AA\ peak flux. In the observations, the weakness of the \oidoub\ emission indicates that it could not give rise on its own to the blue-red asymmetry of \ha\ in \sn. Dust must be present --- it was ignored in this model since it is already too faint. The ionization in the CDS is also too high since the model predicts the presence of \niiauroral, which overlaps with \naid\ emission from the inner ejecta. In a forthcoming study focused on the evolution of \sn\ beyond 1000\,d, we will revisit this last model in the temporal sequence and attempt to cure these deficiencies with insights about the CDS structures and power injection provided by multidimensional hydrodynamics simulations. This current model at 1000\,d should be considered here as preliminary.

The evolution in the year preceding the observations at 1000\,d indicates a significant deceleration of the CDS, in fact much more so than in the evolution from 20 to 700\,d during which the CDS seems to have travelled at a velocity that hardly changed from about 8000\,\kms. This suggests an accumulation of mass within the CDS, perhaps from 0.2 (used in the model throughout) to 0.5\,\msun\ \citep{dessart_wynn_23}. We will return to this point when comparing \sn\ to SN\,1993J.


\begin{figure}
\centering
\includegraphics[width=\hsize]{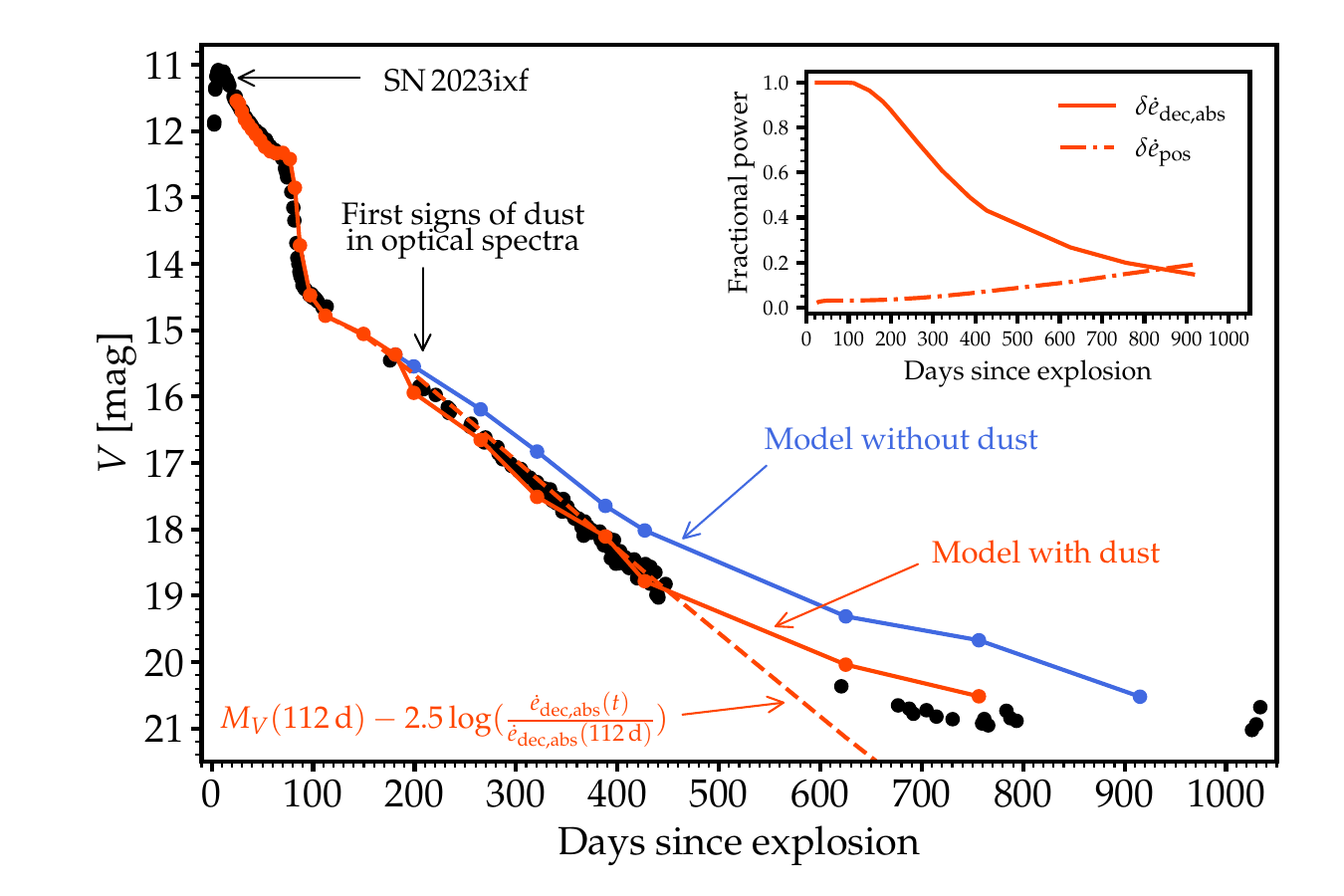}
\caption{Comparison of the $V$-band light curve of \sn\ with the x6p0 model having a time-variable interaction power (blue), as well as with allowance for dust at $\gtrsim$\,200\,d (solid red). All models were presented by \citet{dessart_23ixf_phot_26} as well as in Section~\ref{sect_rt}. The model photometry was corrected for the distance and extinction to \sn. The inset illustrates the evolution of quantities related to the radioactive decay in the model, including the fractional decay power absorbed by the ejecta ($\delta \dot{e}_{\rm dec,abs}$) and the fraction of that power arising from positrons ($\delta \dot{e}_{\rm pos}$). The dashed red line in the main plot indicates the temporal evolution past 112\,d of the model $V$-band magnitude if it arose exclusively from the evolution of the decay power absorbed in the ejecta. Although this curve aligns with the data, it does not represent accurately what takes place in \sn. See Section~\ref{sect_lc} for discussion.
\label{fig_lc}
}
\end{figure}

\section{Photometric properties and comparison to the $V$-band light curve}
\label{sect_lc}

Using the set of models computed from 112 until 998\,d and presented in the preceding section, we can now compare to the full light curve of \sn. One may compare to the bolometric luminosity but the bolometric light curve is not a direct observable. It is often constructed from sparse observations of the electromagnetic spectrum (rarely the UV and IR are observed at all, and even more rarely simultaneously), and that potentially includes external contributions (e.g., from IR emission by external dust). We thus choose to compare to the $V$-band light curve of \sn. It is well sampled, represents a wavelength range that contains a sizable fraction of the flux at most epochs coming from both ejecta and CDS, and is largely free of contamination from external sources. In the modeling presented above, the optical was also one important diagnostic for constraining the interaction (in particular with \ha).

Figure~\ref{fig_lc} shows a comparison of the $V$-band light curve of \sn\ from \citet{zheng_23ixf_25}, augmented with epochs past 450\,d, with the model influenced by interaction power (including results for model ``x6p0 + Pwr($t$)'') from \citet{dessart_23ixf_phot_26} in addition to the models from the previous section), and either with allowance for dust or not. The best-match model (red curve) corresponds to the model with both interaction power and dust past 200\,d. Inherited from \citet{dessart_23ixf_phot_26}, this model matches well the photospheric-phase brightness and duration (as well as the UV to IR spectra of \sn\ during the photospheric phase; see \citealt{dessart_23ixf_phot_26}). It also matches the nebular evolution characterized by a faster drop in $V$ after $\sim 200$\,d. In our models, this knee in the light curve arises from a combination of $\gamma$-ray escape and dust attenuation. Finally, it reproduces the flattening at late times ($\gtrsim$\,600\,d) at about 20.5\,mag. At such late times, the main power source is no longer decay power, which has ebbed due to both its exponential decline and $\gamma$-ray escape (see inset of Fig.~\ref{fig_lc}) but also because of the presence of dust in the regions absorbing this decay power and the emission from the CDS (see Secs.~\ref{sect_208d}--\ref{sect_998d}). The light-curve flattening is instead predominantly due to interaction power, as expected \citep{dessart_late_23}, and as observed in a number of Type II SNe, even for events not originally classified as interacting SNe (e.g., SN\,2017eaw; \citealt{weil_17eaw_20}).

A more careful look at Figure~\ref{fig_lc} reveals that in the absence of dust, the model with interaction power (required to match numerous signatures from UV to IR; Secs.~\ref{sect_208d}--\ref{sect_998d}) would overestimate the $V$-band brightness by a fraction of a magnitude at 200--300\,d, but by $\sim 1$\,mag after about 400\,d. This means that $\gamma$-ray escape would be a sufficient drain of SN power if \sn\ were not interacting. Because \sn\ is instead influenced by interaction at all nebular times (in fact, at all times since shock breakout), dust within the ejecta and CDS is required. As detailed in the preceding sections, we employed various combinations of dust, invoking its presence both in the inner ejecta and in the CDS.


\begin{figure}[h!]
\centering
\includegraphics[width=\hsize]{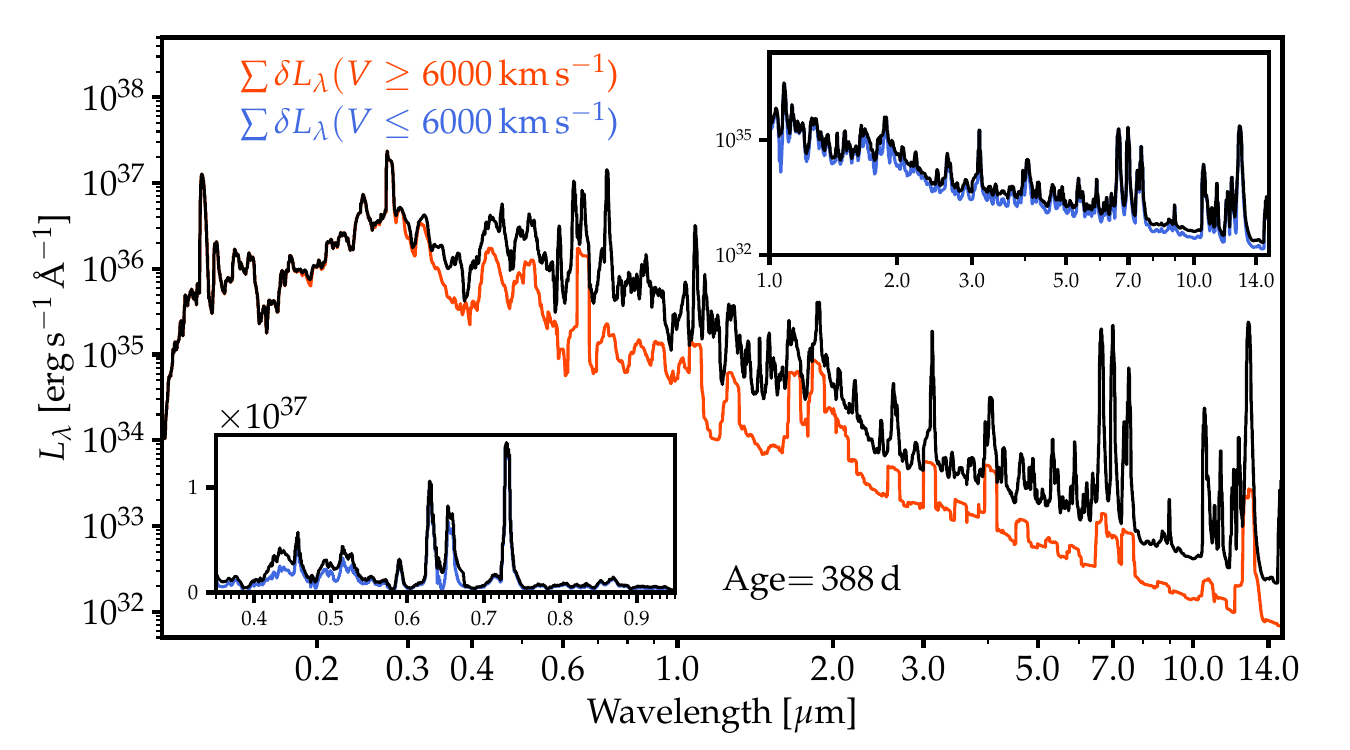}
\includegraphics[width=\hsize]{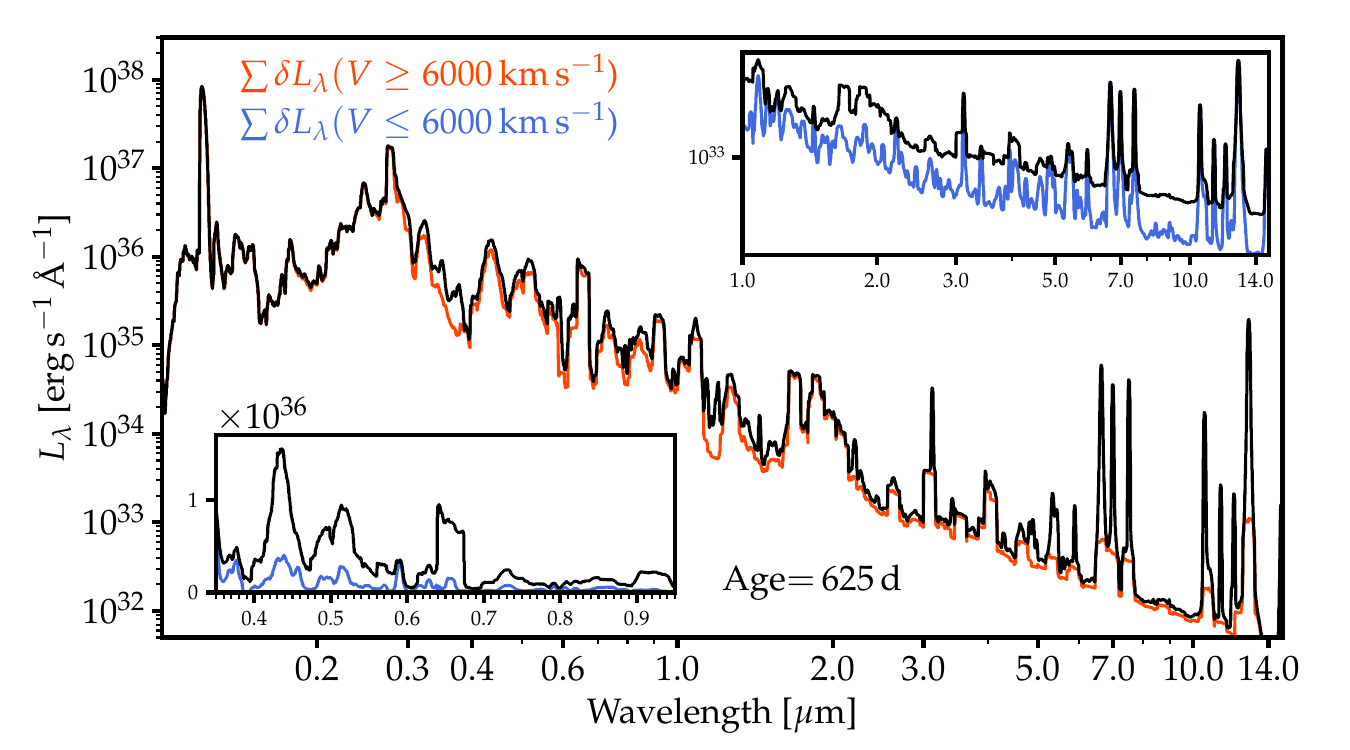}
\caption{Top: Illustration of the spatial origin of the emergent flux in the x6p0 model with interaction power but no dust at 388\,d (see also Figure~\ref{fig_spec_379d}). In this plot, the flux arising from regions below 6000\,\kms\ (blue curve; ``inner ejecta'' emission) is separated from that arising from regions above 6000\,\kms\ (red curve; CDS emission) and compared to the total flux (black). Bottom: Same as top but now for the model at 625\,d and used to compare to observations at 620\,d in Figure~\ref{fig_spec_620d}. In each plot, and for better visibility, only the inner ejecta or the CDS contribution is shown together with the total flux, the other contribution being just the difference.
\label{fig_form}
}
\end{figure}

\section{The hybrid spectral formation at nebular times in Type II SNe interacting with CSM}
\label{sect_form}

Figure~\ref{fig_form} illustrates the morphing of a decay-powered to an interaction-powered SN as time advances in the nebular phase. Specifically, Figure~\ref{fig_form} shows the fractional flux contributions from all regions slower and faster than 6000\,\kms\ in model x6p0 with interaction power at times of 388\,d (see Sec.~\ref{sect_379d} and Fig.~\ref{fig_spec_379d}) and 625\,d (see Sec.~\ref{sect_620d} and Fig.~\ref{fig_spec_620d}). These two regions correspond essentially to the inner ejecta, only powered by radioactive decay, and the CDS, only powered by interaction. As shown by \citet{dessart_late_23}, at such late times the strong UV flux arises essentially exclusively from the power injected in the CDS (or in the vicinity of the CDS if some ambiguity exists between contributions from the reverse and forward shocks). However, at wavelengths longer than $\sim 3500$\,\AA, the flux predominantly stems from the decay-powered inner ejecta at 388\,d, and it is only at very late times (here at 625\,d) that the optical and IR fluxes also arise from the CDS.

In spectral regions where many lines overlap (as in the UV), a pseudocontinuum results and no clear information may be extracted from line profiles, for example with the purpose of inferring their formation region from their Doppler widths. In contrast, in the IR (and to some extent in the optical), lines are more sparsely distributed, and the spatial origin of the lines (i.e., inner ejecta versus CDS) is directly understood from the width of the lines. We thus see, for example, that the forbidden lines from Fe, Co, and Ni in the IR \citep{kotak_04et_09,jerkstrand_04et_12,dessart_ir_25} are narrow and centrally peaked with only a weak broad base, indicating that these lines are primarily arising from the metal-rich inner ejecta, with only a negligible contribution from the outer CDS and its solar-like composition.

Figure~\ref{fig_form} makes no account of the impact of dust, which would complicate matters. At 388\,d, dust may be present in both inner ejecta and CDS and cause a complex, depth-dependent and wavelength-dependent impact. However, at 625\,d, any dust in the inner ejecta would likely annihilate the already weaker emission from the inner ejecta, limiting the SN spectrum-formation region to the CDS, which is also influenced by dust in \sn\ \citep{medler_23ixf_25,wynn_sed_25,singh_23ixf_26}.


\begin{figure*}
\centering
\includegraphics[width=\hsize]{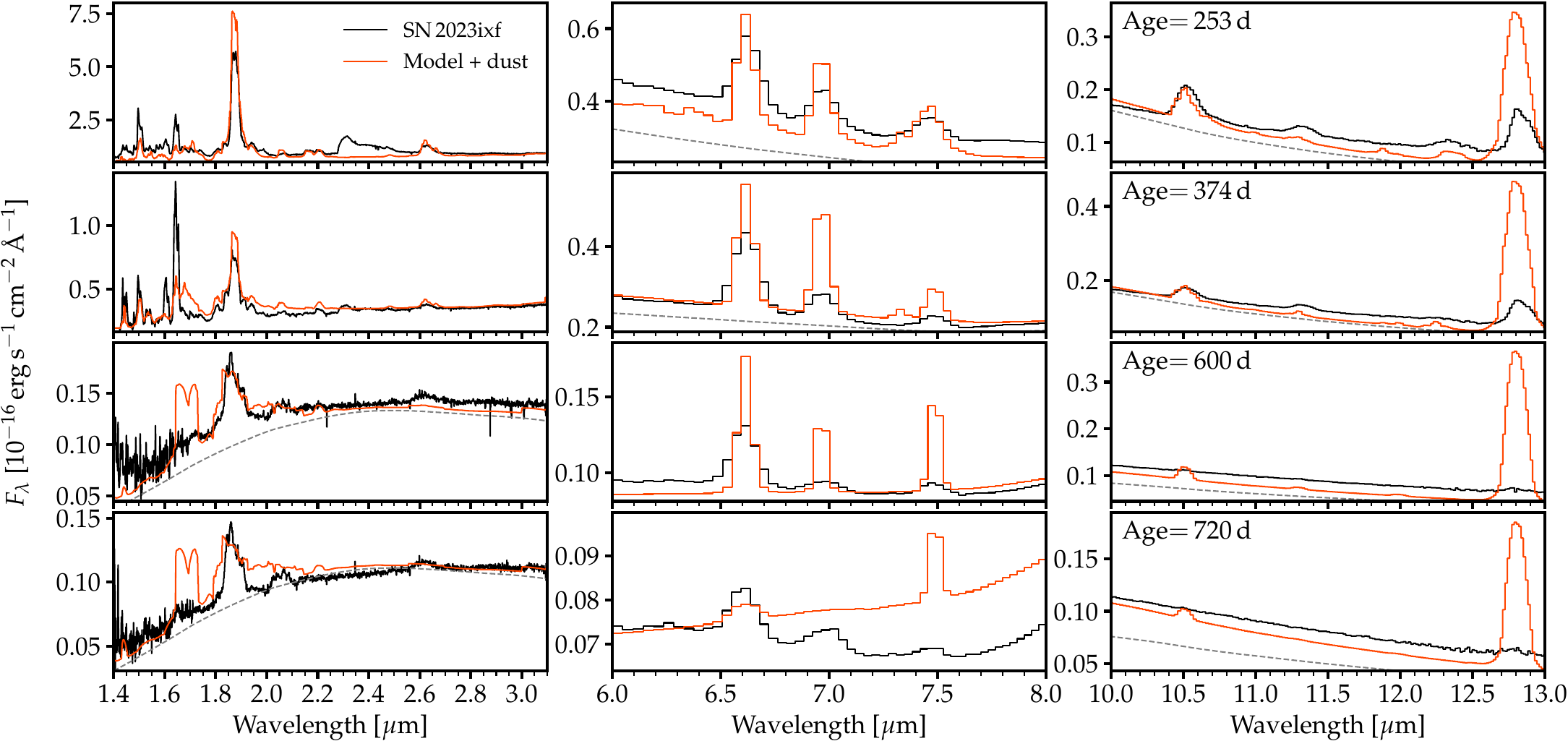}
\caption{Comparison between the observations of \sn\ (black) and the model with interaction power and allowance for dust (red) discussed in Section~\ref{sect_rt}. Each row corresponds to one epoch, and each column corresponds to a wavelength range, with the NIR at left, the 7\mic\ region in the center (i.e., lines of \nkiimir, \ariimir, and a blend of H\one\,7.458\mic\ and [Ni\one]\,7.505\mic), and the 10\mic\ region at right (i.e., primarily lines of [Co\two]\,10.520\mic\ and \neiifs). The data were corrected for redshift and reddening. The model was scaled to the distance of \sn, and was also rebinned to the data in order to compare observation and model at the same resolution.
\label{fig_ir}
}
\end{figure*}

\section{Infrared emission lines in SN\,2023ixf}
\label{sect_ir}

Nebular-phase observations of Type II SNe in the MIR are rare. They were obtained for SN\,1987A with the Kuiper Airborne Observatory \citep{wooden_87A_ir_93} and later for other SNe II with the Spitzer Space Telescope (\citealt{kotak_04dj_05,kotak_05af_06,kotak_04et_09}; \citealt{fabbri_dust_11}; \citealt{meikle_04dj_11}). With the coming in operation of JWST, a few nearby Type II SNe have been observed in great detail, including SN\,2024ggi \citep{baron_24ggi_25,dessart_24ggi_25,mera_24ggi_26} and SN\,2023ixf \citep{medler_23ixf_25,derkacy_23ixf_26}. Although the resolution of the spectra was typically not optimal, these observations revealed that Type II SNe produce a myriad of emission lines from neutral and once-ionized metal species like Ne, Ar, Ni, or Fe, and thus complement the information that is typically gathered from optical spectra, which are most notorious for the presence of Mg\one]\,4571\,\AA, \naid, \oidoub, and \caiidoub, but little else. In this section, we thus compare in more detail the best-matching model x6p0 presented in Section~\ref{sect_rt} with the IR observations of \sn.

Figure~\ref{fig_ir} compares the best-match model presented in Section~\ref{sect_rt} with the IR observations of \sn\ at 253, 374, 600, and 720\,d after explosion \citep{medler_23ixf_25}. In the NIR range shown at left, the main lines in order of increasing wavelength are Mg\one\,1.502, [Fe\two]\,1.644\mic, Fe\two\,1.687\mic, H\one\,1.875\mic, [Ni\two]\,1.939\mic, He\one\,2.058\mic, H\one\,2.166\mic, Na\one\,2.206\mic, and H\one\,2.625\mic. The 7\mic\ region shown in the middle column contains three main lines due to \nkiimir, \ariimir, and a blend of H\one\,7.458\mic\ and [Ni\one]\,7.505\mic. The 10\mic\ region contains mostly [Co\two]\,10.520\mic\ and \neiifs. The broad bump at 2.3\mic, which is associated with the CO first overtone \citep{spyromilio_co_87A_88,liu_dalgarno_95}, is not predicted by the model since molecules were ignored in the \cmfgen\ calculation.

Our dust model does not perfectly fit the background flux from \sn\ --- in the absence of dust the model predicts an IR flux about a factor of 1000 fainter at 10\mic\ (see Sec.~\ref{sect_rt}). The model line widths are comparable to those observed in most cases. In the 7\mic\ region, the very low resolution permits a rough check on profile morphology, and in the 10\mic\ region, the model \neiifs\ is somewhat too broad, suggesting an overestimate in the velocity of Ne-rich material. But in the NIR, where the resolution is much higher, the line widths are well matched by the model. The largest discrepancies in emission lines develop at later times (at $\gtrsim$\,620\,d), in particular with the Fe\two\ emission at 1.687\mic, \ariimir, and \neiifs, all overestimated by the model (\neiifs\ is overestimated at all four nebular epochs). Optical-depth effects are small at such long wavelengths. The disappearance of these lines in \sn\ might indicate an extra coolant in those metal-rich regions, perhaps related to dust formation within them. Clumping, which is neglected here, could also lead to a greater recombination from the inner regions. Because dust obscures the inner-ejecta emission in the optical, it is difficult to gauge such effects from other lines in different spectral regions.

\begin{figure*}[h!]
\centering
\includegraphics[width=\hsize]{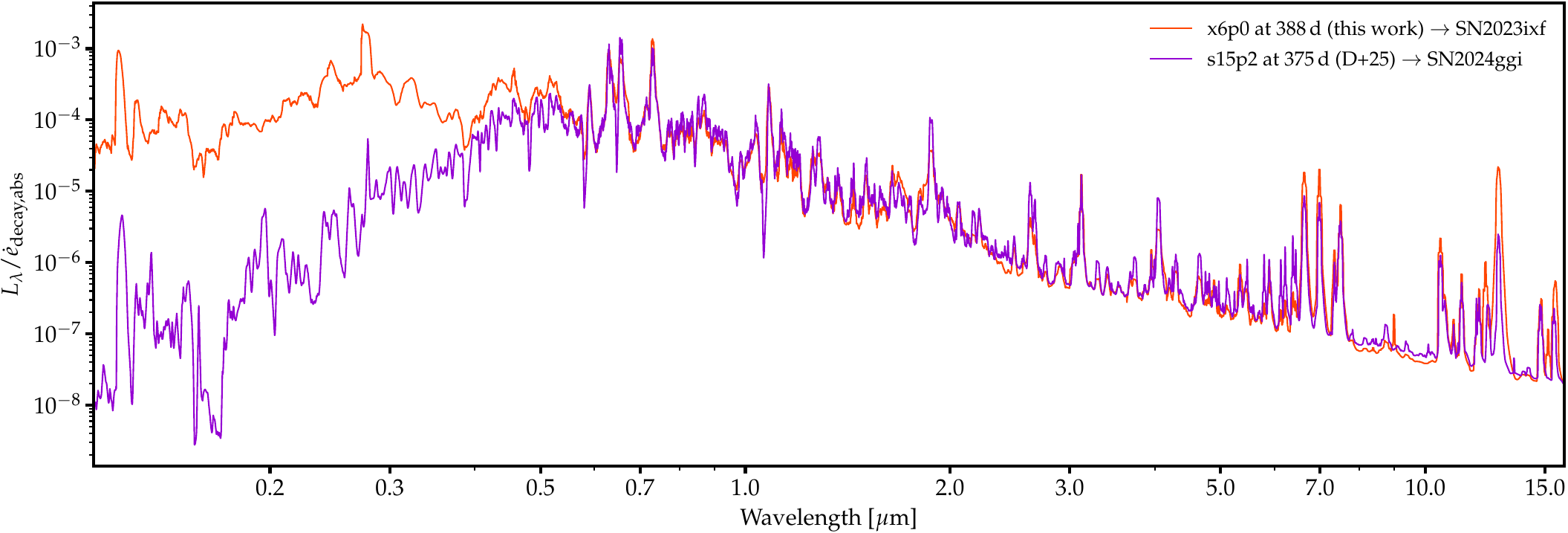}
\includegraphics[width=\hsize]{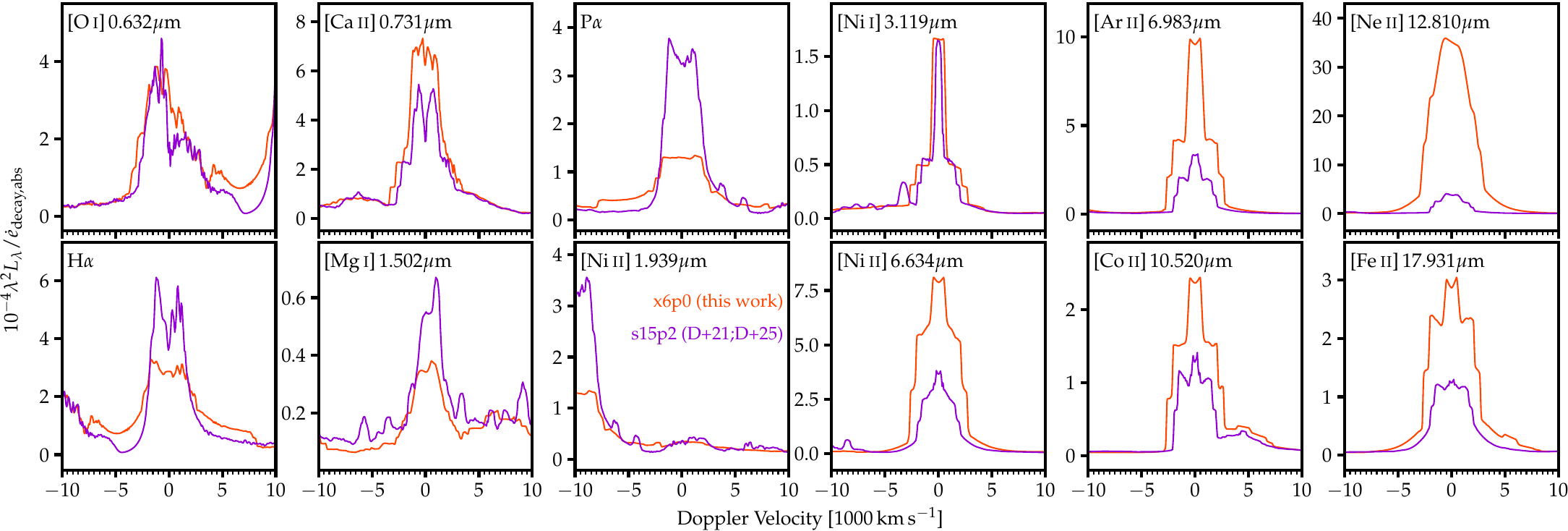}
    \caption{Top: Spectral comparison between the noninteracting model s15p2 at 375\,d from \citet{dessart_sn2p_21}, used for comparison with SN\,2024ggi by \citet{dessart_24ggi_25}, with the  interacting model x6p0 at 388\,d used in this work for \sn. Here, the emergent luminosity has been scaled by the total decay power absorbed in each ejecta model to illustrate how interaction power influences primarily the UV. Bottom: Zoom-in on a number of permitted and forbidden emission lines in the optical and IR ranges and shown versus Doppler velocity.
\label{fig_comp_x6p0_s15p2}
}
\end{figure*}

Figure~\ref{fig_comp_x6p0_s15p2} shows a comparison at $\sim$\,380\,d of the present interacting model x6p0 for \sn\ with the noninteracting s15p2 model \citep{sukhbold_ccsn_16,dessart_sn2p_21} that was later used for modeling the Type II SN\,2024ggi \citep{dessart_24ggi_25} and for which it yielded a satisfactory match from the optical to the IR. This comparison is useful to highlight the importance of interaction, its impact from the UV to the IR, and the variations in IR properties for two different ejecta having essentially the same metal yields. In the top row, the model luminosity versus wavelength is shown, but scaled by the decay power absorbed in each model (which results from the different $\gamma$-ray escape arising from the different \ekin/\mej). Evidently, presented with this normalization, the two models line up beyond 6000\,\AA\ (spectral region powered primarily by radioactive-decay power) and differ dramatically below 6000\,\AA\ and in the UV since interaction power is only introduced in the model x6p0. This implies (and confirms) that the interaction power is strongly channeled into the blue part of the optical and the UV ranges. Interestingly, numerous lines in the UV are present in both models (e.g., \lya), but a factor 10 or 100 stronger in the interacting model. The bottom part of Figure~\ref{fig_comp_x6p0_s15p2} shows in more detail the differences in emission lines. A few lines exhibit an extra, broad boxy emission from the CDS (e.g., \ha\ and P$\alpha$), but otherwise all optical and IR lines at 380\,d primarily arise from the inner, decay-powered ejecta. The differences in expansion rate (i.e., driven by the difference in \mej), lead to strong variations in many lines. These variations reflect the greater $\gamma$-ray escape in model x6p0 (less power absorbed in H-rich material weakens H\one\ lines) but also the lower ejecta density (which strengthens forbidden lines from metals in the IR). The \neiifs\ line is about a factor of ten stronger in model x6p0, and further enhanced by the greater Ne ionization in model x6p0. All these variations in metal-line strength are driven by changes in \ekin/\mej\ rather than composition.

\begin{figure*}[h!]
\centering
\includegraphics[width=\hsize]{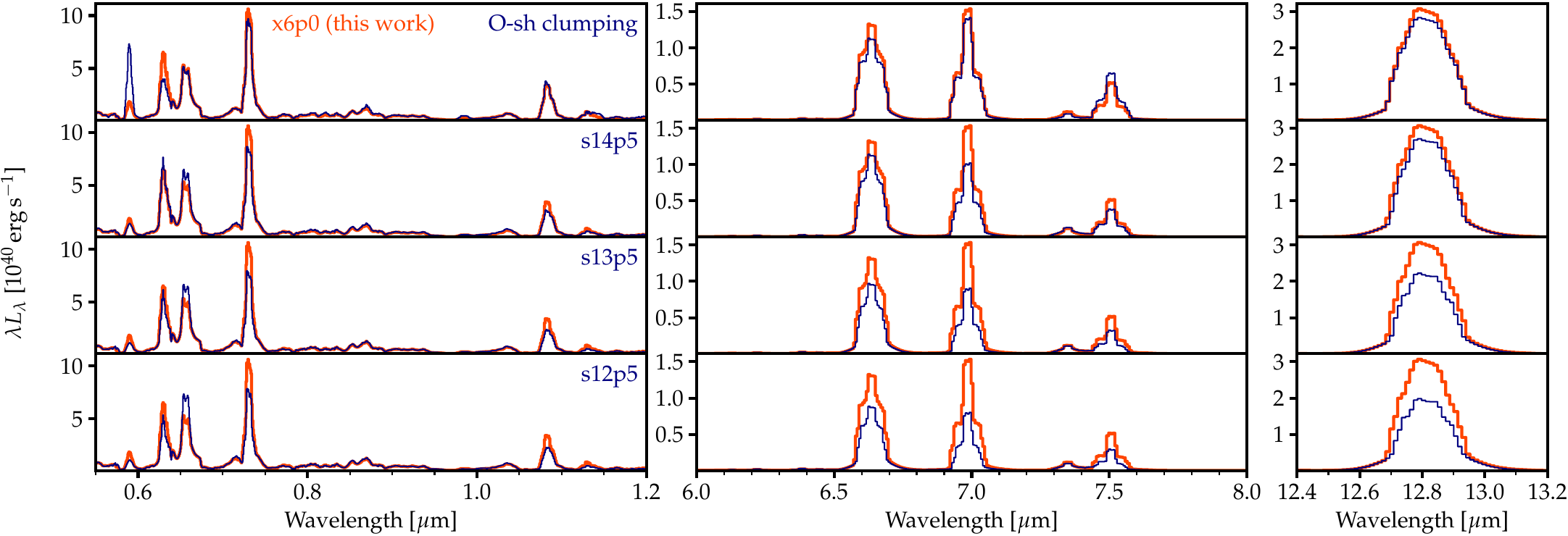}
\caption{Variation of spectral properties in the optical (left), the 7\mic\ region (center), and \neiifs\ (right) resulting from changes in the clumping of the O-rich material (top row) or progenitor mass (lower three rows). All models are at 388\,d and employ the same interaction power of $10^{40}$\,\ergs\ and the same ejecta structure (e.g., density versus velocity) as model x6p0.
\label{fig_dep_ir}
}
\end{figure*}

We can also explore how the results for the interacting model x6p0 would differ with the introduction of clumping or a modulation of its metal yields (Fig.~\ref{fig_dep_ir}). Hence, starting from the ejecta structure (i.e., density, composition etc) of model x6p0, we modified one component and examined the resulting impact on the spectra, focusing on the optical, the 7\mic\ region, and \neiifs\ (left, center, and right columns of Fig.~\ref{fig_dep_ir}, respectively). In our model, introducing a factor of ten clumping of the O-rich material (i.e., clumping the material within the O-rich shells of the shuffled-shell structure shown in Figure~\ref{fig_init}), which is equivalent to introducing a volume filling factor of 10\% for this material, leads to a negligible change in the spectrum apart from a boost of the \naid\ line (it hardly changes the strength of \neiifs). Modifying the progenitor mass from 15.2\,\msun\ to 14.5, 13.5, and 12.5\,\msun\ (using the models of \citet{sukhbold_ccsn_16} and \citet{dessart_sn2p_21}, but with a rescaled \nifs\ mass to 0.05\,\msun), leads to modest variations of a few 10\,\% for all these line diagnostics. The drop in O or Ne abundance with lower progenitor mass does lead to a reduction in \oidoub\ or \neiifs, but the change for the latter is too small to cure the discrepancy with \sn. Overall, these various masses between 12.5 and 15.2\,\msun\ would be compatible with the observations of \sn. The discrepancy with, for example, \neiifs, likely arises from a peculiar Ne abundance or points to a different structure (e.g., asymmetry, or presence of Ne at smaller velocities and higher density). Despite the offsets, these IR lines carry a negligible fraction of the total SN luminosity \citep{dessart_ir_25} --- see also the next section and the illustration of the cumulative flux with increasing wavelength from UV to IR.


\begin{figure*}[h!]
\centering
\includegraphics[width=\hsize]{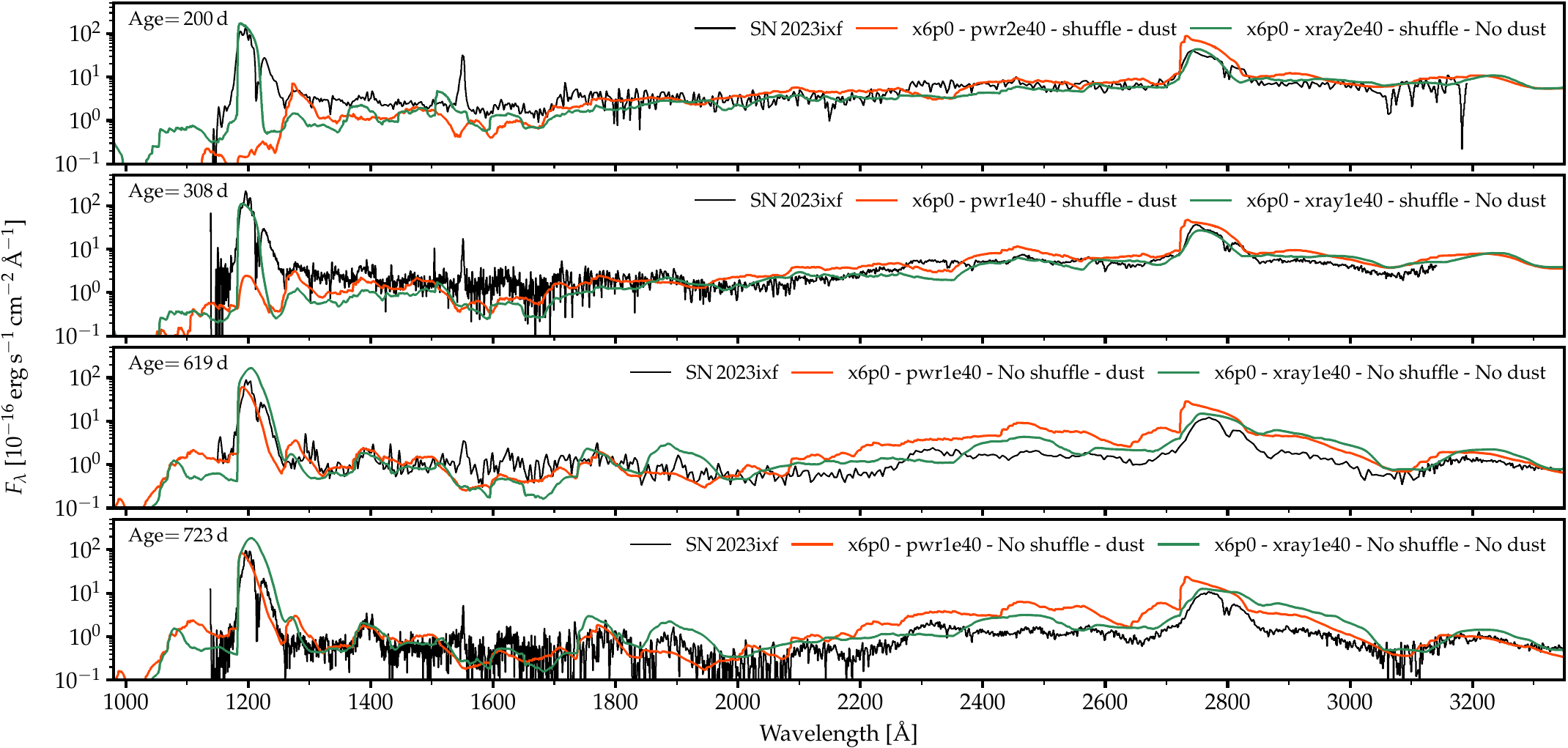}
    \caption{Comparison of the UV observations of \sn\ at 200, 308, 619, and 723\,d from \citet{bostroem_uv_25} and the model x6p0 with power injected within the CDS as high-energy electrons (red) or in the form of X-rays (green).  The latter model ignores dust, although its effect on the UV spectrum is small. Observations and spectra have been rebinned to a fixed resolution of 100\,\kms.
\label{fig_uv_prop}
}
\end{figure*}

\section{UV properties of \sn: Insights from models}
\label{sect_uv}

In this section, we focus on the properties of \sn\ and our models in the UV range, extending the analysis of \citet{bostroem_uv_25} with more simulations and explorations on dependencies and sensitivity of model assumptions. Figure~\ref{fig_uv_prop} compares the HST UV observations at 200, 308, 619, and 723\,d with the models presented in Section~\ref{sect_rt}. The plotting is done in the log to reveal the weaker features, which may be only a hundredth of the \lya\ peak flux but might provide information on diagnostic lines from C, N, O, or intermediate-mass elements (IMEs). However, our simulations predict a UV spectrum entirely controlled by metal-line emission and absorption (the term ``blanketing'' is less appropriate at nebular times since there is no continuum flux to blanket), primarily from once-ionized Fe and secondarily from Ni or Cr. It was not clear from \citet{dessart_csm_22} and \citet{dessart_late_23} whether the UV properties of their models were compromised by the limited number of levels for iron-group elements and because of the neglect of higher ionization states for all C, N, O, and IMEs (they treated C\one\ and C\two, N\one\ and N\two, etc.). However, here, with a huge model atom for metals and ionization states \one\ to \four\ treated for all C, N, O, and IMEs,\footnote{One remaining oversight was to ignore N\five, although its ionization potential is significantly higher than all other ions included. We also find little signature of higher ionization like O\three, both in the model and in the observations. At present, it is not clear whether the 1240\,\AA\ emission is due to N\five\ or comes from \lya.} we continue to find a dominance of Fe line emission and absorption.

The only notable exceptions are the presence of clear lines with \lya, \mgiiuv, and \civuv.
In \sn, \lya\ is the strongest UV line at 200--700\,d, followed by \mgiiuv. Both are broad, strong, and systematically skewed in favor of flux appearing blueward of the rest wavelength (this attenuation of the red side significantly decreases at the later epochs and the \lya\ emission broadens). In our simulations, these lines arise from the outermost parts of the CDS and owe their width to the fast expansion of the CDS. The skewness is due to the optical depth of both the ejecta and CDS, which leads to a greater attenuation of line photons produced from within or interior to the CDS, or from the backside of the ejecta as seen from a distant observer. This attenuation is thus present regardless of dust for the times considered here (i.e., when the CDS is optically thick to UV radiation anyway).

The models also predict the presence of the strong \civuv\ line in all simulations with X-rays (see also discussion in Sec.~\ref{sect_pwr} and Fig.~\ref{fig_xray}), although its shape differs from that observed \citep{bostroem_uv_25}. It is observed as a narrow feature centered on the rest wavelength and thus compatible with emission from X-ray photoionized gas in the unshocked CSM ahead of the SN. In the model, it is produced by the same process of X-ray photoionization but within the outermost parts of the CDS  --- by design the model employs a homologous flow and thus ignores any slow-moving material ahead of the CDS (as there must be in nature).

\begin{figure}
\centering
\includegraphics[width=\hsize]{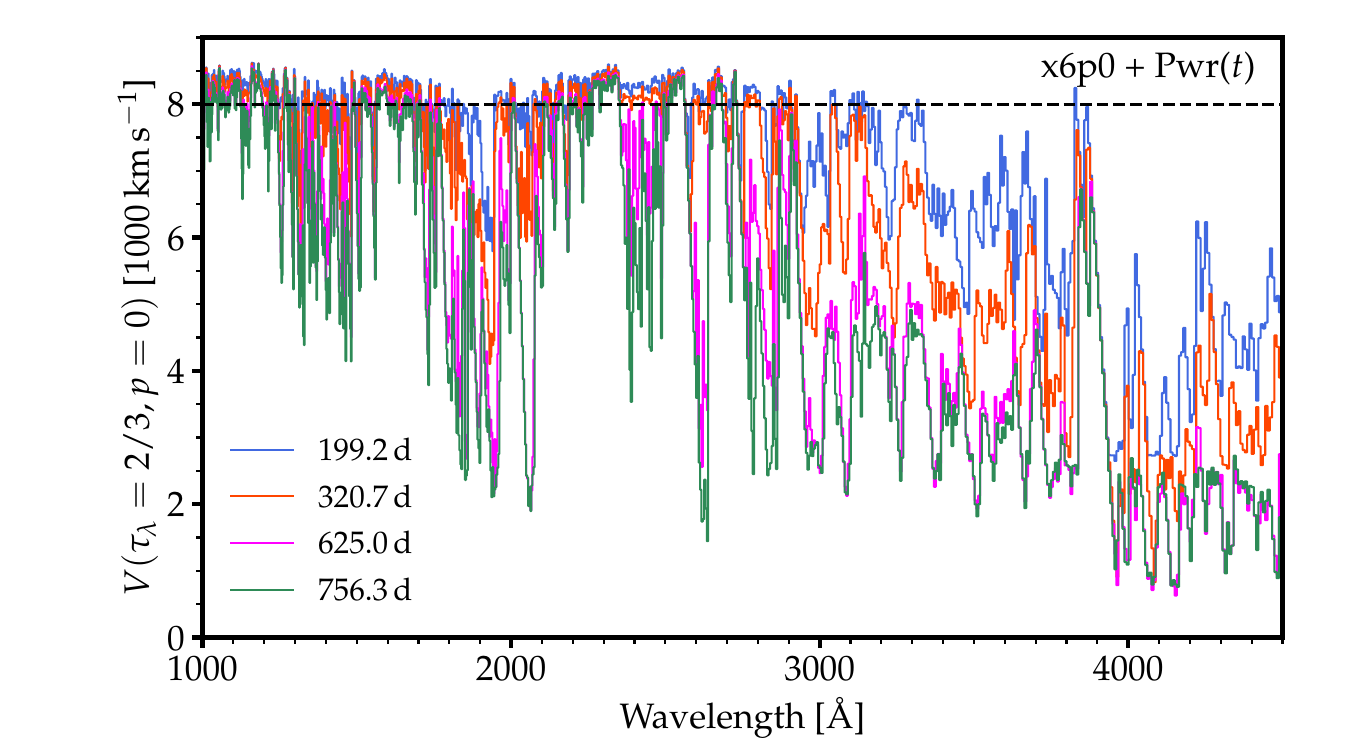}
    \caption{Wavelength dependence of the ejecta velocity $V(\tau_\lambda=2/3)$ at which the radially (i.e., impact parameter $p=$\,0) inward integrated total optical depth is 2/3 for model x6p0 with interaction power and shown as a function of time from about 200 to 760\,d. A rebinning to a resolution of 600 (i.e., 500\,\kms) was applied to reduce the jaggedness of the curves. The abscissa covers from the UV, which is overall optically thick but with localized opacity holes, to the optical, which is less optically thick but with clear opacity bumps associated with lines (mostly from Fe). The ejecta are optically thin to electron scattering at all epochs shown. The dashed-black line indicates the CDS velocity.
\label{fig_vtau_uv}
}
\end{figure}

Figure~\ref{fig_vtau_uv} illustrates the optical-depth effects that persist in the interacting model x6p0 during the nebular phase. It shows that the effective photospheric velocity (as obtained by accounting for all opacity sources) remains located within the CDS essentially throughout the UV, with only gaps of lower opacity, whereas this photospheric velocity is much lower in the optical but with gaps and bumps still present due to lines and in particular from Fe\two\ or Ca\two\,H\&K. The recession of the photospheric velocity with time is also significant in the optical, but hardly present in the UV. The attenuation of the flux from within the CDS, from its limbs, and from the receding part of the CDS naturally follow. Some caution is needed to interpret this figure, though. Under optically thick conditions (e.g., during the so-called photospheric phase), the photosphere is the median location in the ejecta for the emergence of photons. At nebular times, this notion breaks down as emission arises from optically thin regions by definition, at least in terms of electron-scattering optical depth. Obviously, the role of lines is critical, especially when considering the UV range.

\begin{figure}
\centering
\includegraphics[width=\hsize]{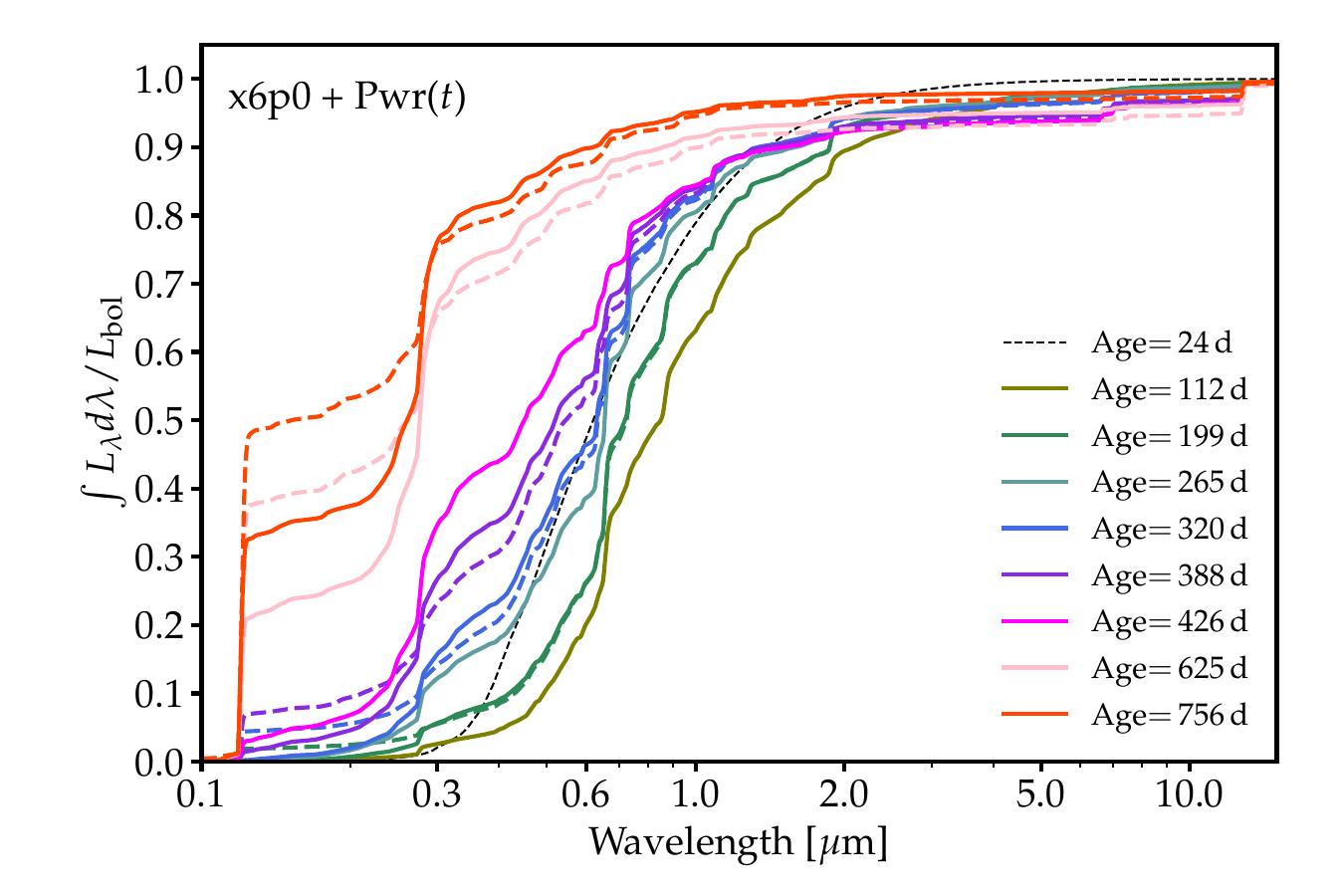}
    \caption{Illustration of the cumulative flux integrated from 0.1 to 15\mic\ for model x6p0 with interaction power. The dashed gray line shows the result at 24\,d (from \citealt{dessart_23ixf_phot_26}). The thick colored lines correspond to the case wherein shock power is injected as high-energy electrons within the CDS, covering the nebular-phase evolution from 112 to 756\,d. The thick dashed lines correspond to model counterparts wherein shock power was injected as X-rays.
\label{fig_cum_flux}
}
\end{figure}

Figure~\ref{fig_cum_flux} documents further how the spectral energy distribution evolves in time and should be considered alongside Fig.~\ref{fig_vtau_uv} to assess optical-depth effects on the emerging radiation. Despite the strong interaction that prevailed in \sn\ at early times in the post-SN~IIn phase \citep{bostroem_23ixf_24,nayana_23ixf_25,zimmerman_23ixf_24}, the fractional flux in the UV at 24\,d was just a few percent of the total (it remains at that level throughout the photospheric phase). However, the UV flux strengthens progressively in the nebular phase, with marked jumps at \lya\ and \mgiiuv\ (and its vicinity). In the models with X-rays, \lya\ contains 50\% of the total flux at $\sim$\,800\,d, as obtained previously by \citet{dessart_late_23}. This reinforces the importance of far-UV observations and \lya\ for assessing the properties of CSM interaction.

As is shown above in Figure~\ref{fig_vtau_uv}, the entire UV flux emerges from around the CDS, and thus one expects that the UV properties will strongly depend on the adopted CDS structure. Unfortunately, the CDS structure is expected to be more complex than adopted here (a spherical shape with a Gaussian profile in velocity), even if allowed to be clumpy. It should be broken into dense clumps and tenuous interclump material, and probably turbulent \citep{chevalier_snr_92,chevalier_blondin_95}. Rather than changing the 1D structure of the CDS in \cmfgen, we have varied the way we inject the shock power. Above, we demonstrated that injecting X-rays makes a significant difference, in particular to \lya\ (e.g., Sec.~\ref{sect_pwr}). This is thought to arise not from a boost of the emission itself, but rather from the rise in ionization that inhibits Mg\two\ cooling outside of the CDS. Since hydrogen has no other ion to which to ionize, \lya\ is far less sensitive to this ionization change. And whatever flux is not carried away by the Mg\two\ line, it is taken care of by \lya. Furthermore, this ionization rise shuts off absorption by metal lines, which is quite strong at 200\,d.

Taking the interacting x6p0 model computed at selected epochs between 199 and 756\,d, we have explored other changes that confirm this (all are shown in the Appendix in Fig.~\ref{fig_vdep}). We found that reverting to the model atom of \citet{dessart_csm_22} has little impact at 200\,d through most of the UV except with the reduced blanketing in the \lya\ region, which is boosted as a result. Shifting the velocity centroid for the injection of shock power from 8000 to 8050, 8100, and 8200\,\kms\ (which is equivalent to injecting power in CDS regions away from the maximum density at 8000\,\kms) tends to boost the far-UV flux and \lya. The reason is that this raises the temperature from the outermost grid regions and shuts off the blanketing. The same effect was obtained by trimming the grid so that it stops just outside of the CDS (i.e., at 8200 rather than 8600\,\kms). This change cancels any absorption that was taking place outside the CDS, and this absorption, mostly impacting \lya, was substantial. The difference between the present simulations and those of \citet{dessart_late_23} are in part related to the different CDS structure employed, something we realized late in the project (see Fig.~\ref{fig_comp_x6p0_d23}). However, by the time the simulations have reached 700\,d, these shortcomings disappear, primarily because of the reduction in optical depth. Obviously, these statements apply to the tests we have done, which included relatively modest variations in the treatment of shock power.


\begin{figure*}[h!]
\centering
\includegraphics[width=\hsize]{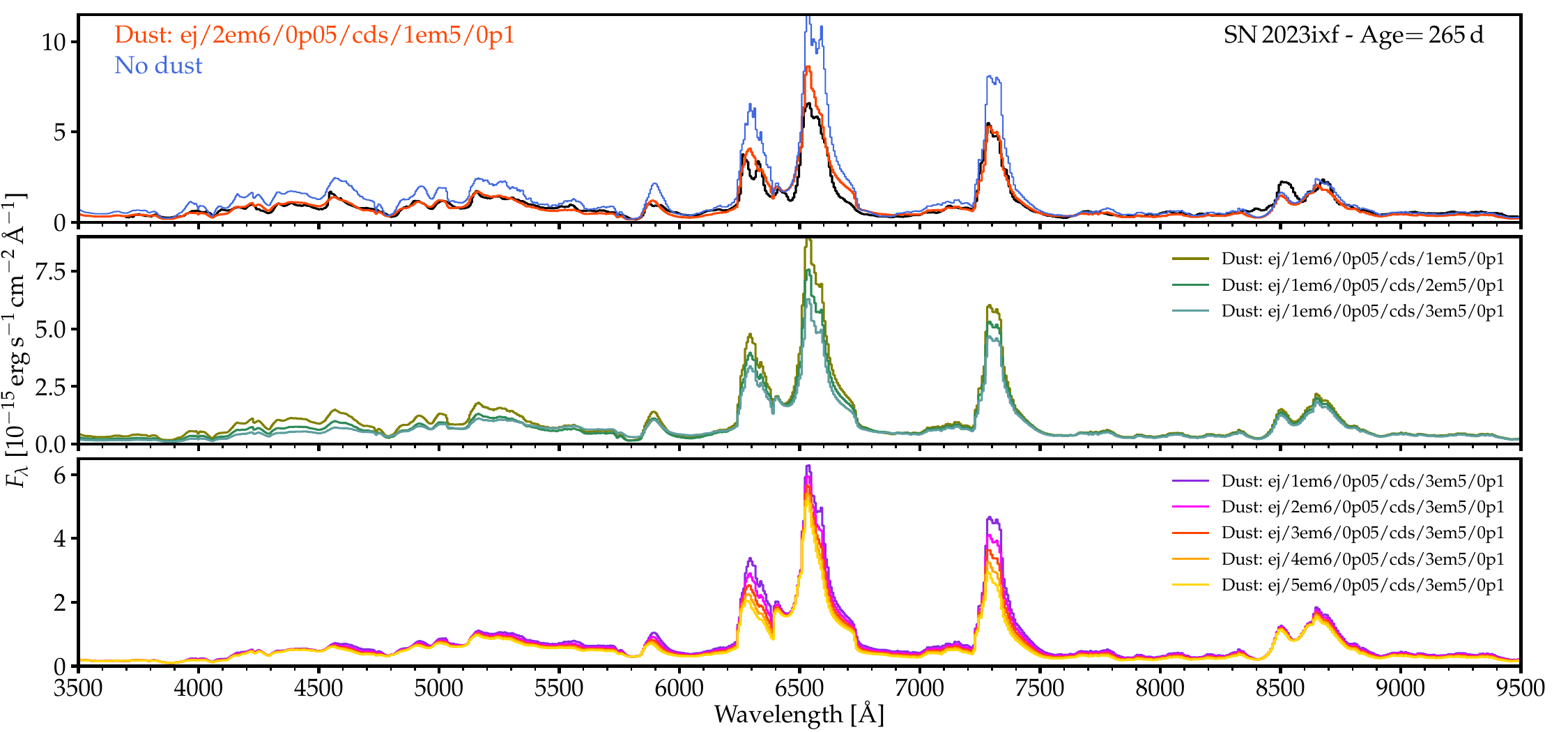}
\caption{Exploration of the impact on the emergent spectrum of various amounts of dust in the inner ejecta and in the CDS for the interacting model x6p0 at 265\,d (see also Sec.~\ref{sect_265d} and Fig.~\ref{fig_spec_265d}). In the top panel, the best-match model with dust is compared to \sn\ (the dust-free counterpart is shown in blue), whereas the other two panels illustrate the spectral variations resulting from changes in the dust mass in the CDS (middle) or in the inner ejecta (bottom). Similar explorations at other epochs are shown in Figure~\ref{fig_dust_other}. The labels give the ejecta (``ej'') and CDS (``cds'') dust properties, with the mass in \msun\ (1em6 stands for $10^{-6}$\,\msun) and the temperature in $10^4$\,K (0p1 stands for 1000\,K).
\label{fig_dust_265d}
}
\end{figure*}

\section{Dust evolution in \sn: results from models}
\label{sect_dust}

Visible signatures of the presence of dust in core-collapse SN ejecta may be a blue-red asymmetry in emission0line profiles (typically observed in the optical), a brightness reduction (in the UV and optical), or a brightness excess (typically in the IR), as inferred in SN\,1987A after 500\,d post-explosion \citep{lucy_dust_89}. Apart from SN\,1987A, essentially all SNe II are located beyond several Mpc and become undetectable after 500\,d owing to the exponential decline of radioactive-decay power (even steeper with $\gamma$-ray escape), preventing a similar analysis of their dust properties --- they are too faint for any observation or analysis to be performed. The only exception to this are transients in which an extra power source, in essentially all cases CSM interaction (one notable exception is the stripped-envelope SN\,2012au; \citealt{milisavljevic_12au_18}), allows for continued monitoring, sometimes for many years after explosion. This has been possible in a number of Type II SNe such as SN\,2004et and 2017eaw  \citep{weil_17eaw_20,shahbandeh_jwst_23,pearson_17eaw_25}, in some Type IIb SNe like SN\,1993J \citep{matheson_93j_00a,bevan_dust_17,szalai_93J_25}, in the transitional Type Ib SN\,2014C \citep{milisavljevic_14C_15,tinyamont_14C_25}, or here with \sn\ \citep{medler_23ixf_25,wynn_sed_25,singh_23ixf_26}. Many transients that eventually formed dust were classified as interacting SNe directly at discovery (see, e.g., the sample from \citealt{niculescu_duvaz_dust_22}). Hence, these observations of dust are perhaps not saying much about dust formation in core-collapse SNe but more about dust formation in the peculiar configuration of ejecta interaction with CSM.

In Section~\ref{sect_rt}, we presented multiepoch radiative-transfer models for \sn, covering from the UV to the IR whenever possible. The first signatures of dust appeared at 200\,d with the NIR excess reported by \citet{park_23ixf_25}. However, with our consistent modeling of the full ejecta and CDS emission across the electromagnetic spectrum, we demonstrated that a global reduction of the optical flux also requires dust --- neglecting dust yields a systematic excess flux in the optical range of several 10\,\%. To make the model flux consistent with the observed flux at $\sim$\,200\,d, we introduced $10^{-6}$\,\msun\ of dust in the inner ejecta but more importantly $5 \times 10^{-6}$\,\msun\ of dust with a temperature of 1330\,K in the CDS. With this choice, both the NIR emission and the optical attenuation are roughly matched (Fig.~\ref{fig_spec_208d}).

As time progressed, more dust was added to the inner ejecta and to the CDS, as there was a greater need to reduce the optical emission as well as introduce a mild reduction of the flux appearing longward of the rest wavelength of most lines forming in the inner ejecta (e.g., the narrow components of \oidoub, \ha, or \caiidoub). With observations at 265\,d, we find that in addition to the dust we introduced in the inner ejecta and CDS, there must be an additional cold component. Furthermore, with the decreasing brightness of the emission from the inner ejecta (because of $\gamma$-ray escape and dust attenuation), it becomes harder to quantify the amount of dust in the inner ejecta at later epochs. To summarize the results obtained in Section~\ref{sect_rt}, the inferred dust mass below 3000\,\kms\ in \sn\ was found to rise from 1, 2, 5, 10, 20, to $100 \times 10^{-6}$\,\msun\ at epochs 208, 265, 300, 379, 442, and 620\,d. We also inferred a CDS mass of 5, 10, 20, 7, 30 and 100 at epochs 208, 265, 329, 379, 442, and 620\,d. These values were obtained with a mixture of Si-rich and C-rich dust in the proportion 1 to 4.5 (see Sec.~\ref{sect_setup}). The results are in qualitative agreement with those of \citet{singh_23ixf_26}, although their inference is based on modeling the blue-red asymmetry of the \ha\ profile and the IR excess. Our approach is instead on a full model of the ejecta and interaction and covers from the UV to the IR. Probably the most striking result from our modeling is that CDS dust is essential to match the optical brightness evolution of \sn, with a very early formation at 200\,d relative to the corresponding epoch of 500\,d in SN\,1987A.


Figure~\ref{fig_dust_265d} illustrates the distinct influence of dust formation in the inner ejecta or in the CDS of the model x6p0 with interaction power for \sn\ at 265\,d (illustrations for other epochs are shown in the Appendix in Fig.~\ref{fig_dust_other}). In the top panel, we show the observations together with the best-match model including dust and the model counterpart without dust. The other two panels show the influence of varying the dust content in the CDS or in the inner ejecta. In this example, varying the CDS mass from 1 to $3 \times 10^{-5}$\,\msun\ leads to a global reduction of the emission from the decay-powered part of the ejecta, which is all interior to the CDS. This acts as a uniform blanket, but with a greater impact for emission arising from larger radii (emission below 5500\,\AA) relative to emission arising from smaller radii (Fe emission beyond 6000\,\AA) because of the different covering fraction of the dusty CDS on the underlying emission (i.e., a tiny volume or a large volume). The greater attenuation at shorter wavelengths also results from the greater dust absorption below 6000\,\AA\ (Fig.~\ref{fig_dust}). A blue-red asymmetry is clearly visible on the broad, boxy component of \ha, with the flux remaining essentially unchanged in between \oidoub\ and \ha\ but dropping significantly as one progresses toward the red edge of \ha. When one varies instead the dust mass within the inner 3000\,\kms\ of the ejecta at that time, the strongest visible impact appears on \naid, \oidoub, \ha, and \caiidoub, with both a global reduction of the emission and a strong deficit on the red side. For lines like Mg\one]\,4571\,\AA, its formation within the same region as numerous Fe\two\ lines can also cause some blue-red asymmetry --- at least some blue-red asymmetry is present in our dust-free models (see, e.g., Fig.~\ref{fig_spec_379d}).


\begin{figure}
\centering
\includegraphics[width=0.9\hsize]{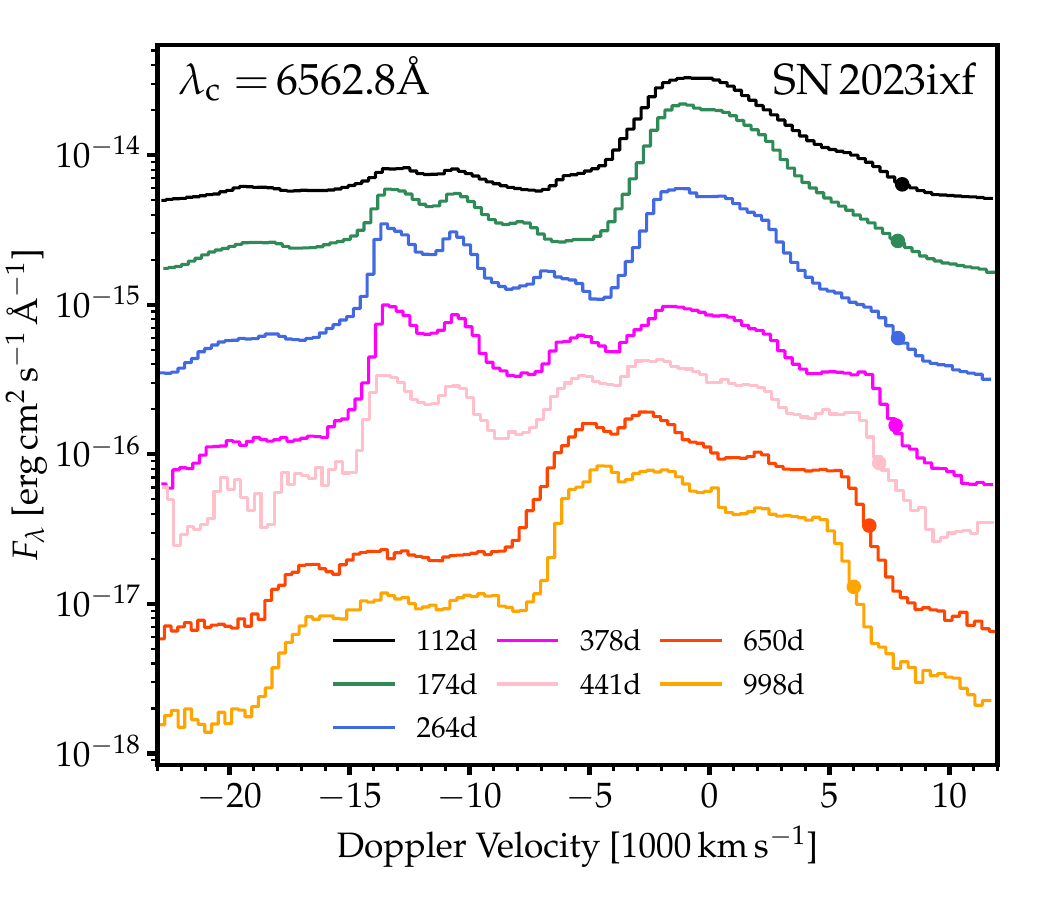}
\includegraphics[width=0.9\hsize]{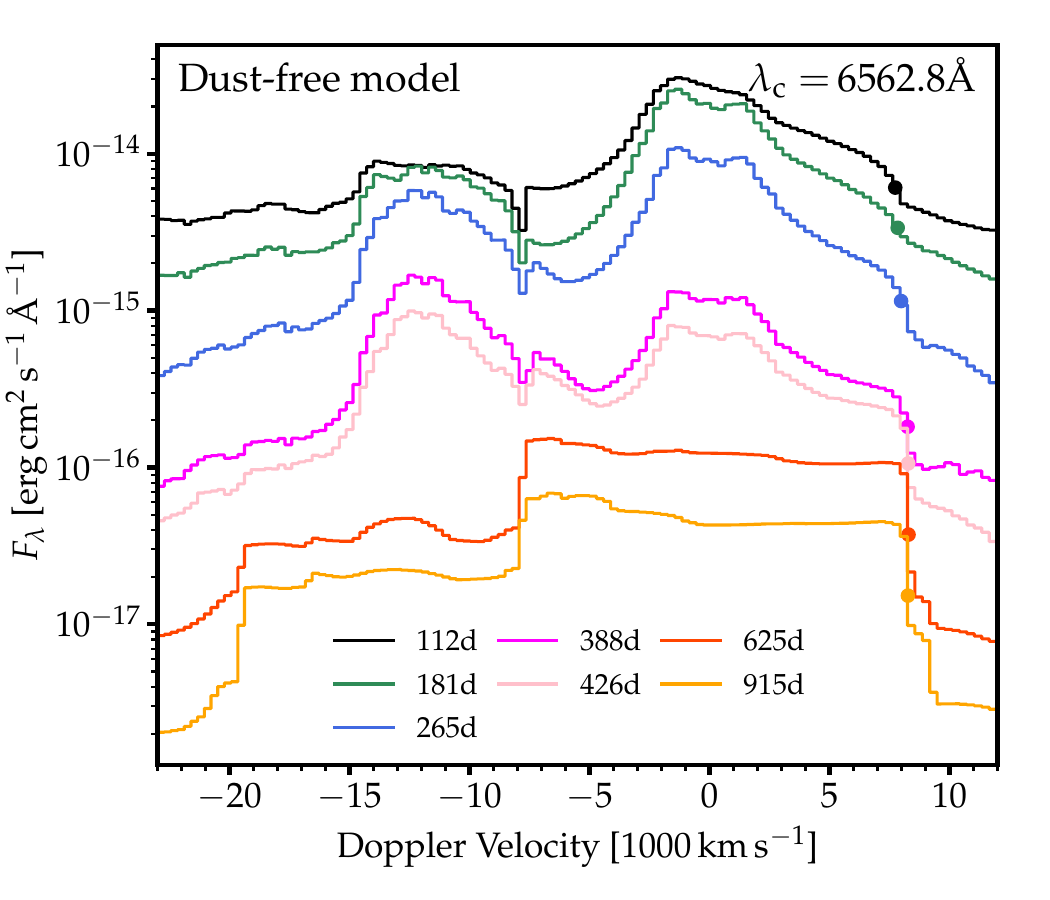}
\includegraphics[width=0.9\hsize]{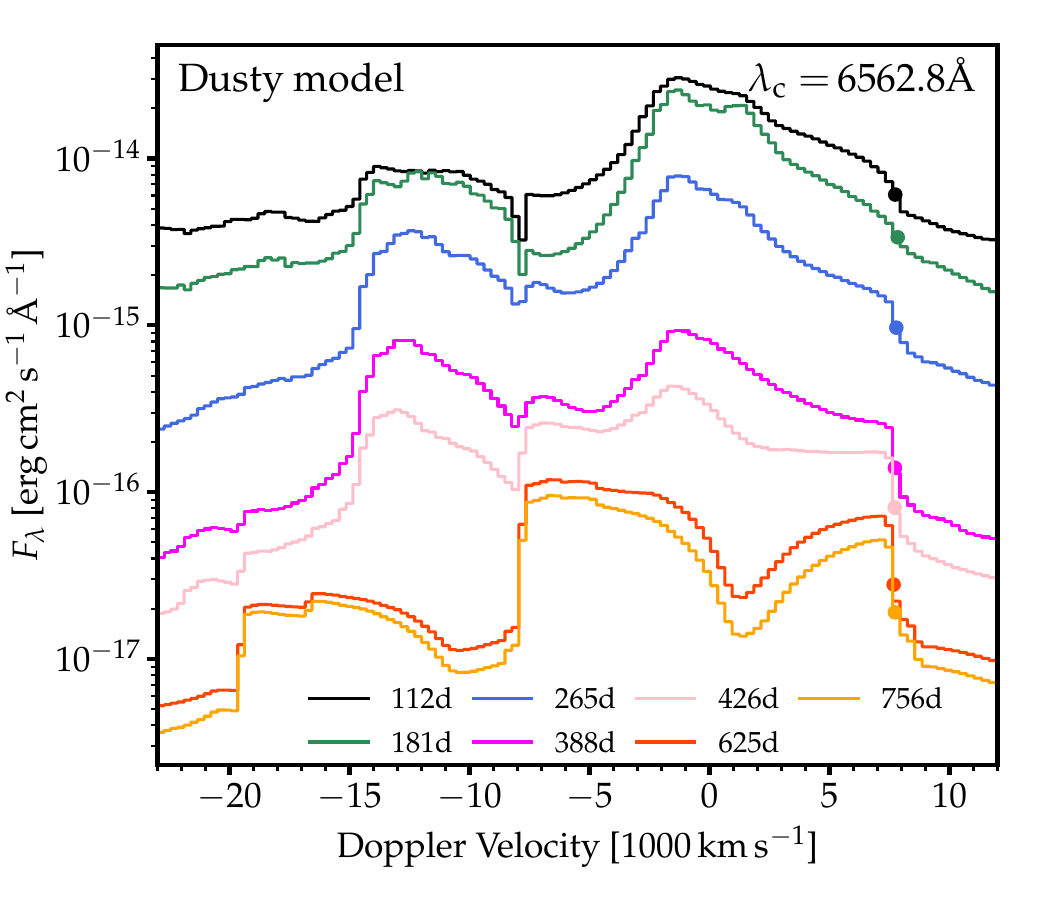}
\caption{Evolution of the region encompassing the \oidoub\ and \ha\ profiles from 112 until 998\,d for \sn\ (top) as well as the best-matching models from Section~\ref{sect_rt} including the dust-free (middle) and dusty (bottom) counterpart (not all epochs coincide). The filled dots indicate the location where the flux is half that at the red edge of the \ha\ emission. The rest wavelength of \ha\ corresponds to 0\,\kms. Observations have been corrected for redshift and reddening and models have been scaled to the distance of \sn. The logarithm of the flux is shown for better visibility. See Section~\ref{sect_ha} for discussion.
\label{fig_evol_ha}
}
\end{figure}
\begin{figure}
\centering
\includegraphics[width=0.9\hsize]{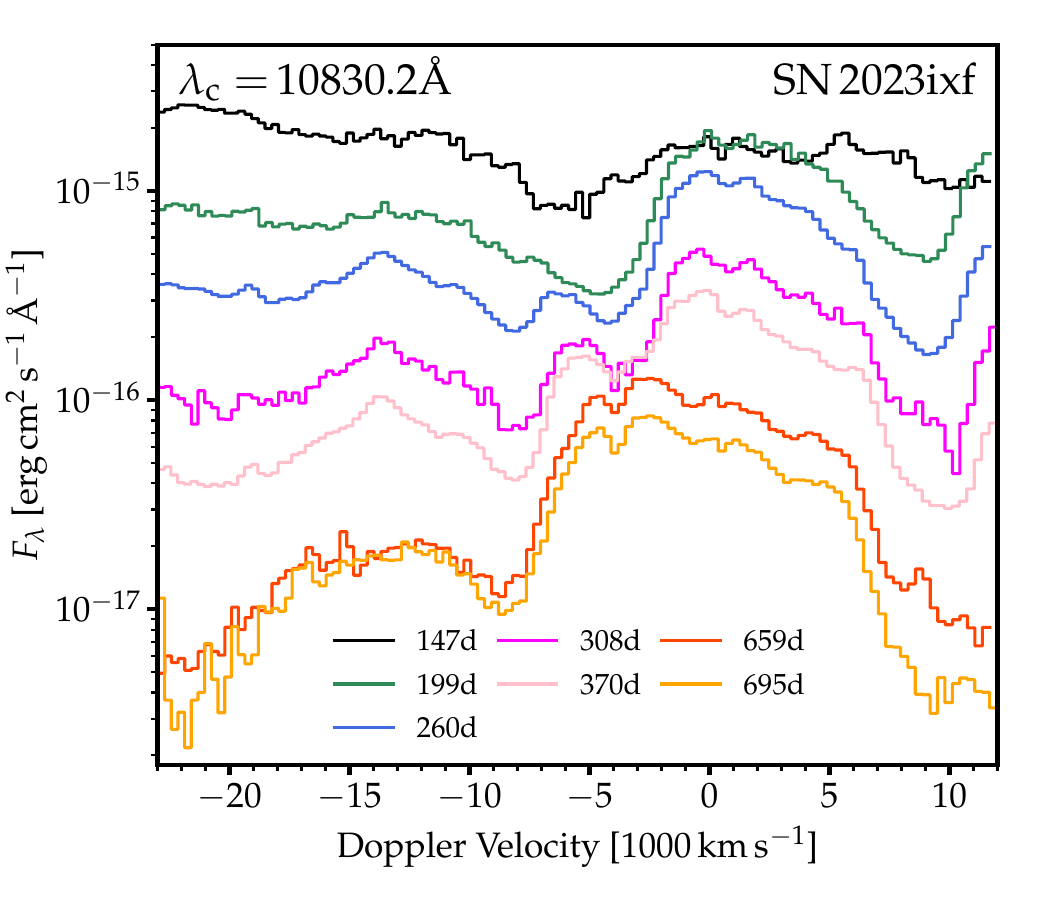}
\includegraphics[width=0.9\hsize]{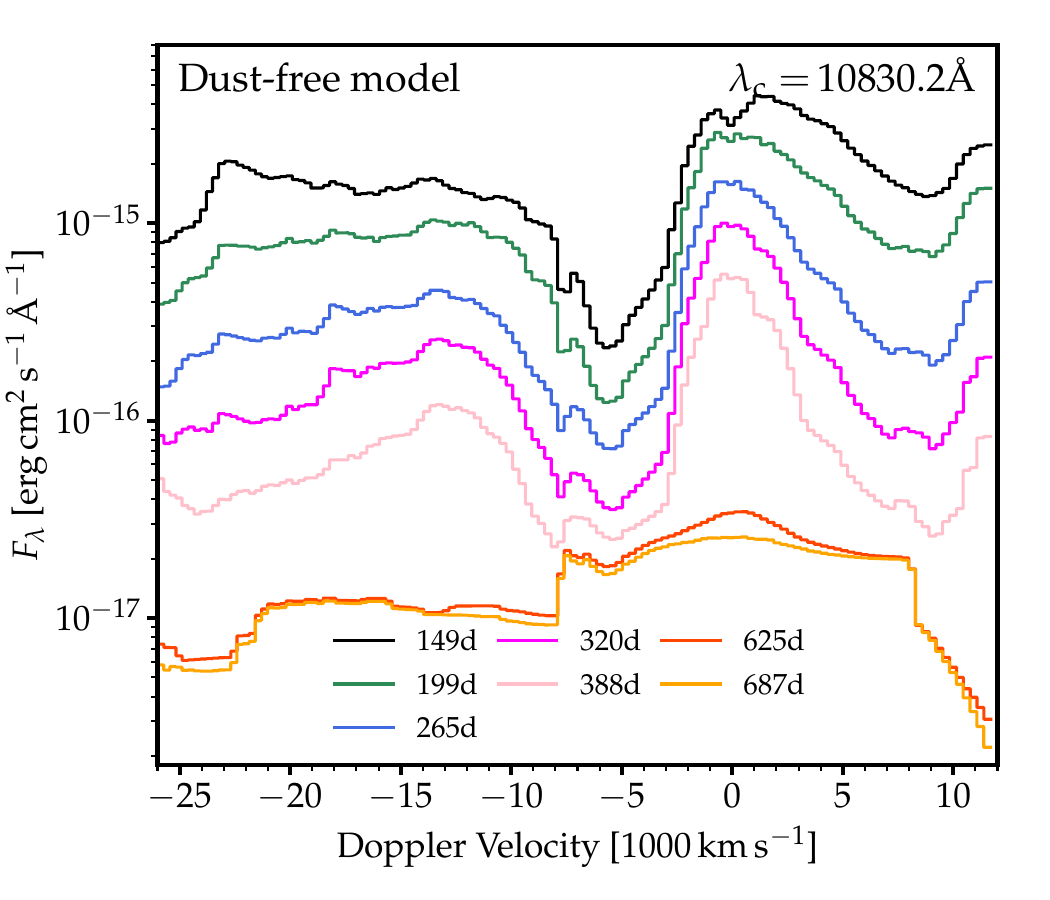}
\caption{Same as Figure~\ref{fig_evol_ha} but now for the evolution of the He\one\,10,830\,\AA\ region. Only the dust-free model is shown (bottom panel).
\label{fig_evol_hei}
}
\end{figure}

\section{Evolution of \oidoub\ and \ha\ profiles}
\label{sect_ha}

Figure~\ref{fig_evol_ha} illustrates in more detail the evolution of the \oidoub\ and \ha\ region of \sn\ from 112 until 998\,d, together with the roughly contemporaneous best-matching models shown in Section~\ref{sect_rt} with and without dust. On this adopted log-scale, the presence of excess emission on the red side of \ha\ is more obvious, together with the presence of \oidoub\ as early as 112\,d. Initially, though, both transitions have strong emission arising from the inner, decay-powered parts of the ejecta and the emission is primarily narrow (i.e., a few 1000\,\kms\ wide). As time progresses, the broad component comes to dominate, and this occurs with a delay for \oidoub. The presence of broad, boxy \ha\ was at the origin of the emission in between \oidoub\ and \ha\ (in SNe without a CDS nor interaction, such as SN\,1999em, the flux goes essentially to zero between these to emission features; \citealt{leonard_99em}). With the simultaneous presence of broad, boxy \oidoub\ and \ha, the \ha\ profile now produces a bump on its blue side, which yields a blue-red asymmetry unrelated to dust.

\begin{figure*}
\centering
\includegraphics[width=0.8\hsize]{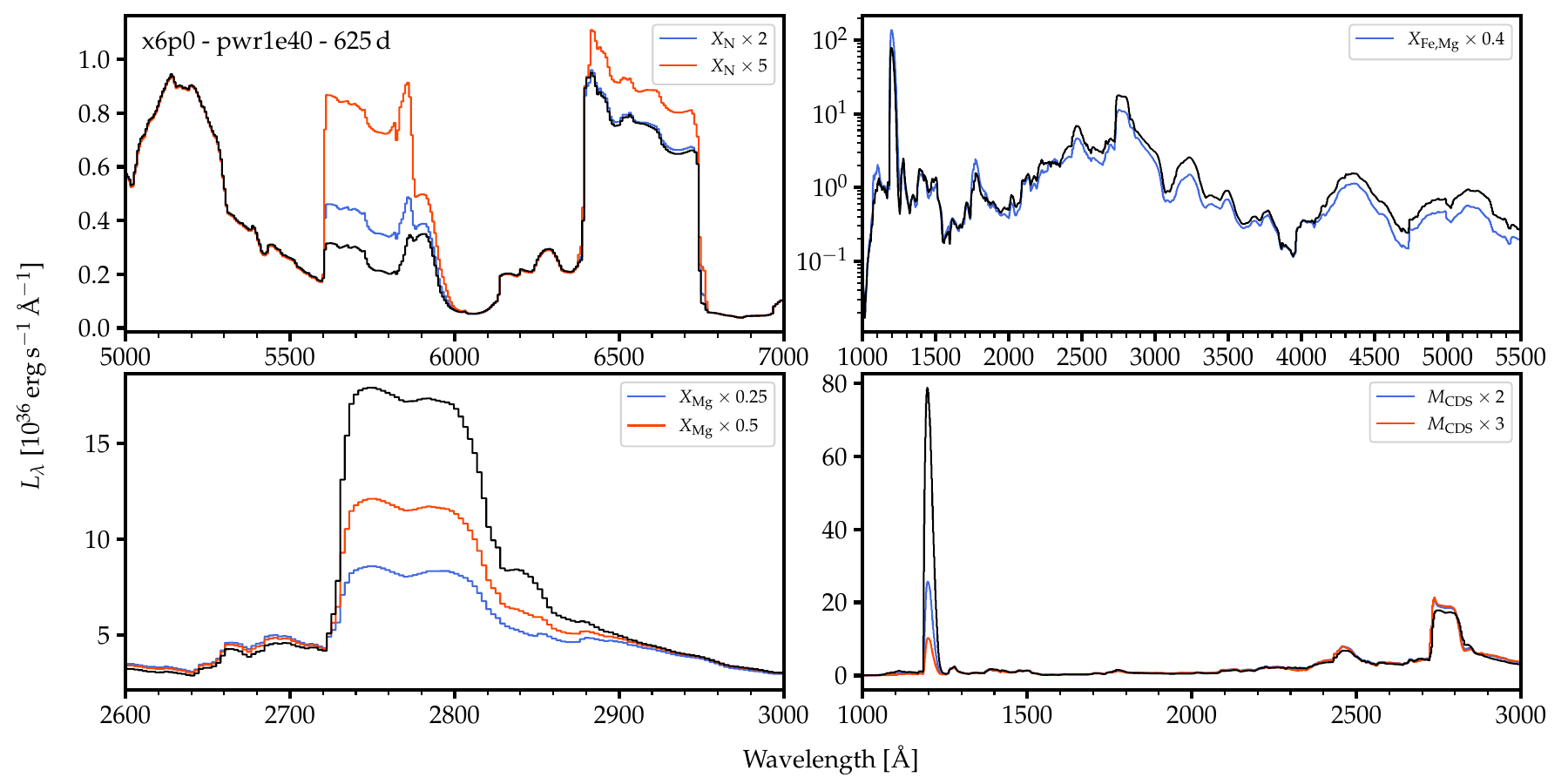}
\caption{Influence of changes in N, Mg, or Fe composition (top row and bottom-left panels) and CDS mass (bottom right) on the spectral properties of the dust-free x6p0 model with an interaction power of $10^{40}$\,\ergs\ at 625\,d. All spectra were rebinned to a resolution of 1000 (i.e., 300\,\kms). See Section~\ref{sect_dep} for discussion.
\label{fig_dep_620d}
}
\end{figure*}

Another feature clearly evident in Figure~\ref{fig_evol_ha} is the reduction of the \ha\ profile width, suggesting CDS at 8000\,\kms\ at 112\,d and at $\sim 6500$\,\kms\ at 998\,d (a filled dot indicates this velocity as the midpoint in flux in the red ledge of \ha). This is a very significant deceleration suggesting a growth in CDS mass by a few 0.1\,\msun\ over that timespan \citep{dessart_wynn_23}. In our modeling, we have neglected this aspect, which may in part explain some of the deficiencies of our models, such as the underestimate of the \ha\ flux. The problem is, however, more complicated that this since the optical appearance reflects also the CDS dust properties and what changes occur in the UV, in particular for \lya.


\section{Evolution of the He\one\,10830\,\AA\ region}
\label{sect_hei}

Figure~\ref{fig_evol_hei} is a counterpart of Figure~\ref{fig_evol_ha} but for the \heinir\ region, which is a complicated blend of numerous transitions. Initially, the line is blended with \pag, but a broad component seems to be present at all times and strengthens considerably after 300\,d. In the (dust-free) model, the broad, boxy component eventually forms but at much later times (after 600\,d), and remains much weaker than observed by a factor of a few. The reason for the mismatch is currently unclear. Similarly, on the blue side of \heinir, several lines are present and overlap at 150\,d (i.e., contributions from S\two, Fe\one, and  N\one), but broad emission due primarily to [S\two]\,1.032\mic\ dominates at late times. This evolution illustrates the drastic change in the nature of SN radiation, with the progressive dominance of emission from the CDS and the appearance of new coolants.


\section{Dependencies on CDS composition and mass}
\label{sect_dep}

Figure~\ref{fig_dep_620d} illustrates the spectral sensitivity to changes in ejecta or CDS composition and structure in the model at 620\,d (Fig.~\ref{fig_spec_620d}). At this time, essentially all the emission arises from the CDS because the interaction power dominates over the absorbed decay power in the inner ejecta but also owing to the presence of dust in those inner regions. In this experiment, we scale the abundance of a given species in the outer ejecta and CDS. This scaling progressively departs from unity beyond about 4500\,\kms. It corresponds to adjusting the metallicity of the progenitor star when scaling Fe or Mg or invoking dredge-up from the outer H-burning zone when adjusting the N abundance. A renormalization is applied after the abundance scaling, either to all species for Fe or Mg (the change in abundance is very small), or to He only when scaling N.

Figure~\ref{fig_dep_620d} indicates that scaling up the nitrogen abundance (top-left panel) impacts two lines and specifcally \niidoub, which overlaps (in fact nearly coincides) with \ha, and \niiauroral, which overlaps only partly with \naid\ as well as with some Fe\two\ emission. Estimating the nitrogen contribution to the strong \ha\ component is difficult, but \niiauroral\ (present in our models but apparently absent in \sn) offers a better way --- this line is observed in SN\,1993J (see next section). Variations in Fe abundance impact the UV and optical range, but varying the Mg abundance is largely limited to changing the strength of \mgiiuv. Modulations in Fe or Mg emission in the UV are compensated by the opposite variation in \lya\ so that the total flux remains unchanged. Finally, we explored the impact of increasing the CDS mass, all else being the same. This reduces the \lya\ flux which instead comes out at longer wavelength, with an increase in Fe and \ha\ emission in the optical.

These explorations are another motivation to build hydrodynamically consistent models of the ejecta and CDS when modeling in detail transients like \sn. We suspect a number of deficiencies in the models we presented in this work result from adopting ejecta and CDS structures that are not quite right (although not inadequate given the satisfactory matches for many epochs).


\begin{figure}[h!]
\centering
\includegraphics[width=\hsize]{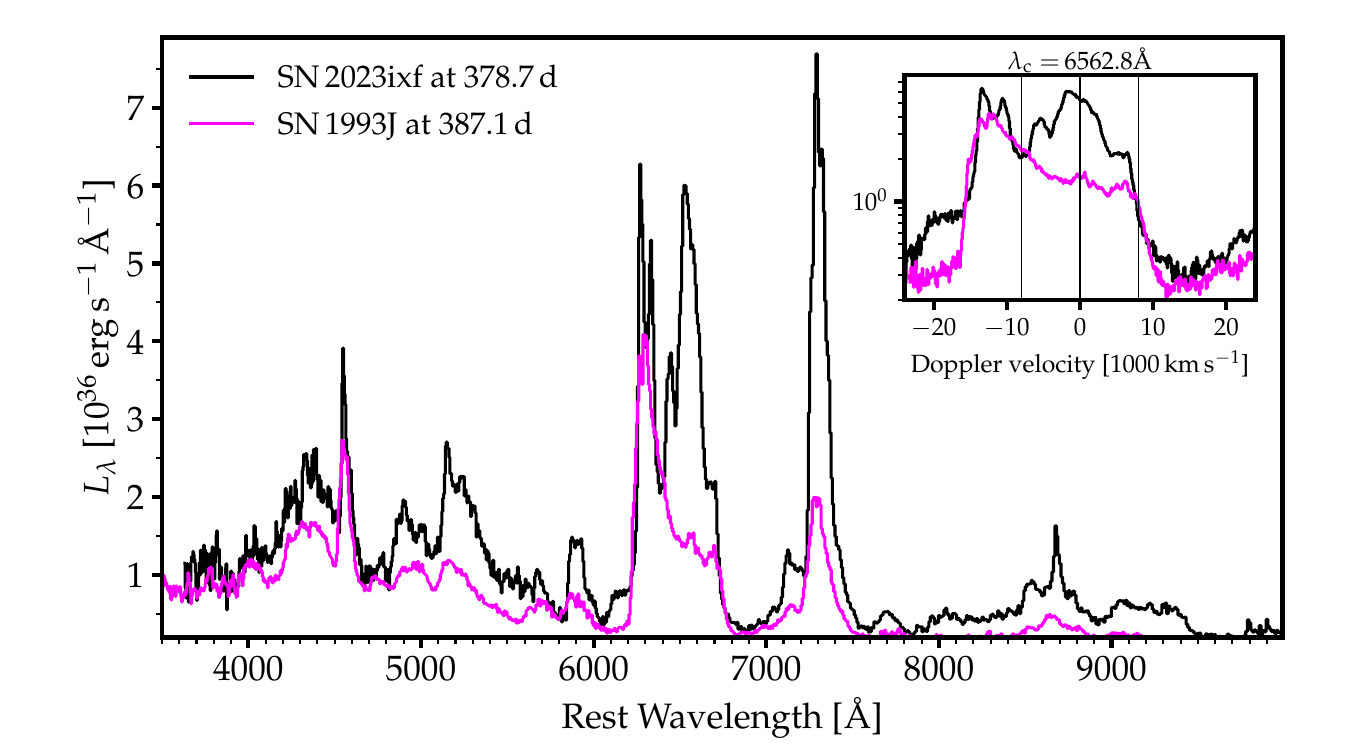}
\includegraphics[width=\hsize]{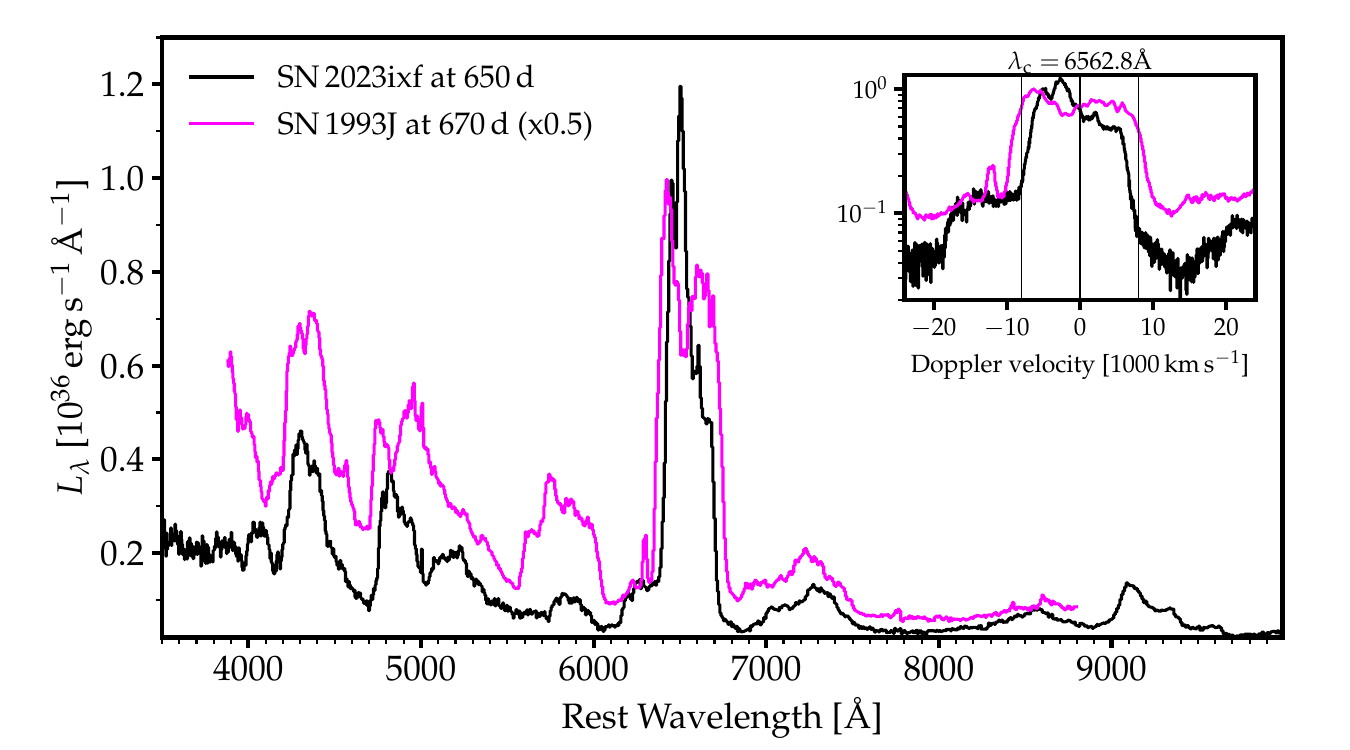}
\includegraphics[width=\hsize]{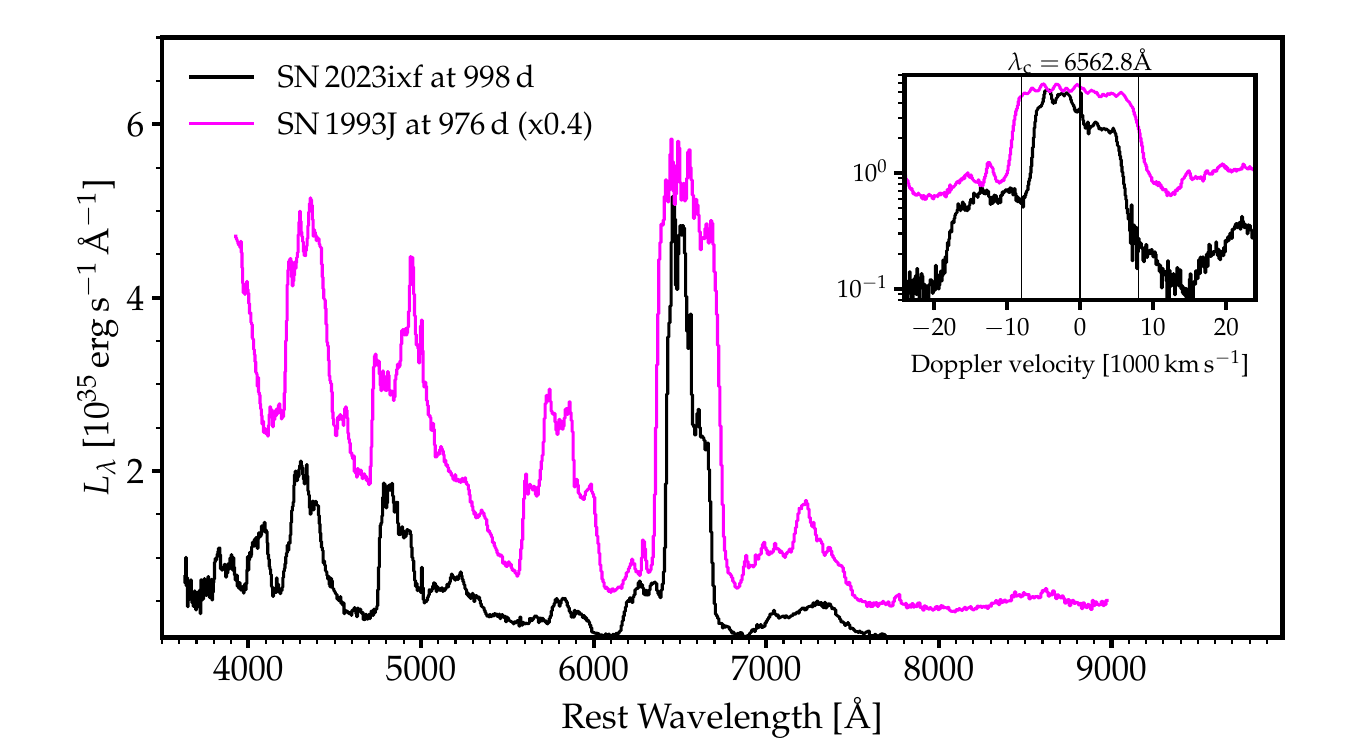}
\caption{Comparison of \sn\ with SN\,1993J at about 380, 660, and 980\,d after explosion. Each spectrum, which agrees within a few 0.01\,mag of the corresponding observed $V$-band magnitude, is shown as luminosity after correcting for the distance and extinction, as well as the redshift. Additional scaling is applied to SN\,1993J at the last two epochs for better visibility. The inset provides a zoom-in view of the \ha\ region (luminosity shown in log space), with vertical bars indicating the Doppler velocity of $\pm$\,8000\,\kms.
\label{fig_93J_23ixf}
}
\end{figure}

\section{Comparison with SN\,1993J}
\label{sect_93J}

In this section, we compare the observations of \sn\ at about 1, 2, and 3\,yr post-explosion with the nearly contemporaneous observations of SN\,1993J. These two transients correspond to very different SNe (Type IIL, IIP, or IIn depending on the nomenclature one wishes to use for \sn\ \citep[e.g.,][]{zheng_23ixf_25}, versus Type IIb for SN\,1993J \citep{filippenko_93J_93,richmond_93J_94}; see, e.g., \citet{filippenko_rev_97} for a review of SN classification), and thus very different ejecta masses and in particular H-rich envelope masses (several solar masses for \sn\ but $\sim 0.1$\,\msun\ for SN\,1993J; \citealt{blinnikov_94_93j,nomoto_93j_93,podsiadlowski_93j_93,woosley_94_93j}). More importantly, \sn\ exhibited strong signatures of interaction with optically thick CSM for about a week after first light, and continuous interaction with optically thin CSM of ever decreasing density thereafter. In contrast, SN\,1993J showed no sign of interaction with optically thick CSM at early times, and no obvious sign of interaction (as far as optical spectra are concerned) with optically thin CSM until about 1\,yr post-explosion. Despite these differences, there are numerous similarities in their radiative properties past 1\,yr.

Figure~\ref{fig_93J_23ixf} compares the optical spectra of \sn\ and SN\,1993J (for the latter, we used a time of explosion of  MJD $=$\,49074, $E(B-V)=$\,0.2\,mag, $z=$\,0.000113, and a distance of 3.63\,Mpc; \citealt{matheson_93j_00a}) at about 370, 670, and 970\,d after explosion (for the last two epochs, the spectra were scaled to match the $V$-band photometry of \citet{richmond_93J_96} and \cite{zhang_93j_late_04}; additional scaling was introduced in the plot to improve the visibility).

The first striking impression from these three panels is that these two SNe share many similarities with the progressive domination of broad, boxy line profiles, with nearly vertical line wings in numerous features. At all times, most of the emission below about 5500\,\AA\ is due to a forest of Fe\two\ lines (with the exception of Mg\one]\,4571\,\AA\ at 370\,d), but this emission is initially mostly from the ejecta before arising nearly exclusively from the CDS in both cases (\citealt{dessart_late_23}; Sec.~\ref{sect_form}). Similarly, the narrowish \oidoub\ at 370\,d eventually disappears before switching to a broad \oidoub\ feature arising exclusively from the CDS --- this is aggravated by the dust absorption of inner ejecta emission and the increasing $\gamma$-ray escape at later times. Emission features that appear narrow at late times likely result from the partial overlap of broad, boxy features (e.g., \niiauroral\ and \naid\ or He\one\,5875\,\AA; \citealt{dessart_late_23}).

The \niidoub\ is likely the main line contributing at 6560\,\AA\ in SN\,1993J at 370\,d (it is a partial contributor in our model of \sn\ at that time), but in both cases, \ha\ becomes the main contributor at later times. Strikingly, the width of that emission feature is essentially the same at 370\,d, but it becomes ever narrower with time in \sn, implying continued sweeping of denser CSM compared with SN\,1993J. Paradoxically, SN\,1993J, a noninteracting Type IIb SN, is more luminous in the optical than \sn, a transient with persisting interaction. This offset in luminosity is likely the result of dust attenuation, largely absent in the case of SN\,1993J within a few years of explosion (e.g., no marked blue-red asymmetry of the \ha\ profile; evidence for dust is clear at later times, decades post explosion; \citealt{bevan_dust_17}; \citealt{szalai_93J_25}).

Another important difference is the higher ionization (or distinct composition) in the spectrum-formation region of SN\,1993J, evident through the conspicuous presence of N\two\ lines and less so of \oidoub, whereas in \sn, N\two\ or He\one\ lines in the optical are weak or missing (though He\one\,10830\,\AA\ appears broad, boxy, and strong at late times; \citealt{wynn_sed_25}) and \oidoub\ emission from the CDS is obvious. Another intriguing feature is the narrow emission likely associated with \oiiidoub\ in SN\,1993J and its absence in \sn. The strength of the optical N\two\ lines is likely boosted by a strong N enrichment in SN\,1993J as inferred from CNO lines in the UV \citep{fransson_cno_93J_98S_05} --- see also Section~\ref{sect_dep}. Such differences in N enrichment between SN\,1993J and \sn\ are natural consequences of the distinct progenitor H-rich envelope properties at the time of explosion, with $\sim$\,0.1\,\msun\ in the former and $\sim$\,5\,\msun\ in the latter. With a CDS mass in both cases of a few 0.1\,\msun, its composition already reflects a contribution from the He/N shell in SN\,1993J whereas a buffer of several \msun\ still separate the CDS from the He/N shell in \sn\ at 1--3\,yr.

The profile of \ha\ has narrowed considerably in \sn, corresponding to a deceleration of the CDS by $\sim$\,1500\,\kms\ between 370 and 998\,d (see Fig.~\ref{fig_evol_ha}), which implies a significant increase of the CDS mass over that timespan --- over the same timespan the CDS has hardly decelerated in SN\,1993J and moves at about 8000\,\kms. Part of this reduction in line width also comes from dust attenuation, both within the CDS, but also from the inner ejecta since it effectively blocks emission from the fastest receding part of the ejecta as seen by a distant observer. This effect is also clearly evident in the width of the broad emission associated with \oidoub.


\section{Conclusions}
\label{sect_conc}

This work is an extension of the modeling presented by \citet{dessart_23ixf_phot_26}, which was focused on the photospheric-phase evolution of \sn, to the nebular-phase evolution at epochs between 112 and 998\,d. The core of our study consists of nonLTE time-dependent calculations with allowance for decay power as well as power from interaction with CSM, including a comparison with photometric and spectroscopic observations of \sn\ from the UV (far-UV and near-UV data from HST), optical (data from Keck and Lick Observatories, as well as Gran Telescopio Canarias), to NIR (data from Keck, MMT, and JWST), to MIR  (data from JWST). Following from \citet{dessart_23ixf_phot_26}, we used the same ejecta model. It corresponds to a 15\,\msun\ progenitor, partially stripped at core collapse and exploding to yield ejecta of 7--8\,\msun\ with $1.2 \times 10^{51}$\,erg and 0.05\,\msun\ of \nifs, and combined with a 0.2\,\msun\ CDS at 8000\,\kms. Apart from the last epochs at 600--1000\,d, all simulations employed a shuffled-shell structure in order to allow for macroscopic mixing without microscopic mixing, since this is critical to obtain reliable predictions for the cooling and thus the radiation arising from the decay-powered part of the ejecta.

This evolution from a few months to a few years of \sn\ follows in many ways that observed for SN\,2017eaw, except in \sn\ the signatures of interaction are present throughout the evolution. In the UV, it takes the form of a nonzero pseudocontinuum flux mostly due to Fe\two\ together with strong emission in \lya\ and \mgiiuv. The fractional flux emerging in \lya\ is observed to increase with time. In the optical, the persistent interaction is inferred from the presence of broad and boxy \ha\ emission, combined with an ever-decreasing narrow component arising from the inner, decay-powered part of the ejecta. After about 2\,yr, the optical radiation from \sn\ arises nearly exclusively from the CDS. In the NIR, the evidence for interaction is the presence of broad lines, and in particular that of \heinir\ (CO emission from the fundamental and first overtone are detected but not modeled here; see \citealt{park_23ixf_25} and \citealt{medler_23ixf_25}). In the MIR, most emission from the gas is due to metal lines forming in the metal-rich inner ejecta, superimposed on a bright continuum arising from internal as well as external dust. All these features, presented in previous work (e.g., \citealt{bostroem_23ixf_24,bostroem_uv_25}; \citealt{medler_23ixf_25}; \citealt{wynn_sed_25}), are well reproduced by our x6p0 model with an interaction power that progressively decreases from $4 \times 10^{40}$\,\ergs\ at 112\,d down to about $10^{40}$\,\ergs\ at 350\,d and essentially constant out to 1000\,d (within the uncertainties of our modeling).

Although the $V$-band light curve flattens only after about 600\,d in \sn, the presence of CSM interaction never ceased after shock breakout. The interaction power decreased considerably over that time, by a factor of $\sim 100$, and the eventual flattening of the light curve simply occurs when the interaction power reprocessed as optical radiation supersedes that from radioactive decay. At that time of 600\,d, our model predicts that $\sim 75$\,\% of the SN flux is emitted below 3000\,\AA, with up to 50\,\% from \lya\ alone, a picture that follows closely the predictions from \citet{dessart_late_23}.

Starting at $\sim 200$\,d, the x6p0 model with interaction power began overestimating the optical flux, something that could only be remedied by invoking dust formation within the CDS. Over the subsequent evolution, dust was introduced in growing quantities to accommodate the necessary attenuation of the optical radiation and the blue-red asymmetry of some line profiles. However, owing to the complexity of obtaining a perfect match to all emission lines, quoted dust masses are only indicative. They are sensible choices but no proof of unicity is presented. For consistency with the IR emission properties and the presence but weakness of the 10\mic\ silicate emission bump, we employed a mixture of Si-rich and C-rich, both of 0.1\mic\ grains, with a relative proportion of 1:4.5. With this choice, the dust mass below 3000\,\kms\ in \sn\ rises from 1, 2, 5, 10, 20, to $100 \times 10^{-6}$\,\msun\ at epochs 208, 265, 300, 379, 442, and 620\,d, whereas the dust mass in the CDS was 5, 10, 20, 7, 30 and $100\,\times\,10^{-6}$\,\msun\ at epochs 208, 265, 329, 379, 442, and 620\,d. With these choices, our model x6p0 with decay power, interaction power, dust, and a growing $\gamma$-ray escape, all combined lead to a satisfactory match to the $V$-band light curve of \sn, as well as the multiepoch optical spectra. Warm CDS dust was employed to match the NIR emission at 200\,d, but it became evident that the amount of dust needed to explain the IR emission exceeds the dust that can be allowed in the ejecta and CDS --- some external dust is needed.

It is not clear whether \sn\ breaks any record on dust formation in SNe since that depends on the subjective assessment of whether it should be classified as an interacting SN. \sn\ certainly interacted throughout its evolution, and thus its formation of dust as early as 200\,d and in the CDS does not make it stand apart from transients such as the interacting SN\,1998S. But these times are indeed much earlier than in SN\,1987A, which formed dust in the inner ejecta at $\gtrsim$\,500\,d \citep{lucy_dust_89}.

As emphasized earlier for dusty models of SN\,1998S \citep{dessart_dust_25}, we found that dust has little impact on the UV emission from \sn\ at 112 to 998\,d. This arises from the fact that the UV is strongly absorbed by the gas present in the ejecta and CDS so that, irrespective of dust, it tends to emerge from the outermost parts of the CDS and thus external to any dust present in the CDS or in the ejecta. Consequently, the flux deficit redward of line center observed in \lya\ or \mgiiuv\ can be explained, and is predicted (at least in our current 1D picture), from dust-free configurations. The addition of dust has a weak impact.

The metal lines present in the optical and IR spectra of \sn\ are compatible with the predictions of our model x6p0 arising from a 15\,\msun\ progenitor. Some discrepancies do arise, however, at late times, with the significant weakening of the IR lines of Ne, Ar, or Ni, whereas these lines are predicted to remain relatively strong (at least detectable) in the model. Explorations revealed that in the x6p0 model, the larger \ekin/\mej\ leads to a slightly higher ionization and stronger \neiifs, for example, relative to the s15p2 model used by \citet{dessart_24ggi_25} to model SN\,2024ggi. Lower-mass progenitors yield only a small reduction in those IR metal lines. For optical diagnostics, dust attenuation adds uncertainty in the determination of those yields.

We also explored the sensitivity of our results to the simplistic treatment of shock power in our simulations. A large fraction of the simulations presented here followed the original approach presented by \citet{dessart_csm_22} and \citet{dessart_late_23} --- that is, by injecting power in the form of high-energy electrons within the CDS and with a prescribed profile. This led to a good agreement throughout the spectrum apart from \lya, which was systematically underestimated relative to observations at all times prior to about 600\,d. Treating this power in the form of X-rays resolved the problem such that a fraction of the flux previously predicted to emerge in \mgiiuv\ was now emerging in \lya. The difference in the two methods was in part the result from adopting a more extended CDS and a larger model atom relative to \citet{dessart_late_23}, which caused here enhanced blanketing by metals as well as enhanced optical depth in \lya. These two effects are inhibited in the model with X-rays because of the ionization surge outside the CDS. Following the temperature and ionization boost on each side of the CDS, the model with X-rays predicts \civuv\ at all times, as well as \oiiidoub\ at times past 600\,d. Although [O\three] is overestimated in strength relative to \sn, its presence is indicative that such high ionization is possible in the vicinity of the reverse shock, as tentatively observed in SN\,1993J at 2--3\,yr \citep{matheson_93j_00a} or in older SNe (see, e.g., \citealt{milisavljevic_12au_18}). This exploration of the treatment of shock power emphasizes the many complications of the problem, and that securing both near-UV as well as far-UV (i.e., \lya) observations of such SNe is decisive.

In our simulations, we adopted a fixed velocity for the CDS at 8000\,\kms. This yields satisfactory matches to observed line widths for most epochs prior to 400\,d, dust contributing to the reduction of the extent of the red edge of, for example, \ha. However, past 400\,d, the model even with allowance for dust overestimates the extent of \ha. We find that the CDS decelerated from about 8000\,\kms\ at 112\,d down to 6500\,\kms\ at 998\,d, suggesting an accumulation of several 0.1\,\msun\ in the process. In the future, a better model for \sn\ will be produced allowing for a more consistent tracking of the dynamical evolution of the CDS as well as a better description of its 3D density structure (in particular, relative to clumping and power injection).


\begin{acknowledgements}

  L.D. acknowledges support from the ESO Scientific Visitor Program for a visit to ESO-Garching
  during the summer 2025.
  W.J.-G. is supported by NASA through Hubble Fellowship grant
  HSTHF2-51558.001-A awarded by the Space Telescope Science Institute,
  which is operated for NASA by the Association of Universities for
  Research in Astronomy, Inc., under contract NAS5-26555.
  K.A.B. is supported by an LSST-DA Catalyst Fellowship; this publication was
  thus made possible through the support of Grant 62192 from
  the John Templeton Foundation to LSST-DA.
  A.V.F.'s research group at UC Berkeley acknowledges
  financial assistance from the Christopher R. Redlich Fund, Gary and
  Cynthia Bengier, Clark and Sharon Winslow, Alan Eustace and Kathy
  Kwan (W.Z. is a Bengier-Winslow-Eustace Specialist in Astronomy),
  Timothy and Melissa Draper, Briggs and Kathleen Wood, Ellyn and Alan
  Seelenfreund (T.G.B. is Draper-Wood-Seelenfreund Specialist in
  Astronomy), and numerous other donors.
  We thank the U.C. Berkeley students who made contributions to the Lick/Nickel
  photometric observations: Andreas Betz, Elliot Dutnall, Daniel Gutierrez,
  Andrew McHaty, Yash Mehta, Elaina Sadler, William Wu, and Emma Yu.
  C.P.G. acknowledges financial support from the Secretary of Universities and Research
  (Government of Catalonia) and by the Horizon 2020 Research and Innovation Programme
  of the European Union under the Marie Sk\l{}odowska-Curie and the Beatriu de Pin\'os 2021
  BP 00168 programme.
  C.P.G. and L.G. recognize support from the Spanish Ministerio de Ciencia
  e Innovaci\'on (MCIN) and the Agencia Estatal de Investigaci\'on (AEI)
  10.13039/501100011033 under the PID2023-151307NB-I00 SNNEXT project, from Centro Superior
  de Investigaciones Cient\'ificas (CSIC) under the PIE project 20215AT016 and the program
  Unidad de Excelencia Mar\'ia de Maeztu CEX2020-001058-M, and from the Departament de
  Recerca i Universitats de la Generalitat de Catalunya through the 2021-SGR-01270 grant.
  Research by  S.V. is supported by NSF grant AST-2407565.

  This work was granted access to the HPC resources of TGCC under the allocation 2024 --
  A0170410554 and 2025 -- A0190416871 on Irene-Rome made by GENCI, France.
  This research has made use of NASA's Astrophysics Data System Bibliographic Services.
   Research at Lick Observatory is partially supported by a gift from Google.
  Some of the data presented herein were obtained at the W. M. Keck
Observatory, which is operated as a scientific partnership among the
California Institute of Technology, the University of California, and
NASA; the observatory was made possible by the generous financial
support of the W. M. Keck Foundation.
  Based in part on observations obtained with the Gran Telescopio Canarias (GTC),
  located at the Observatorio del Roque de los Muchachos (Instituto de Astrof\'isica de Canarias)
  on the island of La Palma, Spain, under programme GTCMULTIPLE2D-25B.

\end{acknowledgements}


\onecolumn
\appendix
\label{sect_appendix}

\section{Observations with the Gran Telescopio Canarias at 998\,d}
\label{sect_gtc_data}

The spectrum obtained on 2026-02-11 (MJD $=$\,61082.237) was acquired with the 10.4\,m Gran Telescopio Canarias (GTC) using the OSIRIS instrument. Observations were carried out with the R1000B grism and a 1.0\,arcsec slit aligned at the parallactic angle, covering a continuous wavelength range of 3630--7500\,\AA\ at a spectral resolution of $\sim$\,1100. The observations consisted of two exposures of 1200~s each, obtained at an average airmass of 1.11.
The data were reduced using a Python-based pipeline built on version 1.11.0 of \texttt{PypeIt} \citep{pypeit, pypeit_zenodo}, following standard procedures including bias subtraction, flat-field correction, wavelength calibration, sky subtraction, and flux calibration. This final spectrum was scaled by a factor of 1.95 to match the $V$-band magnitude of 20.938 obtained at MJD\,61112.

\section{Additional illustrations}
\label{sect_more}

\begin{figure*}[h!]
\centering
\includegraphics[width=0.7\hsize]{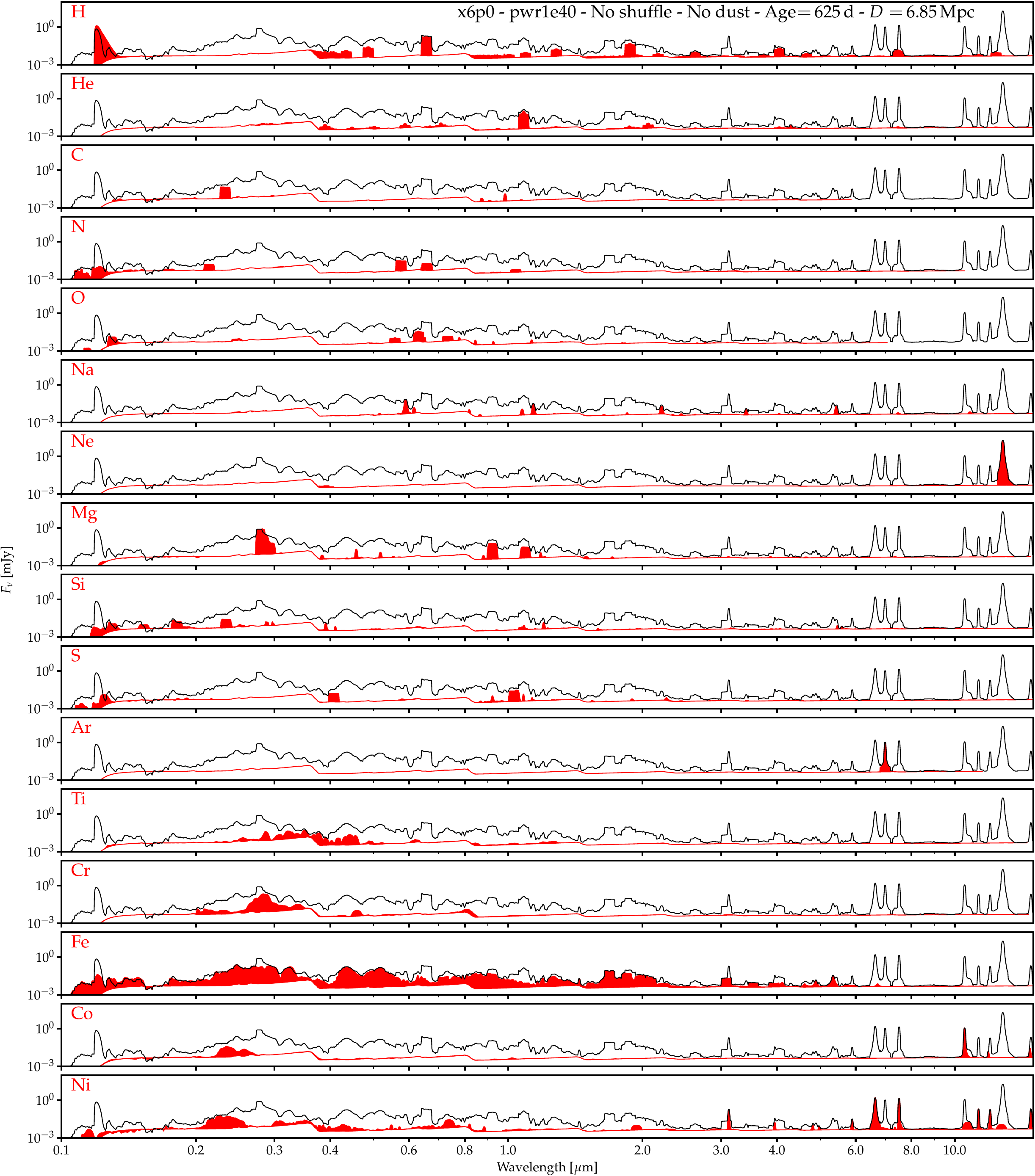}
\caption{
Breakup of species contributions to the total flux in the dust-free model x6p0 with $10^{40}$\,\ergs\ at 625\,d. We present a stack of synthetic spectra and shade in red the flux contribution from bound-bound transitions of important species (see label at top-left in each panel; this contribution also includes the continuum flux). We show the model flux $F_\nu$ (in mJy) versus wavelength (in microns) and scaled to the distance of 6.85\,Mpc.
\label{fig_species_620d}
}
\end{figure*}

\begin{figure*}[h!]
\centering
\includegraphics[width=0.7\hsize]{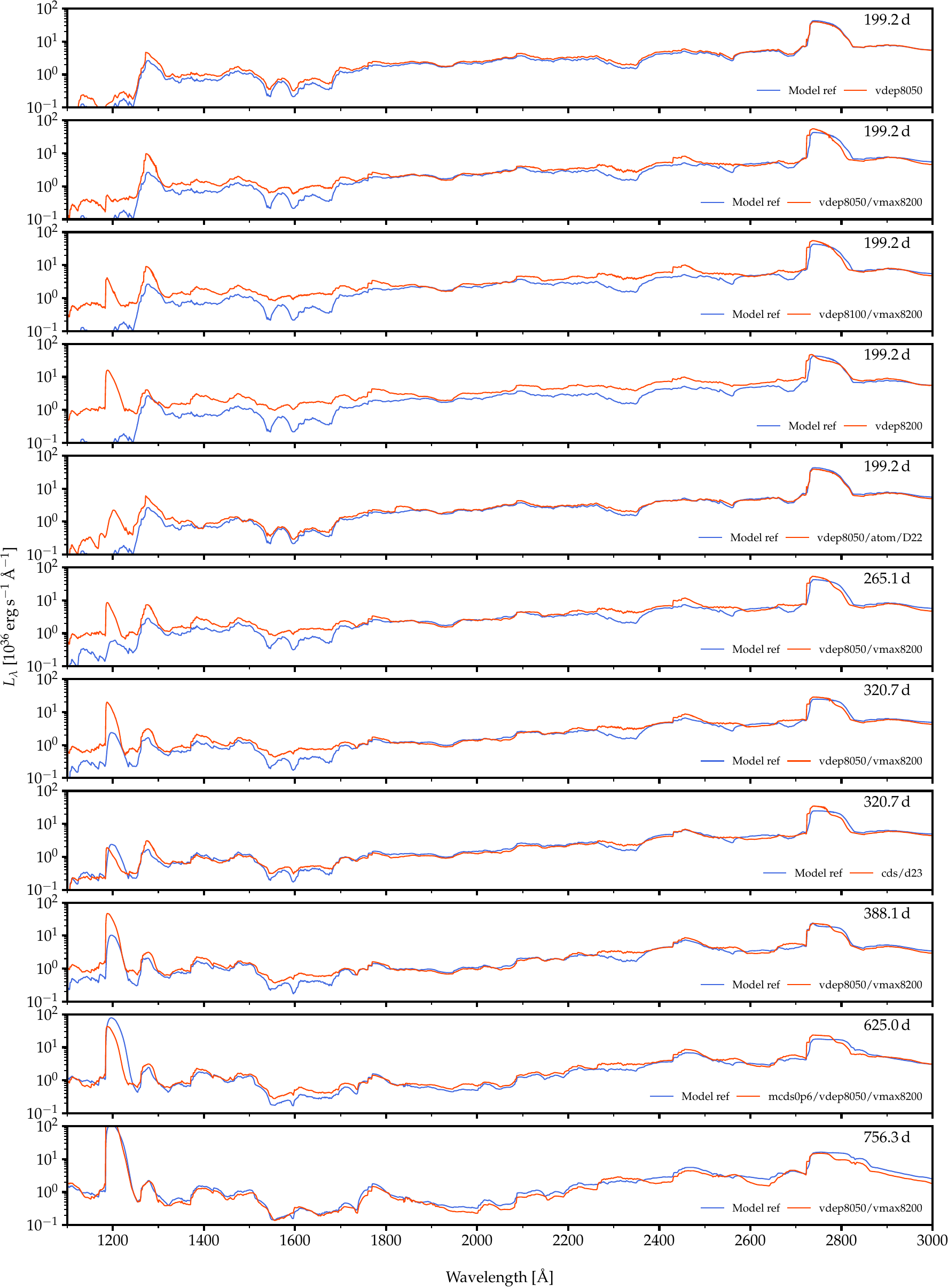}
    \caption{Comparison between the reference, dust-free models presented in Section~\ref{sect_rt} (epochs progressing downward) and variants in which we changed the velocity centroid of the power injection (``vdep''), the maximum grid velocity (``vmax''), the CDS mass (``mcds''), the structure of the CDS (using the configuration of \citealt{dessart_late_23}), or the model atom (``atom/D22'' --- i.e., reverting to that of \citealt{dessart_csm_22}). See Section~\ref{sect_dep} for discussion.
\label{fig_vdep}
}
\end{figure*}

\begin{figure*}[h!]
\centering
\includegraphics[width=0.73\hsize]{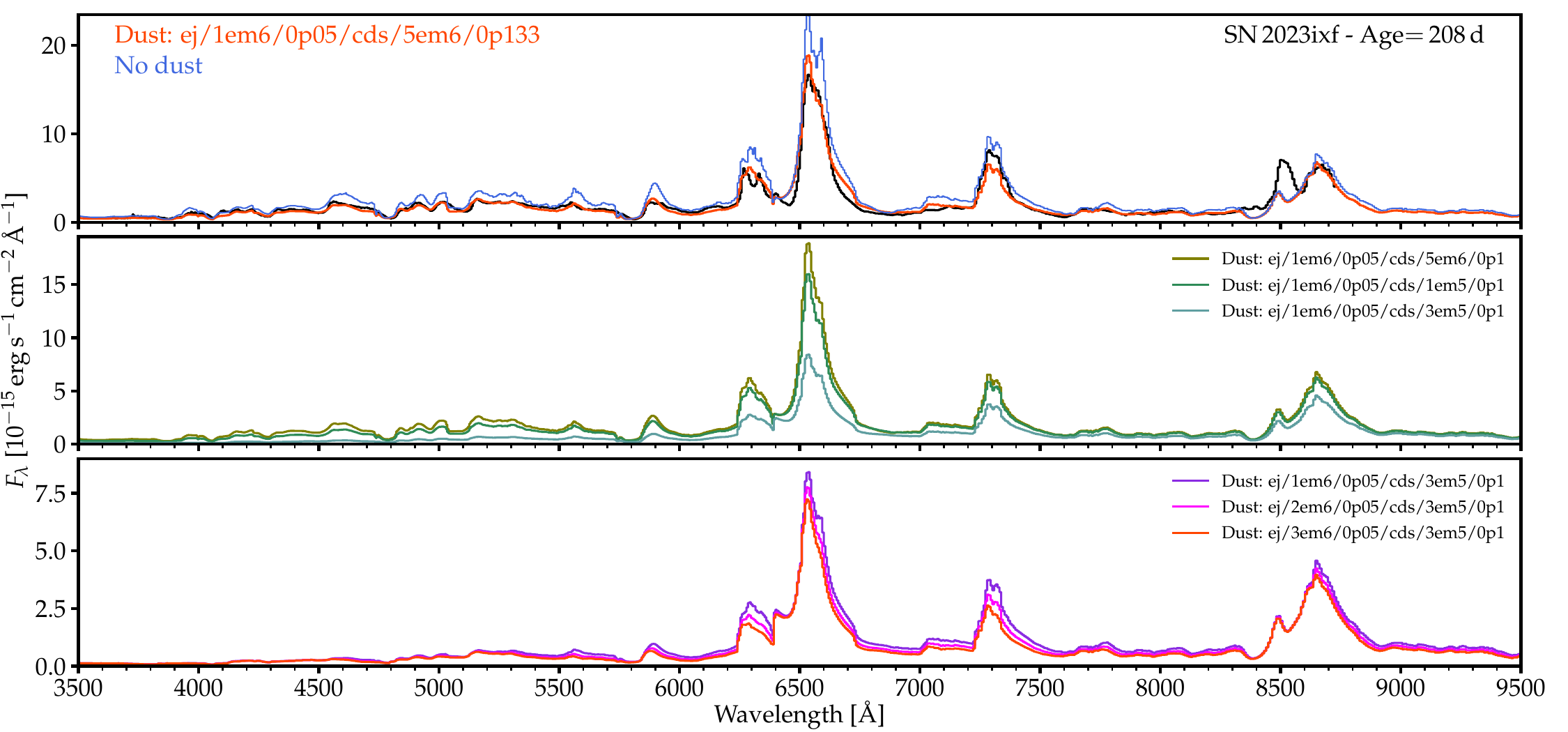}
\includegraphics[width=0.73\hsize]{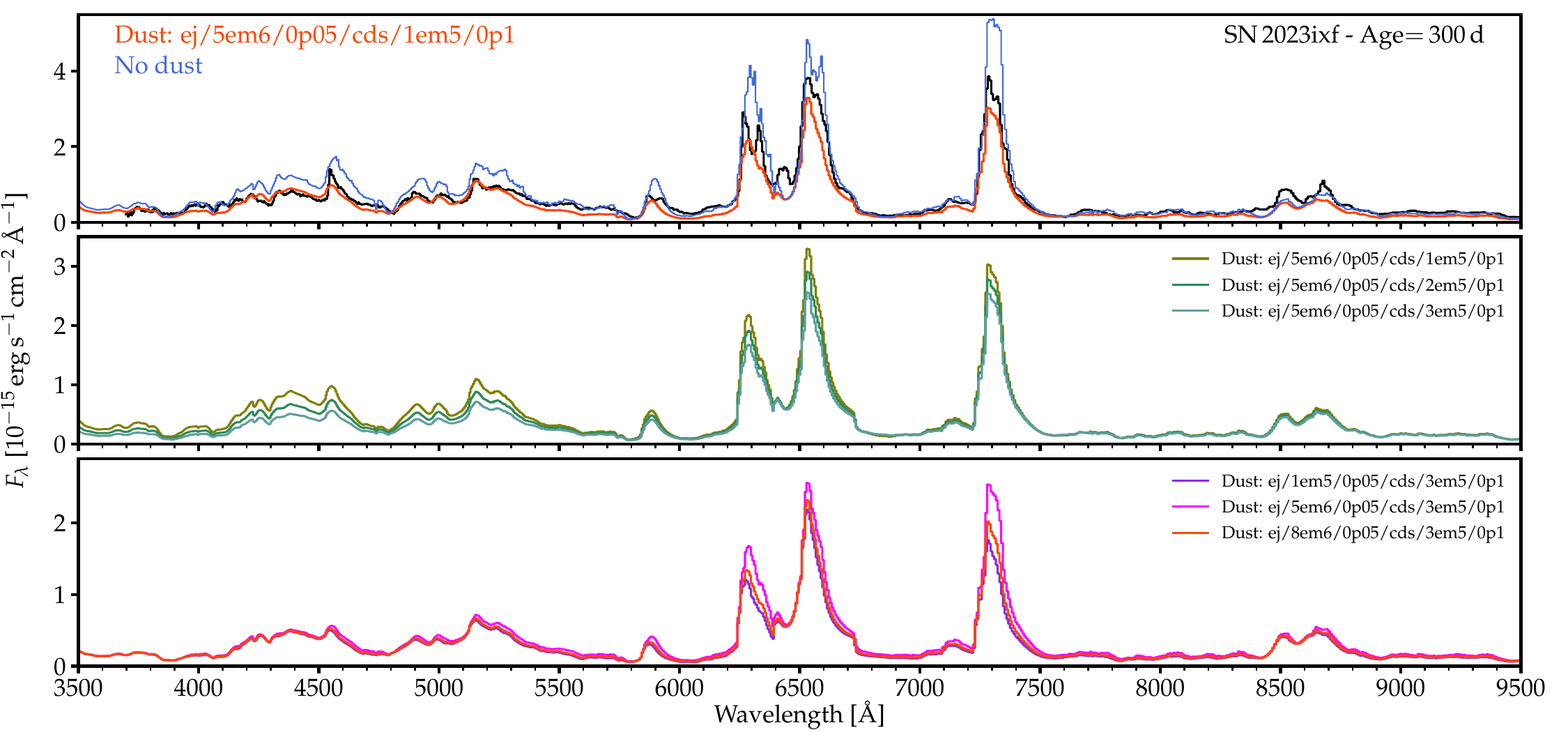}
\includegraphics[width=0.73\hsize]{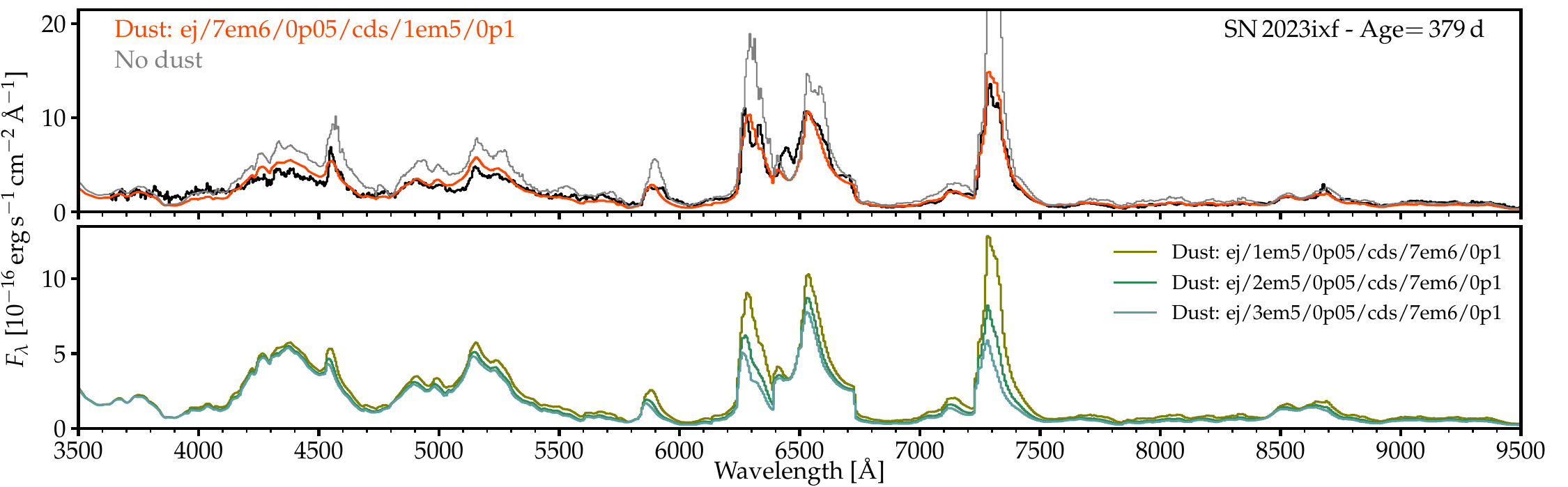}
\includegraphics[width=0.73\hsize]{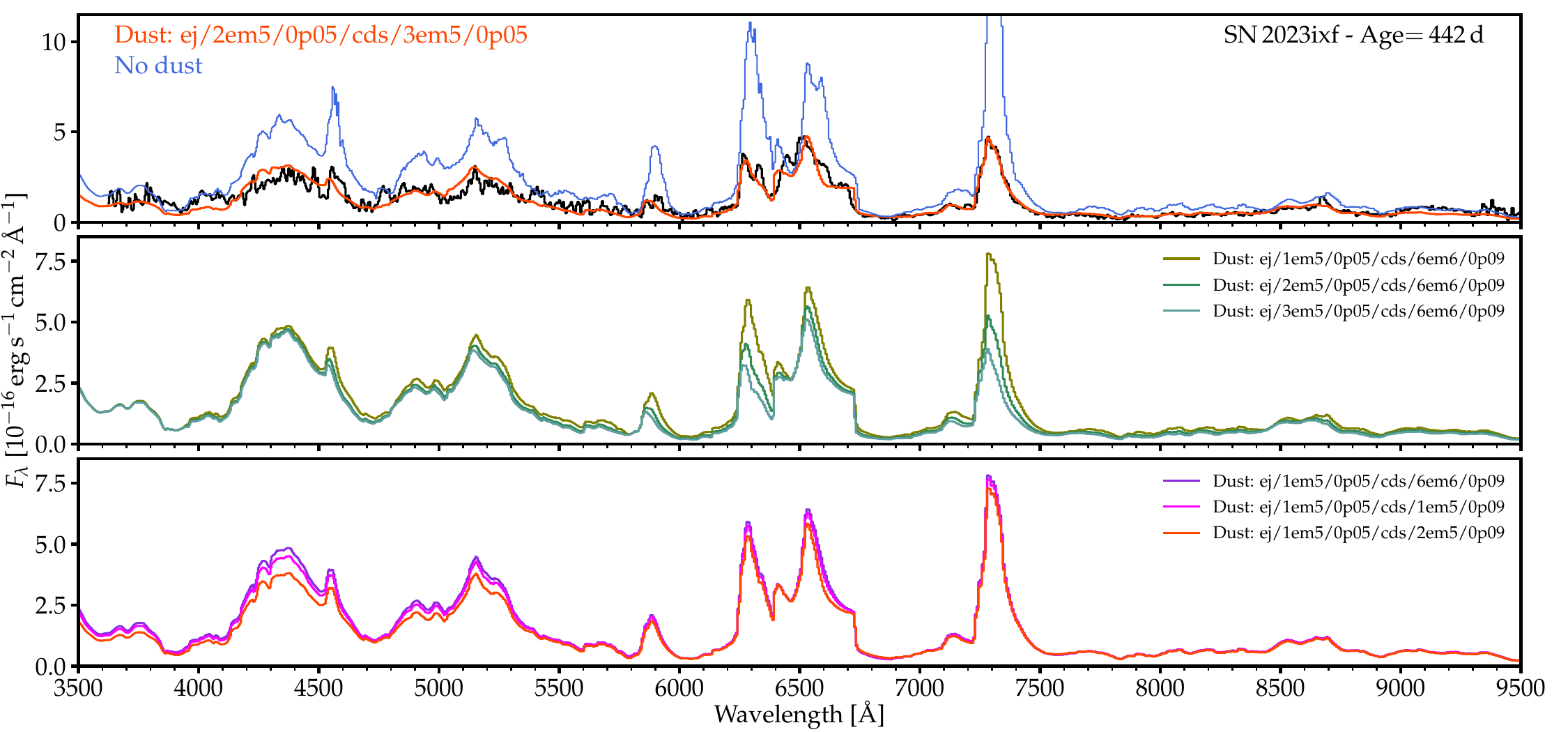}
\caption{Same as Figure~\ref{fig_dust_265d}, but now for epochs of 208, 300, 379, and 442\,d, ordered from top to bottom.
\label{fig_dust_other}
}
\end{figure*}

In this section, we present additional figures to complement the information provided in the main text. In Figure~\ref{fig_species_620d}, we illustrate the contribution to the total flux arising from individual species, accounting for all associated ionization stages treated in the model (see Sec.~\ref{sect_setup}). The spectral range covers from 1000\,\AA\ out to 15\mic. Species included in the model but bearing no significant contribution are excluded from the plot. Figure~\ref{fig_vdep} demonstrates the impact on the UV spectra of model x6p0 when the parameters that treat the injection power or the model atom are varied. We show these results for epochs between 199 and 756\,d. To complement the discussion in Section~\ref{sect_dust}, we also show the sensitivity of adopted dust parameters on the optical spectrum of the x6p0 model with interaction power at 208, 300, 379, and 620\,d in Figure~\ref{fig_dust_other}.

\end{document}